\documentclass[aps,twocolumn,amsmath,amssymb,superscriptaddress,prx,longbibliography]{revtex4-2}

\usepackage{graphicx}
\usepackage{dcolumn}
\usepackage{bm}
\usepackage{xcolor}

\usepackage{ulem}
\usepackage{amsmath,amssymb}

\begin{document}

\title{Opportunities for long-range magnon-mediated entanglement of spin qubits via on- and off-resonant coupling}


\author{Masaya Fukami}
\affiliation{Pritzker School of Molecular Engineering, University of Chicago, Chicago, Illinois, USA}

\author{Denis R. Candido}
\affiliation{Department of Physics and Astronomy, University of Iowa, Iowa City, Iowa, USA}

\author{David D. Awschalom}
\affiliation{Pritzker School of Molecular Engineering, University of Chicago, Chicago, Illinois, USA}
\affiliation{Center for Molecular Engineering and Materials Science Division, Argonne National Lab, Lemont, Illinois, USA}

\author{Michael E. Flatt\'{e}}
\affiliation{Department of Physics and Astronomy, University of Iowa, Iowa City, Iowa, USA}
\affiliation{Department of Applied Physics, Eindhoven University of Technology,  Eindhoven, Netherlands}
\date{\today}


\begin{abstract}
The ability to manipulate entanglement between multiple spatially-separated qubits is essential for quantum information processing. Although nitrogen-vacancy (NV) centers in diamond provide a promising qubit platform, developing scalable two-qubit gates remains a well-known challenge. To this end, magnon-mediated entanglement proposals have attracted attention due to their long-range spin-coherent propagation. Optimal device geometries and gate protocols of such schemes, however, have yet to be determined. Here we predict strong long-distance ($>\mu$m)  NV-NV coupling via magnon modes with cooperativities exceeding unity  in ferromagnetic bar and waveguide structures. Moreover, we explore and compare on-resonant transduction and off-resonant virtual-magnon exchange protocols, and discuss their suitability for generating or manipulating entangled states at low temperatures ($T\lesssim 150$~mK) under realistic experimental conditions. This work will guide future experiments that aim to entangle spin qubits in solids with magnon excitations. 

\end{abstract}

\maketitle
\section{Introduction}

Entanglement and quantum coherence are at the core of quantum information technologies. Among the existing qubit platforms for quantum information processing, nitrogen-vacancy (NV) centers in diamond have attracted significant attention due to their long spin-coherence time, quantum state controllability, and the ability to initialize and readout the spin state optically~\cite{jelezko2004observation,gaebel2006room,hanson2006polarization,hanson2008coherent,fuchs2009gigahertz,bar2013solid,herbschleb2019ultra}. Although there are remarkable applications of NV centers in the areas of quantum sensing and quantum communication~\cite{taylor2008high,sipahigil2012quantum,bernien2013heralded,pfaff2014unconditional,loopfree2015,reiserer2016robust,degen2017quantum,casola2018probing,awschalom2018quantum,mittiga2018imaging,humphreys2018deterministic,bartling2021,Pompili2021}, quantum computation using NV centers remains challenging due to the difficulty of engineering useful long-distance gates, \textit{i.e.} over an optically resolvable distance on the order of micrometers~\cite{jelezko2004observationN,childress2006coherent,neumann2008multipartite,neumann2010quantum,van2012decoherence,dolde2013room} which entangle qubits faster than  decoherence rates. Once this long-distance two-NV gate is established, NV centers will be a scalable platform of quantum computation enabled by their nanoscale localization and on-chip integratability~\cite{toyli2010chip}.

Recently, several potential solutions to this challenge have been proposed by making use of boson modes as an information mediator. While photon-mediated NV-NV entanglement has been experimentally demonstrated over a meter and a kilometer length scales~\cite{bernien2013heralded,pfaff2014unconditional,loopfree2015,humphreys2018deterministic,Pompili2021}, based on indistinguishable single photon detection, its extension to two-qubit gates is still challenging due to its slow entangling rate as a result of its low success probability. It has been proposed, however, that the long-distance two-qubit gates can be realized by harnessing such entangled NV-center pair generation under both single-shot readout and local gates based on the measurement outcome~\cite{Perlin_2018}. This is possible if NV centers have access to quantum memories in the decoherence-free subspace~\cite{lidar1998decoherence}, which survive during the multiple entangling attempts of NV centers that cause decoherence~\cite{reiserer2016robust,Perlin_2018,humphreys2018deterministic,bartling2021,Pompili2021}. Alternatively, as a means for extending NV-NV interaction on a wafer without needing single boson detection and with faster gate operations, hybrid quantum systems have been extensively studied where NV centers interface other bosonic systems~\cite{li2015hybrid,li2016hybrid,lemonde2018phonon,NoriPolariton2018}. In a carbon-nanotube-NV-center hybrid system~\cite{li2016hybrid}, for example, it has been proposed to couple NV centers and phonon modes in a suspended carbon nanotube by injecting an electric current through the nanotube.

\begin{figure}[b]
\includegraphics[scale=1.0]{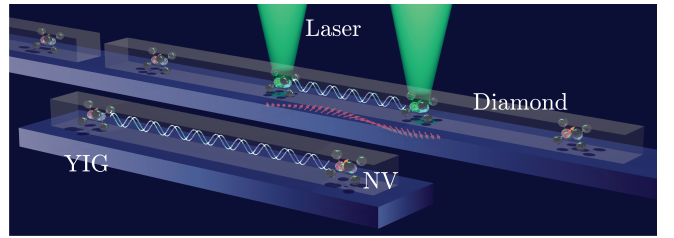}
\caption{Schematic of NV centers in diamond placed on top of an infinitely long magnon waveguide and a finite length magnetic bar made of YIG.}
\label{fig1} 
\end{figure}

Hybrid quantum systems composed of NV centers and magnons in ferromagnets have emerged and attracted attention as another highly promising platform to extend such NV-NV interaction~\cite{trifunovic2013long,flebus2018quantum,flebus2019entangling,muhlherr2019magnetic,zou2020tuning,candido2020predicted,neumanprl2020,rustagi2020,ballestero,wang,solanki}, where NV spins are intrinsically coupled to magnon modes through their dynamical fringe magnetic fields. Taking advantage of virtual-magnon exchange in one-dimensional spin chains~\cite{trifunovic2013long} or transduction of energy quanta in ferromagnetic discs~\cite{candido2020predicted}, NV-NV entanglement has been investigated theoretically~\cite{candido2020predicted,trifunovic2013long}, thus stimulating a variety of experiments on the NV-magnon hybrid system~\cite{wolfe2014off,van2015nanometre,wolfe2016spatially,andrich2017long,du2017control}. Nonetheless, optimal device geometries and gate protocols suitable for entangling separated NV centers have yet to be explored. Moreover, several important practical aspects and entangling schemes of these systems have not been fully addressed theoretically, e.g., realistic ferromagnetic structures, relevant magnetic interactions~\cite{Kalinikos_1986,stancil2009spin,serga2010yig}, finite temperatures, and possible entanglement protocols.

Here we present a practical and realistic hybrid quantum system to engineer NV-NV entanglement over micron length scales via on- and off-resonant magnon excitations at low temperatures ($T\lesssim 150$~mK). The entanglement protocol in this hybrid quantum system is based on the strong coupling of NV spins to the magnon modes in  yttrium-iron-garnet (YIG) nanodevices. Under a realistic geometry and accurately taking into account both dipole and exchange interactions, we obtain strong NV-magnon interactions and high entangling gate to decoherence ratio (GDR) in both an infinitely long YIG waveguide and a finite length YIG bar structure (see Fig.~\ref{fig1}). Especially for the latter, we obtain NV-magnon cooperativity ${\cal C}\gtrsim 10^4$ for on-resonance conditions and NV-NV GDR $\approx 10^3$ under off-resonant magnon excitations for two NV centers separated by more than $2$~$\mu$m. 
This leads to a usefully-fast entangling gate (relative to the qubit decoherence rate) at optically resolvable NV-NV separations. These values of GDR greatly exceed fidelities that were sufficient to demonstrate error correction on other platforms~\cite{errorcorrection}.
All of our results are obtained within a Hamiltonian formalism~\cite{colpa1978diagonalization,nguyen2005spectral}, which allows for semi-analytical expressions for the coupling in terms of the relevant experimental and geometrical quantities.

Finally, we explore and compare the calculated entanglement quality of both on-resonant transduction and off-resonant virtual-magnon exchange entangling gate protocols, which we regard as another major focus in this work. We achieve this comparison by means of a numerical simulation of the Lindblad master equation taking into account two NV centers and a magnon mode near the resonance condition at finite temperatures. More specifically, we analyze and compare the entanglement negativity, fidelity, and degree of the Bell inequality violation for both cases under different parameters of the NV-magnon hybrid system. Notably, our results show that although the off-resonant protocols are robust at temperatures up to $T \approx 150$~mK due to the absence of magnon occupation decay, the transduction protocol outperforms it due to its faster gate operations at lower temperatures if the magnon damping parameter is sufficiently small $\alpha \lesssim (\Delta\omega/\omega_\mu)(1/4g_\mu T_2^{*})[\pi/(\pi-1)]$, with magnon frequency $\omega_\mu$, NV center coherence time $T_2^{*}$, NV-magnon detuning frequency $\Delta \omega$, and NV-spin-magnon-mode coupling $g_\mu$.  Our calculations and analysis serve as a guide for future experiments to engineer on-chip long-distance entangling gates between NV centers mediated by magnons in ferromagnetic nanostructures.

In this article, we begin in Sec.~II with the description of the Hamiltonian formalism for the dipole-exchange magnons coupled to NV centers. In Sec.~III we calculate the full magnonic properties of a YIG  waveguide interacting with NV centers. We obtain the NV-NV coupling strength, the entanglement rate, and the gate to decoherence ratio under the off-resonant NV-magnon interaction condition. Similarly, in Sec.~IV we first calculate the magnonic properties of a finite length YIG bar. Secondly, we evaluate both NV-magnon on-resonant coupling strength and its cooperativity as well as the NV-NV coupling strength under the off-resonant condition. We provide for the latter the entanglement rate and the gate to decoherence ratio. Finally, in Sec.~V we present a complete comparison between the transduction and virtual-magnon-exchange protocols in detail under different system parameters and physical conditions.

\section{Hamiltonian formalism of dipole-exchange magnons and NV-magnon interaction}
\label{secII}

Here we outline the Hamiltonian formalism of dipole-exchange magnons coupled to NV centers providing a complete and accurate  treatment of both magnetic dipole and quantum exchange interactions between the spins in YIG waveguides and bars with finite cross section. This is crucial in our study as the NV centers have eigenfrequencies typically on the order of gigahertz, thus interacting with the so-called dipole-exchange magnons in ferromagnets~\cite{Kalinikos_1986}; using simpler, less accurate magnon dispersion relations as in Ref.~[\onlinecite{trifunovic2013long}] leads to a substantial overestimation of the NV-magnon coupling. As illustrated in Fig.~\ref{fig1}, we consider hybrid quantum devices where NV centers are placed on top of the YIG structures. Whereas multiple NV centers can be placed on top of the infinitely long YIG waveguide in a scalable fashion as shown in Fig.~\ref{fig1}, in the following calculations we only focus on coupling two NV centers. The total Hamiltonian of our hybrid system is written as $\mathcal{H}=\mathcal{H}_\text{NV}+\mathcal{H}_\text{m}+\mathcal{H}_\text{int}$, where $\mathcal{H}_\text{NV}$ is the NV Hamiltonian, $\mathcal{H}_\text{m}$ is the magnon Hamiltonian, and $\mathcal{H}_\text{int}$ is the interaction Hamiltonian,
\begin{eqnarray}
&&\mathcal{H}_\mathrm{NV}=\sum_{i=1,2} D_{\mathrm{NV}}\left(\hat{n}_{\mathrm{NV}} \cdot \mathbf{S}_{\mathrm{NV}_{i}}\right)^{2}+\gamma \mu_{0} \mathbf{S}_{\mathrm{NV}_{i}} \cdot \mathbf{H}_{\mathrm{ext}},\\
&&\mathcal{H}_{\mathrm{m}}=-\mu_{0} \int d \mathbf{r} \mathbf{H}_{\mathrm{ext}} \cdot \mathbf{M} (\mathbf{r})+\frac{\mu_{0}}{2} \int d \mathbf{r} \alpha_\mathrm{ex}(\mathbf{r}) \nabla \mathbf{M}: \nabla \mathbf{M}\nonumber\\
&&\ \ \ \ \ \ \ \ +\frac{\mu_{0}}{2} \int d \mathbf{r} d \mathbf{r}^{\prime}(\nabla \cdot \mathbf{M}(\mathbf{r})) G\left(\mathbf{r}-\mathbf{r}^{\prime}\right)\left(\nabla^{\prime} \cdot \mathbf{M}\left(\mathbf{r}^{\prime}\right)\right),\label{Hmag}\\
&&\mathcal{H}_{\mathrm{int}}=\sum_{i=1,2} \gamma \mu_{0} \mathbf{S}_{\mathrm{NV}_{i}} \cdot \left.\nabla \int d \mathbf{r}^{\prime} G\left(\mathbf{r}-\mathbf{r}^{\prime}\right) \nabla'\cdot \mathbf{M}\left(\mathbf{r}^{\prime}\right)\right|_{\mathbf{r}=\mathbf{r}_{i}}.\nonumber\\\label{Hint}
\end{eqnarray}
Here, $D_\text{NV}=2\pi\times2.877\text{ GHz}$ is the zero-field splitting of the NV center, $\hat{n}_\text{NV}$ is the unit vector along the NV main symmetry axis, $\mathbf{S}_{\text{NV}_i}$ is the spin-$1$ operator of the NV center labeled by $i\in\{1,2\}$, $\gamma=2\pi\times28\text{ MHz/mT}$ is the absolute value of the electron gyromagnetic ratio, $\mu_0$ is the vacuum permeability, $\bf{H}_\text{ext}$ is the external magnetic field, $\bf{M}(\bf{r})$ is the magnetization with the constraint $|\mathbf{M}(\mathbf{r})|=M_s(\mathbf{r})=M_s\mathcal{F}(\mathbf{r})$, $M_s=245.8\ \mathrm{mT}/\mu_0$ is the YIG saturation magnetization, $\mathcal{F}(\mathbf{r})=1$ ($0$) inside (outside) the ferromagnetic structure, ${\alpha_\mathrm{ex}}(\mathbf{r})={\alpha_\mathrm{ex}}\mathcal{F}(\mathbf{r})$, ${\alpha_\mathrm{ex}}=\lambda^2_\text{ex}=D_\text{ex}/\gamma\mu_0M_s$ is the exchange-length squared, $D_\text{ex}=5.39\times 10^{-2}\ \gamma\ \mathrm{mT}\ \mu\mathrm{m}^2$ is the YIG exchange constant, the double-dot product is defined as $\nabla{\bf{M}}:\nabla{\bf{M}}=\partial_a M_b\partial^a M^b$, ${\bf{r}}_i$ is the position of $\text{NV}_i$, $G(\mathbf{r}-\mathbf{r}')=1/4\pi|\mathbf{r}-\mathbf{r}'|$ is the Green's function, and we set $\hbar=1$. We note that the first term in Eq.~(\ref{Hmag}) is the Zeeman energy, the second term is the exchange energy, and the third term is the magnetic dipole energy. Inclusion of both the second and the third term in Eq.~(\ref{Hmag}) results in the dipole-exchange magnons in ferromagnets.

\section{Infinitely long ferromagnetic waveguide}
\label{secIII}
\begin{figure}[t]
\includegraphics[scale=1]{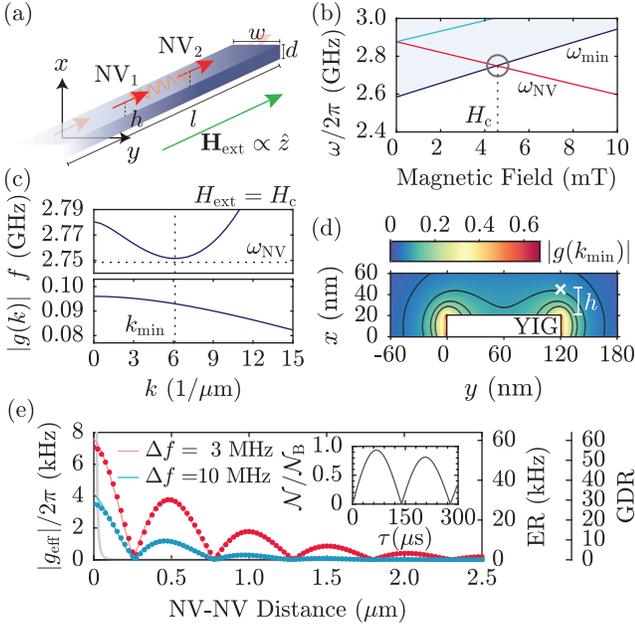}
\caption{(a) Schematic and coordinates of NV centers placed on top of an infinitely long YIG waveguide with applied external magnetic field $\bf{H}_\text{ext}$. (b) NV center's transition frequencies and magnon spectrum as a function of external field $H_\text{ext}$ for $d=20$~nm and $w=120$~nm. Shaded area represents continuum of magnon modes. The lowest magnon frequency $\omega_\text{min}$ and the NV transition frequency $\omega_\text{NV}$ of $|g\rangle\leftrightarrow|e\rangle$ are detuned by $\Delta f=3\text{ MHz}$ at $H_\text{ext}=H_\text{c}$.  (c) Dispersion relation $f(k)=\omega_{k,(0,0)}/2\pi$ of magnons and the dimensionless coupling $g(k)=g(\bm{\rho},k)$ between magnons and the NV center at $H_\text{ext}=H_\text{c}$. The NV center is positioned at $\bm{\rho}=(x,y)=(d+h,w)$ with $h=25\text{ nm}$ [see the white cross mark in (d)]. The minimum frequency $\omega_\text{min}$ and its respective wavenumber $k_\text{min}$ are shown. (d) Spatial density plot of the dimensionless coupling $g(k_\text{min})$ at $H_\text{ext}=H_\text{c}$ with contours at $|g(k_\text{min})|=0.05,0.1,0.15$ and $0.2$. (e) Effective NV-NV coupling strength $g_\text{eff}$ [Eq.~(\ref{EffInt})] as a function of the NV-NV distance under $\Delta f=3\text{ MHz}$ and $\Delta f=10\ \mathrm{MHz}$. The gray curve shows the coupling due to the direct magnetic dipole-dipole interaction between NV centers. The entanglement rate and the gate to decoherence ratio are shown on the right axis for $T_2^*=1\text{ ms}$. Inset shows the time $\tau$ evolution of the entanglement negativity at $T=0$ from the initial state $|g\rangle_1|e\rangle_2$ scaled by the Bell state negativity $\mathcal{N}_\text{B}$.
}
\label{fig2} 
\end{figure}

Here we consider the case of an infinitely long YIG waveguide with thickness, width, and length given by $d$, $w$, and $l(\rightarrow\infty)$, respectively. The external magnetic field is applied along the YIG waveguide, $\mathbf{H}_\text{ext}=H_\text{ext}\hat{z}$, and NV centers are positioned at height $h$ from its top surface [see illustration in Fig.~\ref{fig2}(a)]. The equilibrium magnetization is $\mathbf{M}_0(\mathbf{r})=M_s\hat{z}\mathcal{F}(\mathbf{r})$, for which its contribution in the interaction Hamiltonian Eq.~(\ref{Hint}) vanishes. The NV main symmetry axis is set to be parallel to the external magnetic field, $\hat{n}_\text{NV}=\hat{z}$, for geometrical simplicity. We further define the deviation from the equilibrium magnetization $\delta\mathbf{M}(\mathbf{r})=\mathbf{M}(\mathbf{r})-\mathbf{M}_0(\mathbf{r})\approx\mathbf{m}(\mathbf{r})-[|\mathbf{m}(\mathbf{r})|^2/2M_s(\mathbf{r})]\hat{z}$, where $\mathbf{m}(\mathbf{r})=m_x(\mathbf{r})\hat{x}+m_y(\mathbf{r})\hat{y}$ is a small two-dimensional magnetization deviation. The linearized magnetization dynamics~\cite{shindou2013topological} are governed by the Hamiltonian equation of motion for $m^-(\mathbf{r})= [2\gamma M_s(\mathbf{r})]^{1/2}a(\mathbf{r})$ and $m^+(\mathbf{r})= [2\gamma M_s(\mathbf{r})]^{1/2}a^*(\mathbf{r})$ using the magnon Hamiltonian $\mathcal{H}_\mathrm{m}$ up to  quadratic order in the complex canonical variables $a(\mathbf{r})$ and $a^*(\mathbf{r})$, where we have performed the Holstein-Primakoff approximation~\cite{stancil2009spin} and $m^{\pm}(\mathbf{r})=m_x(\mathbf{r})\pm i m_y(\mathbf{r})$.

To obtain the normal magnon mode frequencies and the dynamical fringe field spatial profiles, we diagonalize the magnon Hamiltonian Eq.~(\ref{Hmag}) by expanding the complex canonical variables assuming totally unpinned surface spins, i.e.,
\begin{equation}
a(\mathbf{r})=\int\frac{dk}{2\pi}e^{ikz}\sum_{nm}f_n^X(x)f_m^Y(y)a_{k,(n,m)}.
\end{equation}
Here, the basis functions are
\begin{eqnarray}
f_n^X(x)&=&\left[\frac{2\mathcal{F}^X(x)}{(1+\delta_{n,0})d}\right]^{\frac{1}{2}}\cos(\kappa_n^Xx),\\ f_m^Y(y)&=&\left[\frac{2\mathcal{F}^Y(y)}{(1+\delta_{m,0})w}\right]^{\frac{1}{2}}\cos(\kappa_m^Yy),
\end{eqnarray}
where $\kappa_n^X=n\pi/d$, $\kappa_m^Y=m\pi/w$, $\mathcal{F}^X(x)=\Theta(x)\Theta(d-x)$, $\mathcal{F}^Y(y)=\Theta(y)\Theta(w-y)$,  and $\Theta$ is the Heaviside step function. As we consider the case where both the thickness and the width of the YIG waveguide are small, we restrict our discussion to the magnon mode subspace with $(n,m)=(0,0)$, which presents uniform magnetization deviations across the $x$-$y$ plane and gives the lowest energy magnon band in the dispersion relation.%

After writing $\mathcal{H}_\mathrm{m}$ up to the quadratic order in the complex canonical variables, applying the Bogoliubov transformation, and promoting the complex canonical variables to the quantum creation and annihilation operators, we obtain the diagonalized Hamiltonian (see Appendix B1)
\begin{eqnarray}
\mathcal{H}_\mathrm{m}=\int\frac{dk}{2\pi}\omega_{k,(0,0)}\beta^\dag_{k,(0,0)}\beta_{k,(0,0)},
\end{eqnarray}
where $\omega_{k,(0,0)}$ is the magnon energy and $\beta_{k,(0,0)}$ ($\beta_{k,(0,0)}^{\dagger}$) is the magnon annihilation (creation) operator satisfying $[\beta_{k,(0,0)},\beta^\dag_{k',(0,0)}]=2\pi\delta(k-k')$.

The coupling strength between magnon modes and NV centers can be obtained by applying the same Bogoliubov transformation in the interaction Hamiltonian Eq.~(\ref{Hint}). As we focus on external magnetic field values $\gamma H_\text{ext}<D_\text{NV}$, the NV center's ground state and the first excited state are $|g\rangle=|S_\text{NV}^z=0\rangle$ and $|e\rangle=|S_\text{NV}^z=-1\rangle$, respectively. Up to the linear order in magnon creation and annihilation operators and using the rotating wave approximation ($|\omega_{k,(0,0)}-\omega_{\rm{NV}} | \ll \omega_{k,(0,0)}+\omega_{\rm{NV}}$), we obtain the interaction Hamiltonian (see Appendix B2)
\begin{eqnarray}
\mathcal{H}_\mathrm{int}=\frac{\sqrt{\omega_M\omega_d}}{\sqrt{w/d^2}}\sum_{i=1,2}\int\frac{dk}{2\pi}g({\bm{\rho}}_i,k)\sigma_{\mathrm{NV}_i}^+\beta_{k,(0,0)}e^{ikz_i}+\mathrm{H.c.},\nonumber\\
\label{Hint0}\end{eqnarray}
in the NV centers' subspaces spanned by $\{|g\rangle_{i},|e\rangle_{i} \}$, where $\omega_M=\gamma\mu_0M_s$,  $\omega_d=\mu_0\gamma^2/d^3$, $g({\bm{\rho}}_i,k)$ is the dimensionless coupling between the NV center spin and the $k$-magnon mode, ${\bm{\rho}}_i$ is the $\mathrm{NV}_i$'s position in the $x$-$y$ plane, $\sigma^+_{\mathrm{NV}_i}=|e\rangle_i  \langle g|$, and $\sigma^-_{\mathrm{NV}_i}={(\sigma^+_{\mathrm{NV}_i})^{\dagger}}$. The virtual-magnon-mediated NV-NV interaction can be obtained via the Schrieffer-Wolff transformation~\cite{bravyi2011schrieffer} as $\mathcal{H}^{\mathrm{NV}-\mathrm{NV}}_{\mathrm{eff}}=-\left(g_\text{eff}\sigma_{\mathrm{NV}_{1}}^{+} \sigma_{\mathrm{NV}_{2}}^{-}+\text{H.c.}\right)$ with (see Appendix B3)
\begin{eqnarray}\label{GeffWG}
g_\text{eff}=\frac{\omega_{M} \omega_{d}}{w/d^2} \int \frac{d k}{2 \pi}\left|g(k)\right|^{2} \frac{\exp [ik(z_1-z_2)]}{\omega_{k,(0,0)}-\omega_{\mathrm{NV}}},\label{EffInt}
\end{eqnarray}
where $g_\mathrm{eff}$ is the effective NV-NV coupling strength, $\omega_\mathrm{NV}=D_\mathrm{NV}-\gamma H_\mathrm{ext}$ is the transition frequency of $|g\rangle\leftrightarrow|e\rangle$, and we write $g(k)=g({\bm{\rho}}_i,k)$ assuming ${\bm\rho}_1={\bm{\rho}}_2$. The above expression is valid when $(\omega_{M}\omega_{d}d^2/2\pi w)\int dk |g(k)|^2(\omega_{k,(0,0)}-\omega_{\rm{NV}})^{-2} \ll 1$. We note that this effective coupling strength $g_{\rm{eff}}$ for the off-resonant configuration does not depend on the temperature, as it is independent of the initial magnon number state $|n_{\rm{m}} \rangle$ (\textit{i.e.} from second order perturbation theory) even though the NV-magnon coupling strength matrix element is proportional to $\sqrt{n_{\rm{m}} +1}$ (see Appendix B4).

In Fig.~\ref{fig2}(b) we plot the NV center's transition frequencies and magnon mode frequencies as a function of the external magnetic field $H_\text{ext}$, where we have assumed $(d,w)=(20\text{ nm}, 120\text{ nm})$ for the waveguide dimensions~\cite{wang2019spin}. As we take the limit where the length of the YIG waveguide is infinity ($l\rightarrow\infty$), the magnon mode frequencies form a continuum with its minimum denoted as $\omega_\text{min}$. At field $H_\text{ext}=H_c$, the NV center's lower transition frequency $\omega_\text{NV}$ is detuned from the magnon dispersion minimum $\omega_\text{min}$  by $\Delta\omega=\omega_\text{min}-\omega_\text{NV}=2\pi\Delta f=2\pi\times 3\text{ MHz}$. Figure~\ref{fig2}(c) shows the magnon dispersion relation near $\omega_\text{min}$ and the wavenumber dependence of the dimensionless coupling strength $g(k)$ at $H_\text{ext}=H_\text{c}$, ${\bm \rho}_i=(d+h)\hat{x}+w\hat{y}$, and $h=25\text{ nm}$ [see the cross marker in Fig.~\ref{fig2}(d)]. The coupling strength also depends on the spatial position of the NV center relative to the YIG waveguide, which is shown in Fig.~\ref{fig2}(d). As the dynamical fringe magnetic field generated by a single magnon is confined near the YIG device, the coupling strength is larger if the NV center is positioned near the YIG waveguide.

Under the off-resonant condition shown in Fig.~\ref{fig2}(c), the NV centers on top of the YIG waveguide interact to each other via the exchange of virtual magnons. In Fig.~\ref{fig2}(e), we plot the effective NV-NV coupling strength $g_\text{eff}$ [Eq.~(\ref{GeffWG})] as a function of the NV-NV distance \hbox{$\delta z=|z_1-z_2|$} for both $\Delta f=3$~MHz and $\Delta f=10$~MHz cases represented by the red and blue dots, respectively. The coupling decays rapidly with detuning, which allows the entangling interaction to be switched off by increasing the external magnetic field from $H_{\rm{ext}}=H_c$ by $\approx$ $0.1$ mT. We show that the calculated coupling strength is well explained by the analytical formula
\begin{equation}
g_\text{eff}\approx\frac{\omega_M\omega_{\bar{d}}}{\Delta\omega}|g(k_\text{min})|^2\cos(k_\text{min}\delta z)e^{-\delta z/\xi_0}
\end{equation}
as shown by the solid red and blue curves in Fig.~\ref{fig2}(e), where $\xi_\text{0}=\sqrt{D_\text{ex}/\Delta\omega}$ is the spin correlation length and $\omega_{\bar{d}}=\mu_0\gamma^2/(\xi_0 wd)$. The entangling gate rate $\mathrm{ER}=4g_\mathrm{eff}/\pi$ and the gate to decoherence ratio $\mathrm{GDR}=4g_\mathrm{eff}T_2^*/\pi$ are shown on the right axis, where a coherence time $T_2^*=1\text{ ms}$ of the NV center is used~\cite{herbschleb2019ultra}. As we obtain $\mathrm{GDR}>10$ for $1\ \mu\text{m}$ separated NV centers, we predict a useful and practical entangling gate. 

To show that this system can manipulate the NV-NV entanglement, we perform a simulation using the Lindblad master equation. In the inset of Fig.~\ref{fig2}(e) we plot the entanglement negativity~\cite{vidal2002computable} $\mathcal{N}$ at $T=0$ as a function of the NV-NV interaction time after the preparation of the initial spin state in $|g\rangle_1|e\rangle_2$, where the negativity is normalized by the Bell state's negativity $\mathcal{N}_\mathrm{B}$. As we obtain $\mathcal{N}>0$, we clearly demonstrate that the NV centers are entangled. If multiple NV centers are placed on top of the YIG waveguide (see Fig.~\ref{fig1}), neighboring two-NV gates can thus be performed by locally changing the external magnetic field around the two NV centers to shift their transition frequencies relative to the minimum magnon mode frequency in the range $\Delta\omega>0$. Alternatively, local electric field~\cite{electricfieldNV2011} or strain~\cite{PhysRevLett.113.020503} can be used to shift NV centers' transition frequencies to avoid applying a local magnetic field at the underlying YIG location, the effect of which is discussed in Appendix K.

In Fig.~\ref{fig3} we plot the NV-NV entanglement rate and the gate to decoherence ratio as a function of the waveguide thickness $d$ for different waveguide dimensions and NV centers' heights $h$. We assume a fixed NV-NV distance of $1\ \mu\text{m}$, $(x_i,y_i)=(d+h,w)$, and  $\Delta\omega=2\pi\times3\text{ MHz}$. The red (blue) solid curve shows the waveguide thickness $d$ dependence of the $\mathrm{ER}$ and the $\mathrm{GDR}$ under the fixed aspect ratio $w/d=6$ at $h=25\text{ nm}$ ($5\text{ nm}$), and the red (blue) dashed curve shows the dependence where the waveguide width is kept constant with $w=120\text{ nm}$ at $h=25\text{ nm}$ ($5\text{ nm}$). From these graphs we see that in order to make the entangling gate faster, one can either have the NV center closer to the YIG waveguide (diminishing $h$) or make the waveguide's cross-sectional area smaller. As for placing NV centers in proximity to the YIG waveguide, we note the common challenge of making high coherence NV centers near the diamond surface due to the surface noise known in the area of NV-based quantum sensing~\cite{ohno2012engineering}.

\begin{figure}[t]
\includegraphics[scale=1]{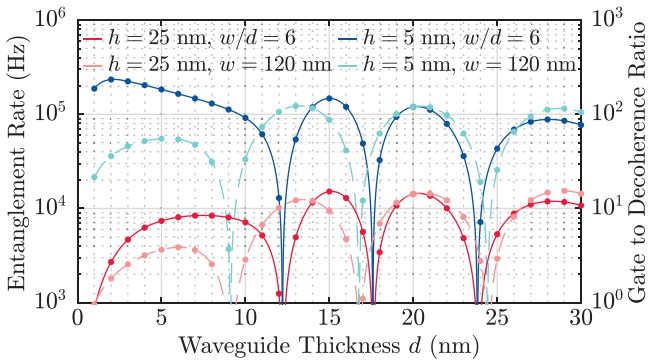}
\caption{The entanglement rate and the gate to decoherence ratio between two NV centers separated by $1\ \mu\text{m}$ as a function of the waveguide thickness $d$. NV centers are placed on the YIG waveguide as drawn in Figs. 2(a) and 2(d). Red curves and blue curves are calculated for $h=25 \text{ nm}$ and $5 \text{ nm}$, respectively. Solid curves and dashed curves are calculated for a fixed aspect ratio $w/d$ and width $w$ of the waveguide, respectively. Sharp dips correspond to the nodes in the oscillation of $g_\text{eff}$ as shown in Fig.~\ref{fig2}(e). Calculation is performed for detuning $\Delta\omega/2\pi=3 \text{ MHz}$.
}
\label{fig3} 
\end{figure}

\section{Finite length ferromagnetic bar}
\label{secIV}
In this section we show that the NV-magnon coupling strength can be strongly enhanced under the magnon confinement effect of a finite length ferromagnetic bar. As the magnon mode frequencies are discretized for this case, the system allows us to control the NV levels to be on- and off-resonant to the magnon levels. Here, the interaction Hamiltonian Eq.~(\ref{Hint0}) can be transformed into the form of the Jaynes-Cummings model~\cite{candido2020predicted,raimond2006exploring}, and the entangling gate schemes used in both quantum optics and circuit quantum electrodynamics can now be implemented in our hybrid quantum system~\cite{sillanpaa2007coherent,ansmann2009violation,manovitz2017fast}.

We first obtain the NV-magnon interaction Hamiltonian for a finite length YIG bar using a similar procedure as done in Sec.~III. For that, we first take the equilibrium magnetization to be $\mathbf{M}_0=M_s\mathcal{F}(\mathbf{r})\hat{z}$ and approximate the $x,y$ component of the resulting static demagnetization field in Eq.~(\ref{Hmag}) to be negligible compared to its $z$ component. Although there is also a finite static demagnetization field contribution in the interaction Hamiltonian~Eq.~(\ref{Hint}), we verified that its value is small under the geometry parameters and NV center positions we consider.

Accordingly, we diagonalize the magnon Hamiltonian through the following expansion of the complex canonical variable 
\begin{equation}
a(\mathbf{r})=\sum_{nmp}f_n^X(x)f_m^Y(y)f_p^Z(z)a_{(nmp)},
\end{equation}
where the $z$-directional basis function is \begin{equation}
f_p^Z(z)=\left[\frac{2\mathcal{F}^Z(z)}{(1+\delta_{p,0})l}\right]^{\frac{1}{2}}\cos(\kappa_p^Zz),
\end{equation}
$\kappa_p^Z=p\pi/l$, and $\mathcal{F}^Z(z)=\Theta(z)\Theta(l-z)$. As we consider the case with $d,w\ll l$, we restrict our discussion to the magnon mode subspace with $(n,m)=(0,0)$. Considering $z$-directional modes with $p=0,1,\cdots,N$, where $p=N$ labels the highest $z$-directional wavenumber mode to be taken into account, and keeping terms up to the quadratic order in the complex canonical variables, we obtain a $2(N+1)\times2(N+1)$ non-diagonal quadratic boson Hamiltonian. After applying the Bogoliubov transformation with the paraunitary matrix~\cite{colpa1978diagonalization,shindou2013topological} and promoting the complex canonical variables to the quantum creation and annihilation operators, we obtain (see Appendix C1)
\begin{equation}
\mathcal{H}_\mathrm{m}=\sum_{p=0,1,\cdots}\omega_{(00p)}\beta^\dag_{(00p)}\beta_{(00p)}.
\end{equation}
In a similar way as in Sec.~III, the NV-magnon interaction Hamiltonian can be mapped into the form of the Jaynes-Cummings model~\cite{candido2020predicted,raimond2006exploring} (see Appendix C2)
\begin{equation}
\mathcal{H}_\mathrm{int}=\sum_{i=1,2}\sum_{\mu=(00p)}g_\mu(\mathbf{r}_i)\sigma_{\mathrm{NV}_i}^+\beta_\mu+\mathrm{H.c.},
\end{equation}
where $g_\mu(\mathbf{r}_i)\propto\sqrt{\omega_M\omega_{dwl}}$ [$\omega_{dwl}=\mu_0\gamma^2/(dwl)$] is the coupling strength between the NV center spin and the $\mu$-magnon mode in the unit of energy. {As the magnon creation operator ${\beta_\mu^{\dagger}}$ applied to the magnon number state $|n_{\mu} \rangle$ gives rise to a factor of $\sqrt{n_{\mu}+1}$, we expect the on-resonant NV-magnon configuration to have $\sqrt{n_{\mu}+1}$ faster energy-transfer oscillations between the NV-center spin and the $\mu$-magnon mode. However, at finite temperature, which can be thought of as a statistical mixture of different magnon-number states, these different-period oscillations will average out incoherently. Therefore, finite temperature does not improve the quality of NV-NV entanglement via magnon modes even though the mean magnon number $\langle n_{\mu}\rangle$ is larger, indicating that  magnon-mediated NV-NV entanglement needs to be performed at low temperatures $T \lesssim 150 $~mK (see Sec.~V).}

\begin{figure}[t]
\includegraphics[scale=1]{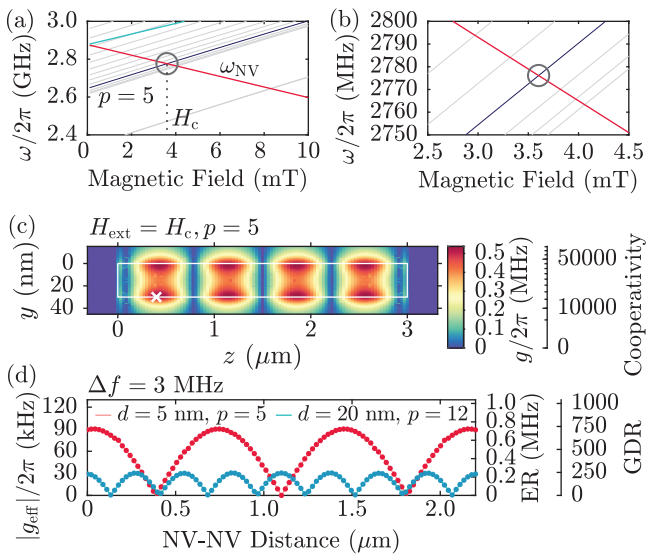}
\caption{(a) NV center's transition frequencies and magnon spectrum as a function of external field $H_\text{ext}$ for $(d,w,l)=(5\text{ nm},30\text{ nm},3\ \mu\text{m})$. The dark gray and red lines represent frequencies $\omega_{(00,p=5)}$ and $\omega_\mathrm{NV}$, respectively. (b) Zoom-in of the crossing region between $\omega_{(005)}$ and $\omega_\text{NV}$. (c) Spatial plot of the coupling strength $g=g_{(005)}$ at $H_\text{ext}=H_\text{c}$ and $h=5\text{ nm}$. The white rectangle delimits the bar dimension, and the white cross mark represent the position of $\text{NV}_1$ referred in (d). The corresponding cooperativity ${\cal C}_{(005)}$ is shown on the right axis. (d) Effective NV-NV coupling strength $g_{\rm{eff}}$ between two NV centers as a function of the NV-NV distance, where $\text{NV}_1$ and $\text{NV}_2$ are placed at $\mathbf{r}_1=(d+h)\hat{x}+w\hat{y}+(400\ \mathrm{nm})\hat{z}$ and $\mathbf{r}_2=\mathbf{r}_1+\delta z \hat{z}$, respectively. The red (blue) curve is calculated for $(d,p)=(5\text{ nm},5)$ [$(d,p)=(20\text{ nm},12)$]. The entanglement gate rate (ER) and the gate to decoherence ratio (GDR) are shown on the right axis. In both cases aspect ratio is $w/d=6$, length of the magnetic bar is $l=3 \ \mu\text{m}$, and detuning is $\Delta f=\Delta\omega/2\pi=(\omega_{(00p)}-\omega_\text{NV})/2\pi=3 \text{ MHz}$.}
\label{fig4} 
\end{figure}

In Fig.~\ref{fig4}(a) we plot the external magnetic field $H_\text{ext}$ dependence of the discretized magnon mode frequencies of a YIG bar with dimensions $(d,w,l)=(5\text{ nm}, 30\text{ nm}, 3\ \mu\text{m})$. The neighboring magnon mode frequencies are separated from each other by over $2\pi\times 10\text{ MHz}$ for modes with $p\geq5$, as shown in Fig.~\ref{fig4}(b). At field $H_\text{ext}=H_\text{c}$, the NV center's transition frequency $\omega_\text{NV}$ and the magnon mode frequency $\omega_{(005)}$ are on-resonant. We plot in Fig.~\ref{fig4}(c) the spatial distribution of the NV-magnon coupling strength $g_{(005)}$ at $H_\text{ext}=H_\text{c}$ for a fixed NV center height $h=5\text{ nm}$ [see Fig.~\ref{fig2}(a)], and obtain $g_{(005)}\approx 2\pi\times 0.5$~MHz depending on the NV center positions. With the Gilbert damping parameter of YIG $\alpha=10^{-5}$~\cite{tabuchi2014hybridizing} and the coherence time of NV centers $T_2^*=1\text{ ms}$~\cite{herbschleb2019ultra}, we show on the right axis of Fig.~\ref{fig4}(c) the corresponding {single magnon $\mu$-mode} cooperativity\cite{li2015hybrid,NoriPolariton2018} 
\begin{equation}
{{\cal C}_\mu=\frac{|g_\mu(\mathbf{r})|^2}{\alpha\omega_\mu /T_2^{*}}}
\end{equation}
which is a dimensionless measure of the coupling. {We emphasize that because this represents the single-magnon-mode cooperativity, the temperature dependence only appears in $\alpha$ and $T_{2}^*$ which for the purpose of our low-temperatures analysis are assumed to be independent of temperature.} We find ${\cal C}_{(005)}\gtrsim 10^4$ over a wide range of NV center positions, achieving the strong coupling regime for our hybrid quantum system. In contrast to Sec.~III, where we have a translationally invariant infinitely long waveguide, here the position of the NV center along $z$-direction plays a major role in the coupling strength. Our calculations enable us to optimize both the coupling strength and the cooperativity in order to increase NV-NV entanglement efficiency in our system.

The virtual-magnon-mediated NV-NV interaction is calculated in a similar way as in Eq.~(\ref{GeffWG}) under the condition $|g_{\mu}(\bf{r})|\ll |\omega_{\mu}-\omega_{\rm{NV}} |$, and we obtain 
\begin{equation}
g_\mathrm{eff}=\frac{g_{\mu}\left(\mathbf{r}_{1}\right)g_{\mu}^{*}\left(\mathbf{r}_{2}\right)}{\omega_\mu-\omega_\mathrm{NV}}
\end{equation}
with $\mu=(005)$ (see Appendix C3). {In the same way as in the waveguide case, this virtual-magnon-mediated coupling strength is independent of temperature.} Here, the two NV centers are placed at  \hbox{$\mathbf{r}_1=(d+h)\hat{x}+w\hat{y}+(400\ \mathrm{nm})\hat{z}$} [see a cross mark in Fig.~\ref{fig4}(c)] and $\mathbf{r}_2=\mathbf{r}_1+\delta z \hat{z}$, where $\delta z$ is the NV-NV distance along the bar length. In Fig.~\ref{fig4}(d) we plot $g_\mathrm{eff}$ as a function of $\delta z$ for the detuning \hbox{$\Delta\omega=\omega_{(005)}-\omega_\text{NV}=2\pi\times3\mathrm{ MHz}$}, which could be produced by electric field~\cite{electricfieldNV2011}, strain~\cite{PhysRevLett.113.020503} or magnetic field deviation from $H_{\rm{ext}}=H_c$. The corresponding entangling gate rate and the gate to decoherence ratio are shown on the right axis. Surprisingly, useful entangling gates for $2.2\ \mu\text{m}$ separated NV centers with $g_{\rm{eff}}=2\pi\times90$~kHz and $\rm{GDR}>700$ are predicted for this YIG bar system. This makes experiments more accessible in terms of the independent optical initialization and the readout of NV centers than the waveguide case. 

We have also calculated these quantities for a less challenging to fabricate YIG geometry with \hbox{$(d,w,l,h)=(20\text{ nm}, 120\text{ nm}, 3\ \mu\text{m}, 5\text{ nm})$}. The result is plotted as a blue curve in Fig.~\ref{fig4}(d), for which we obtain $\text{GDR}>100$ for the $2.2\ \mu\text{m}$ separated NV centers. This result clarifies the significance of using the YIG bar structures to entangle two NV centers separated by a few micrometers. Moreover, the discretized magnon mode frequencies allows for controlling the NV center frequencies to be on-resonant to one of the magnon mode frequencies, which enables the entanglement of two NV centers via the transduction of energy quanta that we discuss in the next section. We also comment that it would be possible to control the NV-magnon coupling strength via parametric driving of the discretized magnon modes as studied in the cavity quantum electrodynamics~\cite{Leroux2018} (see Appendix I).

\section{Transduction and virtual-magnon exchange protocols}
\label{secV}

\begin{figure}[t]
\includegraphics[scale=1]{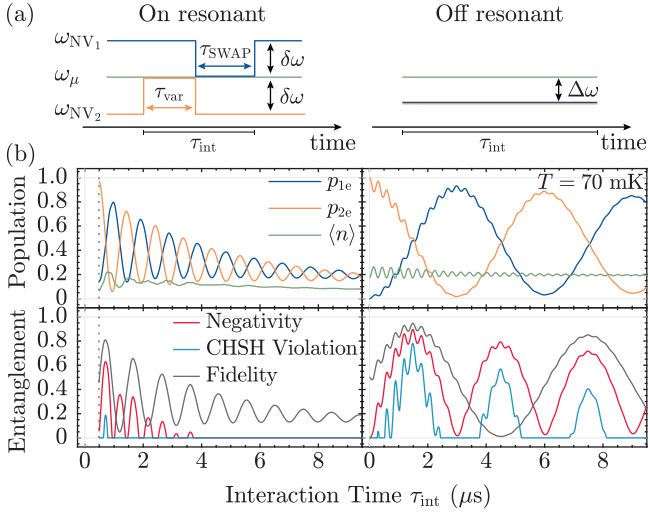}
\caption{(a) Schematic of on-resonant transduction (left) and off-resonant virtual-magnon exchange (right) entanglement protocols. (b) Comparison of the two protocols at $T=70 \text{ mK}$. The top two figures show NV center's excited state population $p_{ie}$ ($i=1,2$) and magnon population  $\langle n \rangle=\langle \hat{n}_\mu\rangle$ [$\mu=(005)$] at the end of the gate operations as a function of the total system interaction time. NV centers are separated by $2.2 \ \mu\text{m}$ on top of the YIG bar [see Fig.~\ref{fig4}(c)]. For the transduction protocol, NV center frequencies are modulated as illustrated in the inset, where each line represents the frequency of NV centers or the magnon mode. The bottom two figures show entanglement measures as a function of the interaction time. The red, sky blue, and gray curves are the entanglement negativity scaled by the Bell-state's negativity, the degree of the Bell inequality violation (violated if the curve is above zero), and the fidelity to the target pure entangled states, respectively.}
\label{fig5} 
\end{figure}

In this section, we explore and compare two entangling gate protocols for our hybrid quantum system,  on-resonant transduction and  off-resonant virtual-magnon exchange. Entanglement via the transduction protocol is simulated by controlling the NV center frequencies independently as illustrated in the left schematic of Fig.~\ref{fig5}(a). For this case, the NV spins are initially prepared in the state $|g\rangle_1|e\rangle_2$, i.e., $\text{NV}_1$ ($\text{NV}_2$) is in its ground (excited) state. We first make $\omega_{\text{NV}_2}$ on-resonant to the $\mu$-magnon mode frequency $\omega_{{\mu}}$ for a certain time $\tau_\mathrm{var}$ during which $\omega_{\text{NV}_1}$ is detuned from $\omega_{{\mu}}$ by ${\delta\omega=}2\pi\times5\text{ MHz}$. Second we swap the $\mathrm{NV}_1$ spin state and the magnon state by making $\omega_{\rm{NV_1}}=\omega_{\mu}$ for the swap gate time $\tau_\mathrm{SWAP}$ during which $\omega_{\mathrm{NV}_2}$ is detuned from $\omega_{\mu}$ by ${\delta\omega}$. The total interaction time in this protocol is $\tau_\mathrm{int}=\tau_\mathrm{var}+\tau_\mathrm{SWAP}$ and is varied by changing $\tau_\mathrm{var}$. The control of the NV centers' transition frequencies can be performed by applying a local magnetic field, electric field~\cite{electricfieldNV2011}, or strain~\cite{PhysRevLett.113.020503}. An alternative possibility of controlling the transition frequencies would be to use a periodic modulation of the external magnetic field~\cite{Oliver1653,XufengPRL2020} (see Appendix J). In contrast, in the virtual-magnon exchange protocol the NV centers' frequencies are both detuned from the $\mu$-magnon mode frequency by $\Delta \omega=\omega_{\mu}-\omega_{\mathrm{NV}_{1,2}}=2\pi\times 3\ \mathrm{MHz}$ [see the right schematic of Fig.~\ref{fig5}(a)]. After the preparation of the NV centers' spin state in $|g\rangle_1|e\rangle_2$, the whole system evolves over the interaction time $\tau_\mathrm{int}$. 

The time evolution of our hybrid quantum system for both protocols is simulated using the Lindblad master equation~\cite{lindblad1976,breuer2002theory,li2015hybrid} at a finite temperature $T$ considering two NV centers and a magnon mode $\mu$,
\begin{eqnarray}
\dot{\rho}=&&-i[\mathcal{H}(t), \rho]+2 \kappa\left(1+\bar{n}_\mathrm{m}^\mathrm{th}\right) \mathcal{D}[a] \rho\nonumber\\
&&+2 \kappa \bar{n}_\mathrm{m}^\mathrm{th} \mathcal{D}\left[a^{\dagger}\right] \rho+\frac{\gamma_{2}}{2}\sum_{i=1,2} \mathcal{D}\left[\sigma^z_{\mathrm{NV}_i}\right] \rho,\label{mainLindblad}
\end{eqnarray}
where $\mathcal{D}[\mathcal{O}] \rho=\mathcal{O} \rho \mathcal{O}^{\dagger}-\frac{1}{2}(\mathcal{O}^{\dagger} \mathcal{O}\rho+\rho\mathcal{O}^{\dagger} \mathcal{O})$, $\kappa=\alpha\omega_\mu$, $\gamma_2=1/T_2^*$, $a=\beta_{\mu}$, $a^\dagger=\beta^\dagger_{\mu}$,  $\bar{n}_\mathrm{m}^\mathrm{th}=(\exp[\omega_\mu/k_\mathrm{B}T]-1)^{-1}$ is the thermal magnon population, $T$ is temperature, $k_\mathrm{B}$ is the Boltzmann constant, and $\rho$ is the density operator. Here, the magnon damping parameter $\kappa=\alpha\omega_{\mu}$ is based on the dissipation term in the  Landau–Lifshitz–Gilbert equation $\left.\partial_t\mathbf{M}\right|_{\mathrm{diss}}=(\alpha/M_{\mathrm{s}})\mathbf{M}\times \partial_t\mathbf{M}$, resulting in $\left.\partial_t\beta_\mu\right|_\mathrm{diss}\approx-\alpha\omega_\mu\beta_\mu$ under the assumption $\partial\omega_{\mu}/\partial H_{\mathrm{ext}}\approx \mu_0 \gamma$, which is verified by Fig.~\ref{fig4}(b) (see Appendix D1). For the magnon mode contribution in the total Hamiltonian $\mathcal{H}(t)$, we only take into account the magnon mode with $\mu=(005)$, as this mode produces the dominant contribution in the NV-NV interaction as well as the magnon induced decoherence of NV centers in both protocols. As the NV center's longitudinal decay rate is much smaller than the transverse decoherence rate~\cite{bar2013solid}, we assume it to be zero in the simulation. The two NV centers are separated by $2.2 \ \mu\text{m}$ along the YIG bar length with $\mathbf{r}_1=(d+h)\hat{x}+w\hat{y}+(400\ \mathrm{nm})\hat{z}$ and $\mathbf{r}_2=\mathbf{r}_1+(l-800\ \mathrm{nm})\hat{z}$. We use the Gilbert damping parameter $\alpha=10^{-5}$ of YIG~\cite{tabuchi2014hybridizing} and the NV center coherence time $T_2^*=1 \text{ ms}$~\cite{herbschleb2019ultra}.

In the upper two panels of Fig.~\ref{fig5}(b), we plot the NV centers' excited state population $p_{ie}$ ($i=1,2$) and the magnon population $\langle n \rangle=\langle \hat{n}_{\mu}\rangle$ [$\mu=(005)$] at the end of the transduction {(on resonant)} and the virtual-magnon exchange {(off resonant)} protocols as a function of the total system interaction time $\tau_\mathrm{int}$ at $T=70 \text{ mK}$. In the lower two panels we plot three different entanglement measures as a function of the interacting time $\tau_\mathrm{int}$ for each protocol. More specifically, we plot the entanglement negativity normalized by the Bell-state's negativity, the degree of the Bell inequality violation, and the fidelity to the target pure entangled states, which are given by the red, sky blue, and gray curves, respectively. The resulting states are entangled if $\mathcal{N}>0$, and one expects to observe the violation of the Clauser-Horne-Shimony-Holt (CHSH) form of Bell inequality if $\text{CHSH Violation}>0$~\cite{horodecki1995violating,bartkiewicz2013entanglement} (see Appendix D1).

In Fig.~\ref{fig5}(b) we first find that the transduction protocol is faster in gate operation as compared to the virtual-magnon exchange protocol. This is because the NV-magnon on-resonant coupling rate $g_{\mu}\approx 2\pi \times 0.5$~MHz is larger than the off-resonant NV-NV coupling rate $g_{\rm{eff}}\approx 2\pi \times 90$~kHz. On the other hand, it is observed that the virtual-magnon exchange protocol results in larger amplitude oscillations in the NV centers' excited state populations and higher fidelity under the parameters and the temperature used in the simulation.  This result is understood by a combination of two factors. First, the virtual-magnon exchange protocol only creates magnons virtually (with magnon population suppressed by $g_\mu/\Delta\omega$ due to the energy mismatch), thus being approximately insensitive to the magnon damping parameter. Secondly, the magnon damping rate $\alpha\omega_\mu$ is faster than the NV center's decoherence rate $1/T_2^*$, and therefore there is more loss of information if a real magnon is excited. Nonetheless, in both protocols we predict entangled states can be manipulated and the violation of the Bell inequality will be observed.

To further compare the two entanglement protocols, we have performed simulations under multiple temperatures and have observed that the virtual-magnon-exchange protocol is more robust at higher temperatures up to $\approx150$~mK (see Appendix D2 and E). Moreover, we show that both protocols do not produce useful entanglement for $T\gtrsim150$~mK due to the NV centers' dephasing from magnon number fluctuations of modes with $\mu\neq(005)$. We have also evaluated the decay contribution due to these magnon modes and have verified that this is negligible for temperatures $T\leq150$~mK for both upper and lower transitions of NV centers (see Appendix H). Interestingly, the transduction protocol improves more drastically at lower temperatures than the virtual-magnon exchange protocol. Based on the zero temperature analysis, we find an inequality for which the transduction protocol performs better (see Appendix D2)
 
\begin{eqnarray}
\alpha \lesssim \frac{\Delta\omega/g_\mu}{4(1-1/\pi)}\frac{1}{\omega_\mu T_2^{*}}.
\end{eqnarray}
For the parameters used in this section, the transduction protocol is shown to outperform the virtual-magnon exchange protocol (with $\Delta \omega=2\pi\times 3\ \mathrm{MHz}$) if $\alpha \lesssim  10^{-7}$. In Appendix D2 we provide phase diagrams in ($\alpha$, $1/T_2^*$)-space for which protocol gives higher fidelity under multiple detuning values. Analytical expressions for the fidelity in the limit $\alpha\omega_\mu/g_\mu\ll1$ and $T_2^{*-1}/g_\mu\ll1$ are also provided. To show that the magnon-mediated entanglement scheme can directly be extended to two-qubit entangling gates, we have also calculated an average gate fidelity $\bar{F}$~\cite{nielsen2002simple} as a square-root-of-$i$SWAP gate for the off-resonance protocol, and have obtained $\bar{F}\approx0.88$ at $T=70$~mK (see Appendix F). 

As for keeping the system at low temperatures $T\lesssim 150$~mK, we note that the laser illumination and microwave irradiation on the system for the initialization, manipulation, and readout of NV centers may cause unwanted heating. Although YIG has been studied under microwave irradiations in superconducting qubit platforms~\cite{Lachance-Quirion425} and color centers have been studied under laser illuminations in dilution refrigerator temperatures $T<100$~mK~\cite{Evans662,PhysRevLett.120.053603,singh2020epitaxial,PhysRevB.102.104114}, it would be important to minimize the average microwave irradiation and laser illumination power on the system to maintain the required low temperatures. Here, of particular interest is the possibility of cooling down the target magnon mode to its ground state in analogy to cavity optomechanics techniques~\cite{PhysRevLett.99.093901,PhysRevLett.99.093902,CoolingMech2011,NVmechcooling2013,NVmechcooling2017}, e.g., via the optomagnonic interaction~\cite{PhysRevLett.121.087205} or via the coupling to NV centers~\cite{NVmechcooling2013,NVmechcooling2017}. For example in Fig.~\ref{fig5}(b), we have observed that the mean magnon occupation number at the end of the on-resonant protocol is smaller than its thermal level [see $\langle n(\tau_{\rm{int}}=0)\rangle$ in the off-resonant protocol], which is reminiscent of the ground-state cooling of magnons and motivates future studies on the alternative cooling methods of the NV-magnon hybrid quantum system.

We also note that the small Gilbert damping parameter $\alpha= 10^{-5}$ used in the current study may be optimistic for small YIG structures as the value is obtained from bulk YIG samples~\cite{tabuchi2014hybridizing}. This is partially due to the nontrivial magnetic behavior at millikelvin temperatures of the gadolinium-gallium-garnet (GGG) substrates on which YIG is typically grown~\cite{kosen2019microwave}, which would be mitigated by employing a free-standing structure~\cite{awschalom2021quantum}, and also due to the impurity relaxation mechanism in YIG~\cite{jermain2017increased}. However, with remarkable advances in recent magnonics research, it has been shown that the damping of thin YIG films can be improved considerably, e.g., with techniques based on a recrystallization of amorphous YIG into single crystals~\cite{Recrystalization2016}. Additionally, we obtain a high cooperativity $\mathcal{C}\approx 500$ even with the larger Gilbert damping parameter $\alpha=10^{-3}$ as calculated from Fig.~\ref{fig4}(c). We have further performed simulations with $\alpha=10^{-3}$ in Appendix G, and find that the entangled state can still be produced at $T=70$~mK for the off-resonant protocol, although further optimization on the detuning frequency is needed to improve the quality of the entanglement in order to avoid the overlap of the NV centers' transition frequencies with the now broader linewidth of the magnon mode resonance (see Appendix G).

\section{Conclusion}
\label{secVI}
We study hybrid quantum systems consisting of NV centers in diamond and magnons in ferromagnetic bar and waveguide structures. Based on the Hamiltonian formalism of the dipole-exchange magnons, we predict useful two-NV entangling gates over $1$-$2\ \mu\text{m}$ NV-NV separations at finite temperatures. Transduction and virtual-magnon exchange protocols of entanglement are explored and compared under realistic experimental conditions. Although the transduction protocol is faster in gate operation, the virtual-magnon exchange protocol results in higher fidelity as the typical Gilbert damping parameter of YIG makes the magnons less coherent than the NV centers. We have obtained entangled state's fidelity $F\approx0.81$ for the transduction protocol and $F\approx0.95$ for the virtual-magnon exchange protocol at $T=70$~mK. The virtual-magnon exchange protocol is also found to be robust against thermal magnon fluctuations, although the transduction protocol outperforms it close to zero temperature for $\alpha\omega_\mu T_2^*\lesssim {(\Delta\omega/g_\mu)}/{[4(1-1/\pi)]}$. Calculations presented in this study help to implement optimal device geometries and entangling gate protocols in future experiments trying to entangle spatially separated NV centers using magnons in ferromagnets.

\section*{Acknowledgement}
This work is supported by the Vannevar Bush Faculty Fellowship ONR N00014-17-1-3026, the U.S. Department of Energy, Office of Basic Energy Sciences, Materials Science and Engineering Division (M.~F, D.~D.~A.), the  U.S. Department of Energy, Office of Basic Energy Sciences under Award Number DE-SC0019250 (D.~C. and M.~E.~F.), and the U.S. Department of Energy, Office of Science, National Quantum Information Science Research Centers (D. D. A.). The authors thank H.-S. Chang, J.C. Karsch, G. Smith, P.C. Jerger, A. Crook, Y. Tsaturyan, L.R. Weiss, and S.E. Sullivan for useful discussions.

\normalem

\appendix

\section{Hamiltonian formalism of dipole-exchange magnons}
\subsection{Model Hamiltonian of the NV-magnon hybrid system}
The total Hamiltonian $\mathcal{H}$ of our hybrid system composed of NV centers and magnons is presented in Sec.~II. We note that the interaction Hamiltonian $\mathcal{H}_\mathrm{int}$ can also be understood in terms of the dipolar tensor $\hat{D}(\mathbf{r}-\mathbf{r}')$:
\begin{eqnarray}
&&\mathcal{H}_{\mathrm{int}}=\sum_{i=1,2} \gamma \mu_{0} \mathbf{S}_{\mathrm{NV}_{i}} \cdot \left.\int d \mathbf{r}^{\prime} \widehat{D}\left(\mathbf{r}-\mathbf{r}^{\prime}\right) \cdot \mathbf{M}\left(\mathbf{r}^{\prime}\right)\right|_{\mathbf{r}=\mathbf{r}_{i}},\\
&&\hat{D}(\mathbf{r}-\mathbf{r}')=-(\nabla\otimes\nabla')G(\mathbf{r}-\mathbf{r}')\nonumber\\
&&\quad\quad\quad\quad\ =\frac{1}{4 \pi}\left(\frac{3}{\left|\mathbf{r}-\mathbf{r}^{\prime}\right|^{5}}\left(\mathbf{r}-\mathbf{r}^{\prime}\right) \otimes\left(\mathbf{r}-\mathbf{r}^{\prime}\right)-\frac{1}{\left|\mathbf{r}-\mathbf{r}^{\prime}\right|^{3}}\right)\ \nonumber\\
&&\quad\quad\quad\quad\quad \ ,\mathrm{when\ }\mathbf{r}\neq\mathbf{r}'.
\end{eqnarray}

The magnetization dynamics governed by the Landau-Lifshitz-Gilbert (LLG) equation (without dissipation) is obtained by the Hamiltonian equation of motion with the following identification of the magnetization and the classical complex canonical variables following the Holstein-Primakoff transformation:
\begin{eqnarray}
&&M_{-}(\mathbf{r})=\sqrt{2 \gamma M_\mathrm{s}(\mathbf{r})} a(\mathbf{r}) f\left(a^{*}(\mathbf{r}) a(\mathbf{r})\right),\\
&&M_{+}(\mathbf{r})=\sqrt{2 \gamma M_\mathrm{s}(\mathbf{r})} a^{*}(\mathbf{r}) f\left(a^{*}(\mathbf{r}) a(\mathbf{r})\right),\\
&&M_{z}(\mathbf{r})=M_\mathrm{s}(\mathbf{r})-\gamma a^{*}(\mathbf{r}) a(\mathbf{r}),
\end{eqnarray}
where $M_{\pm}=M_x\pm i M_y$, $M_z=\sqrt{M_\mathrm{s}^2-M_x^2-M_y^2}$, and $f(x)=\sqrt{1-\gamma x/(2M_\mathrm{s}(\mathbf{r}))}$. Here, $a(\mathbf{r})$ and $a^*(\mathbf{r})$ are the complex canonical variables satisfying $\partial_ta(\mathbf{r})=-i\delta{\mathcal{H}}/\delta a^*(\mathbf{r})$ and $\partial_ta^*(\mathbf{r})=+i\delta{\mathcal{H}}/\delta a(\mathbf{r})$. The relation between $a(\mathbf{r})$, $a^*(\mathbf{r})$ and $M_-(\mathbf{r})$, $M_+(\mathbf{r})$ is carefully chosen such that it satisfies the dissipationless LLG equation $\partial_t \mathbf{M}=-\gamma \mu_{0} \mathbf{M} \times \mathbf{H}_{\mathrm{eff}}$. This is also consistent with the standard sign convention of the time evolution of the creation/annihilation operators $a\leftrightarrow\hat{a}\propto e^{-i\omega t}$ and $a^*\leftrightarrow\hat{a}^\dagger\propto e^{+i\omega t}$. Note that we use $\gamma>0$ as the absolute value of the electron gyromagnetic ratio, so the electron gyromagnetic ratio is $-\gamma$. The Hamiltonian equation of motion gives
\begin{eqnarray}
\partial_{t} M_{x}=-\gamma M_{z} \frac{\delta \mathcal{H}}{\delta M_{y}} ; \quad \partial_{t} M_{y}=+\gamma M_{z} \frac{\delta \mathcal{H}}{\delta M_{x}},\label{LLGcoherent}
\end{eqnarray}
and writing $\mathcal{H}[M_x,M_y]=\mathcal{W}[M_x,M_y,M_z(M_x,M_y)]$ with $M_z(M_x,M_y)=\sqrt{M_\mathrm{s}^2-M_x^2-M_y^2}$, we obtain
\begin{eqnarray}
\partial_{t} M_{x}=\gamma\left[M_{y} \frac{\delta \mathcal{W}}{\delta M_{z}}-M_{z} \frac{\delta \mathcal{W}}{\delta M_{y}}\right],\nonumber\\
\quad \partial_{t} M_{y}=\gamma\left[M_{z} \frac{\delta \mathcal{W}}{\delta M_{x}}-M_{x} \frac{\delta \mathcal{W}}{\delta M_{z}}\right].
\end{eqnarray}
As the effective field is obtained by $\mu_0\mathbf{H}_\mathrm{eff}=-\delta\mathcal{W}/\delta\mathbf{M}$, the dissipationless LLG equation $\partial_t \mathbf{M}=-\gamma \mu_{0} \mathbf{M} \times \mathbf{H}_{\mathrm{eff}}$ is successfully derived.
    
In the following discussions, we apply an external field along the $\hat{z}$ direction of Fig.~\ref{fig2}(a), $\mathbf{H}_\mathrm{ext}=H_\mathrm{ext}\hat{z}$, and for geometrical simplicity we take the NV main symmetry axis to be along $\hat{z}$ axis, i.e., $\hat{n}_\mathrm{NV}=\hat{z}$. We will consider the case where the equilibrium magnetization is uniform across the ferromagnet and parallel to $\hat{z}$, $\mathbf{M}(\mathbf{r})=\mathbf{M}_0(\mathbf{r})=M_\mathrm{s}(\mathbf{r})\hat{z}$. Although in principle we need to obtain the equilibrium magnetization from the energy minimization of $\mathcal{H}_\mathrm{m}$, in the infinitely long waveguide case $\mathbf{M}(\mathbf{r})=M_\mathrm{s}(\mathbf{r})\hat{z}$ holds as the field is applied along the direction where demagnetization factor is zero. In the finite length magnetic bar case, this is still approximately correct as in our setting the length $l$ is much larger than both the width $w$ and the thickness $d$. Under this uniform equilibrium magnetization, {components of the} small deviation from the equilibrium $\delta\mathbf{M}(\mathbf{r})=\mathbf{M}(\mathbf{r})-\mathbf{M}_0(\mathbf{r})$ are given by
\begin{eqnarray}
&&\delta M_-(\mathbf{r})\approx\sqrt{2\gamma M_s(\mathbf{r})}a(\mathbf{r})=m_-(\mathbf{r})\label{HP1},\\
&&\delta M_+(\mathbf{r})\approx\sqrt{2\gamma M_s(\mathbf{r})}a^*(\mathbf{r})=m_+(\mathbf{r})\label{HP2},\\
&&\delta M_z(\mathbf{r})=-\gamma a^*(\mathbf{r})a(\mathbf{r})\approx-\mathbf{m}^2(\mathbf{r})/(2M_\mathrm{s}(\mathbf{r})).
\end{eqnarray}
Here, $\mathbf{m}(\mathbf{r})=m_x(\mathbf{r})\hat{x}+m_y(\mathbf{r})\hat{y}$ is a small two-dimensional magnetization deviation. Now the deviation can be written as $\delta\mathbf{M}(\mathbf{r})\approx\mathbf{m}(\mathbf{r})-(\mathbf{m}^2(\mathbf{r})/2M_\mathrm{s}(\mathbf{r}))\hat{z}$.

\subsection{Simplification of the magnon Hamiltonian}
In the following calculation, we simplify the magnon Hamiltonian $\mathcal{H}_\mathrm{m}$. We write $\mathcal{H}_\mathrm{m}=\mathcal{H}_{\mathrm{Z}}+\mathcal{H}_\mathrm{ex}+\mathcal{H}_\mathrm{dip}$, where $\mathcal{H}_\mathrm{Z}$ is the Zeeman Hamiltonian, $\mathcal{H}_\mathrm{ex}$ is the exchange Hamiltonian, and $\mathcal{H}_\mathrm{dip}$ is the magnetic dipole Hamiltonian given by
\begin{eqnarray}
&&\mathcal{H}_{\mathrm{Z}}=-\mu_{0} \int d \mathbf{r} \mathbf{H}_{\mathrm{ext}} \cdot \mathbf{M}(\mathbf{r}),\\
&&\mathcal{H}_{\mathrm{ex}}=\frac{\mu_{0}}{2} \int d \mathbf{r} \alpha_\mathrm{ex}(\mathbf{r}) \nabla \mathbf{M}(\mathbf{r}): \nabla \mathbf{M}(\mathbf{r}),\\
&&\mathcal{H}_{\mathrm{dip}}=\frac{\mu_{0}}{2} \int d \mathbf{r} d \mathbf{r}^{\prime}(\nabla \cdot \mathbf{M}(\mathbf{r})) G\left(\mathbf{r}-\mathbf{r}^{\prime}\right)\left(\nabla^{\prime} \cdot \mathbf{M}\left(\mathbf{r}^{\prime}\right)\right),\nonumber\\
\end{eqnarray}
where the double-dot product is defined as $\nabla{\bf{M}}:\nabla{\bf{M}}=\partial_a M_b\partial^a M^b$. Firstly, we simplify the Zeeman Hamiltonian and the dipole Hamiltonian. Using $\mathbf{M}(\mathbf{r})=\mathbf{M}_0(\mathbf{r})+\delta\mathbf{M}(\mathbf{r})$, we obtain
\begin{eqnarray}
&&\mathcal{H}_{\mathrm{Z}}=-\mu_{0} \int d \mathbf{r} \mathbf{H}_{\mathrm{ext}} \cdot \delta \mathbf{M}(\mathbf{r})+\text {const.},\\
&&\mathcal{H}_{\mathrm{dip}}=\mathcal{H}_{\mathrm{dem}}+\mathcal{H}_{\mathrm{dip}(2)}+\mathrm{const.},\\
&&\mathcal{H}_{\text {dem }}=-\mu_{0} \int d \mathbf{r} \mathbf{H}_{\mathrm{d}}(\mathbf{r}) \cdot \delta \mathbf{M}(\mathbf{r}) ,\\
&&\mathcal{H}_{\operatorname{dip}(2)}=\frac{\mu_{0}}{2} \int d \mathbf{r} d \mathbf{r}^{\prime}(\nabla \cdot \delta \mathbf{M}(\mathbf{r})) G\left(\mathbf{r}-\mathbf{r}^{\prime}\right)\left(\nabla^{\prime} \cdot \delta \mathbf{M}\left(\mathbf{r}^{\prime}\right)\right).\nonumber\\
\end{eqnarray}
Here, $\mathcal{H}_{\text {dem }}$ is the demagnetization field Hamiltonian, $\mathcal{H}_{\operatorname{dip}(2)}$ is the dipole Hamiltonian that is second order in $\delta \mathbf{M}$, and $\mathbf{H}_\mathrm{d}(\mathbf{r})$ is the demagnetization field defined by
\begin{equation}
\mathbf{H}_{\mathrm{d}}(\mathbf{r}) = \nabla \int d \mathbf{r}^{\prime} G\left(\mathbf{r}-\mathbf{r}^{\prime}\right)\left(\nabla^{\prime} \cdot \mathbf{M}_{0}\left(\mathbf{r}^{\prime}\right)\right).
\end{equation}
For the infinitely long waveguide, we have $\mathbf{H}_\mathrm{d}(\mathbf{r})=0$. For the finite length magnetic bar structure, we approximate $\mathbf{H}_\mathrm{d}(\mathbf{r})\approx H_\mathrm{d}^z(\mathbf{r})\hat{z}$ as the $z$-component is dominant compared to the $x$ and $y$ components. Therefore, we obtain
\begin{equation}
\mathcal{H}_{\mathrm{dem}} \approx-\mu_{0} \int d \mathbf{r} H_{\mathrm{d}}^{z}(\mathbf{r}) \delta M_{z}(\mathbf{r}).
\end{equation}
Up to the quadratic order in $\mathbf{m}(\mathbf{r})$, we obtain
\begin{eqnarray}
&&\mathcal{H}_{\mathrm{Z}} \approx \mu_{0} H_{\mathrm{ext}} \int d \mathbf{r} \frac{\mathbf{m}^{2}(\mathbf{r})}{2 M_\mathrm{s}(\mathbf{r})}+\text {const.} ,\\
&&\mathcal{H}_{\mathrm{dem}} \approx \mu_{0} \int d \mathbf{r} H_{\mathrm{d}}^{z}(\mathbf{r}) \frac{\mathbf{m}^{2}(\mathbf{r})}{2 M_\mathrm{s}(\mathbf{r})} , \\
&&\mathcal{H}_{\operatorname{dip}(2)} \approx \frac{\mu_{0}}{2} \int d \mathbf{r} d \mathbf{r}^{\prime}(\nabla \cdot \mathbf{m}(\mathbf{r})) G\left(\mathbf{r}-\mathbf{r}^{\prime}\right)\left(\nabla^{\prime} \cdot \mathbf{m}\left(\mathbf{r}^{\prime}\right)\right).\nonumber\\
\end{eqnarray}
Using $M_\mathrm{s}(\mathbf{r})=M_\mathrm{s}\mathcal{F}(\mathbf{r})$ and writing $\mathbf{m}(\mathbf{r})=\vec{\mathcal{M}}(\bf{r})\mathcal{F}(\mathbf{r})$, we obtain
\begin{eqnarray}
&&\mathcal{H}_{\mathrm{Z}} \approx \mu_{0} H_{\mathrm{ext}} \int d \mathbf{r} \mathcal{F}(\mathbf{r}) \frac{\vec{\mathcal{M}}^{2}(\mathbf{r})}{2 M_\mathrm{s}}+\mathrm{const.}\label{SimpHZ}  ,\\
&&\mathcal{H}_{\mathrm{dem}} \approx \mu_{0} \int d \mathbf{r} \mathcal{F}(\mathbf{r}) H_{\mathrm{d}}^{z}(\mathbf{r}) \frac{\vec{\mathcal{M}}^{2}(\mathbf{r})}{2 M_\mathrm{s}}\label{SimpHdem}, 
\end{eqnarray}
\begin{widetext}
\begin{eqnarray}
&&\mathcal{H}_{\mathrm{dip}(2)} \approx \frac{\mu_{0}}{2} \int d \mathbf{r} d \mathbf{r}^{\prime}(\nabla \cdot \vec{\mathcal{M}}(\mathbf{r}) \mathcal{F}(\mathbf{r})) G\left(\mathbf{r}-\mathbf{r}^{\prime}\right)(\nabla^{\prime} \cdot \vec{\mathcal{M}}\left(\mathbf{r}^{\prime}\right) \mathcal{F}\left(\mathbf{r}^{\prime}\right)).\label{SimpHdip2}
\end{eqnarray}
Similarly, the exchange Hamiltonian can be written, using $\mathbf{M}(\mathbf{r})=\mathbf{M}_0(\mathbf{r})+\delta\mathbf{M}(\mathbf{r})$, as
\begin{eqnarray}
\mathcal{H}_{\mathrm{ex}}=-\mu_{0} \int d \mathbf{r} \delta \mathbf{M}(\mathbf{r}) \cdot \partial_{\mu}\left[\alpha_\mathrm{ex}(\mathbf{r}) \partial^{\mu} \mathbf{M}_{0}(\mathbf{r})\right]-\frac{\mu_{0}}{2} \int d \mathbf{r} \delta \mathbf{M}(\mathbf{r}) \cdot \partial_{\mu}\left[\alpha_\mathrm{ex}(\mathbf{r}) \partial^{\mu} \delta \mathbf{M}(\mathbf{r})\right]+\mathrm{const.}
\end{eqnarray}
Up to the quadratic order in $\mathbf{m}(\mathbf{r})$, the above equation becomes
\begin{eqnarray}
\mathcal{H}_{\mathrm{ex}} \approx \frac{\mu_{0}}{2} \int d \mathbf{r} \frac{\mathbf{m}^{2}(\mathbf{r})}{M_\mathrm{s}(\mathbf{r})} \partial_{\mu}\left[\alpha_\mathrm{ex}(\mathbf{r}) \partial^{\mu} M_\mathrm{s}(\mathbf{r})\right]-\frac{\mu_{0}}{2} \int d \mathbf{r} \mathbf{m}(\mathbf{r}) \cdot \partial_{\mu}\left[\alpha_\mathrm{ex}(\mathbf{r}) \partial^{\mu} \mathbf{m}(\mathbf{r})\right]+\mathrm{const.}
\end{eqnarray}
Using $M_\mathrm{s}(\mathbf{r})=M_\mathrm{s}\mathcal{F}(\mathbf{r})$, $\alpha_\mathrm{ex}(\mathbf{r})=\alpha_\mathrm{ex}\mathcal{F}(\mathbf{r})$, and writing $\mathbf{m}(\mathbf{r})=\vec{\mathcal{M}}(\bf{r})\mathcal{F}(\mathbf{r})$, we obtain
\begin{eqnarray}
\mathcal{H}_{\mathrm{ex}} &\approx&\frac{\mu_{0} \alpha_\mathrm{ex}}{2} \int d \mathbf{r} \mathcal{F}^{3}(\mathbf{r}) \nabla \vec{\mathcal{M}}(\mathbf{r}): \nabla \vec{\mathcal{M}}(\mathbf{r})+\mathrm{const.}\nonumber ,\\
&=&-\frac{\mu_{0} \alpha_\mathrm{ex}}{2} \int d \mathbf{r} \left(\mathcal{F}^{3}(\mathbf{r}) \vec{\mathcal{M}}(\mathbf{r}) \cdot \nabla^{2} \vec{\mathcal{M}}(\mathbf{r})+\left[\partial_{\mu} \mathcal{F}^{3}(\mathbf{r})\right] \vec{\mathcal{M}}(\mathbf{r}) \cdot \partial^{\mu} \vec{\mathcal{M}}(\mathbf{r})\right)+\text { const., }
\end{eqnarray}
\end{widetext}
where the double-dot product is $\nabla\vec{\mathcal{M}}:\nabla\vec{\mathcal{M}}=\partial_a \mathcal{M}_{b}\partial^a \mathcal{M}^{b}$. Note that the term $\partial_{\mu} \mathcal{F}^{3}(\mathbf{r})$ in the second equation gives a delta-functional contribution peaked at the ferromagnet's boundary. Using the totally-free surface spin condition, $\partial_\mu\vec{\mathcal{M}}=\vec{0}$ on the ferromagnet's boundary, we obtain
\begin{equation}
\mathcal{H}_{\mathrm{ex}} \approx-\frac{\mu_{0} \alpha_\mathrm{ex}}{2} \int d \mathbf{r} \mathcal{F}^{3}(\mathbf{r}) \vec{\mathcal{M}}(\mathbf{r}) \cdot \nabla^{2} \vec{\mathcal{M}}(\mathbf{r})+\text{const.} \label{SimpHex}
\end{equation}
Combining equations (\ref{SimpHZ}), (\ref{SimpHdem}), (\ref{SimpHdip2}), and (\ref{SimpHex}), we obtain
\begin{widetext}
\begin{eqnarray}
&&\mathcal{H}_\mathrm{m}\approx\mu_{0} \int d \mathbf{r} \mathcal{F}(\mathbf{r}) (H_{\mathrm{ext}} +H_{\mathrm{d}}^z(\mathbf{r}) )\frac{\vec{\mathcal{M}}^{2}(\mathbf{r})}{2 M_\mathrm{s}}-\frac{\mu_{0} \alpha_\mathrm{ex}}{2} \int d \mathbf{r} \mathcal{F}^{3}(\mathbf{r}) \vec{\mathcal{M}}(\mathbf{r}) \cdot \nabla^{2} \vec{\mathcal{M}}(\mathbf{r})\nonumber\\
&&\ \ \ \ \ \ \ \ \ \ +\frac{\mu_{0}}{2} \int d \mathbf{r} d \mathbf{r}^{\prime}(\nabla \cdot \vec{\mathcal{M}}(\mathbf{r}) \mathcal{F}(\mathbf{r})) G\left(\mathbf{r}-\mathbf{r}^{\prime}\right)(\nabla^{\prime} \cdot \vec{\mathcal{M}}\left(\mathbf{r}^{\prime}\right) \mathcal{F}\left(\mathbf{r}^{\prime}\right)),\label{SimpHmagAll}
\end{eqnarray}
\end{widetext}
where we dropped the constant shift in energy.

\section{Infinitely long ferromagnetic waveguide}
\subsection{Diagonalization of the magnon Hamiltonian}
To obtain the magnon dynamics and the magnon spatial profiles for the infinitely long ferromagnetic waveguide ($l\rightarrow\infty$), we diagonalize the magnon Hamiltonian Eq.~(\ref{SimpHmagAll}) by expanding $\vec{\mathcal{M}}(\mathbf{r})$ as 
\begin{widetext}
\begin{eqnarray}
&&\vec{\mathcal{M}}(\mathbf{r})=\int \frac{d k}{2 \pi} e^{-i k z} \sum_{n m} \psi_{n}^{X}(x) \psi_{m}^{Y}(y) \sqrt{2 \gamma M_\mathrm{s}} \frac{1}{2}\left[a_{k,(n,m)}^{*}\thinspace \thinspace a_{-k,(n,m)}\right]\left[\begin{array}{l}\widehat{e}_{-} \\ \widehat{e}_{+}\end{array}\right],
\end{eqnarray}
\end{widetext}
\begin{eqnarray}
&&\psi_{n}^{X}(x)=\sqrt{\frac{2}{\left(1+\delta_{n, 0}\right) d}} \cos \left(\kappa_{n}^{X} x\right),\\
&&\psi_{m}^{Y}(y)=\sqrt{\frac{2}{\left(1+\delta_{m, 0}\right) w}} \cos \left(\kappa_{m}^{Y} y\right),
\end{eqnarray}
where $\kappa_n^X=n\pi/d$, $\kappa_m^Y=m\pi/w$, $n,m=0,1,\cdots$, $\widehat{e}_\pm=\hat{x}\pm i\hat{y}$, and $a_{k,(n,m)}$ is the complex canonical variable in the new basis. Note that we have $\mathbf{m}(\mathbf{r})=\vec{\mathcal{M}}(\bf{r})\mathcal{F}(\mathbf{r})$ and in the current geometry $\mathcal{F}(\mathbf{r})=\mathcal{F}^X(x)\mathcal{F}^Y(y)$, where $\mathcal{F}^X(x)=\Theta(x)\Theta(d-x)$, $\mathcal{F}^Y(y)=\Theta(y)\Theta(w-y)$, and $\Theta$ is the Heaviside step function. Recalling $M_\mathrm{s}(\mathbf{r})=M_\mathrm{s}\mathcal{F}(\mathbf{r})$ and using Eqs.~(\ref{HP1}) and (\ref{HP2}), the above expansion corresponds to the following:
\begin{eqnarray}
&&a(\mathbf{r})=\int \frac{d k}{2 \pi} e^{i k z} \sum_{n m} f_{n}^{X}(x) f_{m}^{Y}(y) a_{k,(n,m)},\label{WGexp_a}\\
&&a^{*}(\mathbf{r})=\int \frac{d k}{2 \pi} e^{-i k z} \sum_{n m} f_{n}^{X}(x) f_{m}^{Y}(y) a_{k,(n,m)}^{*},\label{WGexp_astar}\\
&&f_{n}^{X}(x)=\sqrt{\frac{2\mathcal{F}^X(x)}{\left(1+\delta_{n, 0}\right) d}} \cos \left(\kappa_{n}^{X} x\right),\\ &&f_{m}^{Y}(y)=\sqrt{\frac{2\mathcal{F}^Y(y)}{\left(1+\delta_{m, 0}\right) w}} \cos \left(\kappa_{m}^{Y} y\right),
\end{eqnarray}
which are presented in the main text.
After simplification, the magnon Hamiltonian Eq.~(\ref{SimpHmagAll}) becomes,
\begin{widetext}
\begin{eqnarray}
&&\mathcal{H}_\mathrm{m}=\frac{1}{2} \int \frac{d k}{2 \pi} \sum_{n_{1} m_{1} \atop n_{2} m_{2}}\left[a_{k,\left(n_{1},m_{1}\right)}^{*}\thinspace \thinspace a_{-k,\left(n_{1},m_{1}\right)}\right]\left[\begin{array}{ll}A_{k,\left(n_{1} m_{1}\right)\left(n_{2} m_{2}\right)} & B_{k,\left(n_{1} m_{1}\right)\left(n_{2} m_{2}\right)} \\ B_{k,\left(n_{1} m_{1}\right)\left(n_{2} m_{2}\right)}^{*} & A_{k,\left(n_{1} m_{1}\right)\left(n_{2} m_{2}\right)}^{*}\end{array}\right]\left[\begin{array}{c}a_{k,\left(n_{2},m_{2}\right)} \\ a_{-k,\left(n_{2},m_{2}\right)}^{*}\end{array}\right],
\label{HmagFullMatrix}
\end{eqnarray}
with
\begin{eqnarray}
&&A_{k,\left(n_{1} m_{1}\right)\left(n_{2} m_{2}\right)} = \Delta_{k,\left(n_{1} m_{1}\right)} \delta_{\left(n_{1} m_{1}\right)\left(n_{2} m_{2}\right)}+\omega_{M} H_{k,\left(n_{1} m_{1}\right)\left(n_{2} m_{2}\right)}^{00},\label{s40}\\
&&B_{k,\left(n_{1} m_{1}\right)\left(n_{2} m_{2}\right)} = \omega_{M} H_{k,\left(n_{1} m_{1}\right)\left(n_{2} m_{2}\right)}^{01},\\
&&\Delta_{k,(n m)}=\omega_{H}+D_{\mathrm{ex}} K_{k,(n m)}^{2},
\end{eqnarray}
where $\omega_M=\gamma\mu_0M_\mathrm{s}$, $\omega_H=\gamma\mu_0H_\mathrm{ext}$, $K_{k,(n m)}^{2} = k^{2}+\left(\kappa_{n}^{X}\right)^{2}+\left(\kappa_{m}^{Y}\right)^{2}$, $D_\mathrm{ex}=\alpha_\mathrm{ex}\omega_M$, and
\begin{eqnarray}
&&H_{k,\left(n_{1} m_{1}\right)\left(n_{2} m_{2}\right)}^{00}=\frac{1}{2}\left(H_{k,\left(n_{1} m_{1}\right)\left(n_{2} m_{2}\right)}^{X X}+H_{k,\left(n_{1} m_{1}\right)\left(n_{2} m_{2}\right)}^{Y Y}+i\left(H_{k,\left(n_{1} m_{1}\right)\left(n_{2} m_{2}\right)}^{X Y}-H_{k,\left(n_{1} m_{1}\right)\left(n_{2} m_{2}\right)}^{Y X}\right)\right), \\
&&H_{k,\left(n_{1} m_{1}\right)\left(n_{2} m_{2}\right)}^{01}=\frac{1}{2}\left(H_{k,\left(n_{1} m_{1}\right)\left(n_{2} m_{2}\right)}^{X X}-H_{k,\left(n_{1} m_{1}\right)\left(n_{2} m_{2}\right)}^{Y Y}-i\left(H_{k,\left(n_{1} m_{1}\right)\left(n_{2} m_{2}\right)}^{X Y}+H_{k,\left(n_{1} m_{1}\right)\left(n_{2} m_{2}\right)}^{Y X}\right)\right).
\end{eqnarray}
Here, $H_{k,\left(n_{1} m_{1}\right)\left(n_{2} m_{2}\right)}^{X X}$, $H_{k,\left(n_{1} m_{1}\right)\left(n_{2} m_{2}\right)}^{X Y}$, $H_{k,\left(n_{1} m_{1}\right)\left(n_{2} m_{2}\right)}^{Y X}$, and $H_{k,\left(n_{1} m_{1}\right)\left(n_{2} m_{2}\right)}^{Y Y}$ are given by
\begin{eqnarray}
&&H_{k,\left(n_{1} m_{1}\right)\left(n_{2} m_{2}\right)}^{X X}=\int d \boldsymbol{\rho}_{1} d \boldsymbol{\rho}_{2}(\partial_{x_1}\varphi^{XY}_{n_1 m_1}(\boldsymbol{\rho}_{1}))\frac{K_0(|k(\boldsymbol{\rho}_{1}-\boldsymbol{\rho}_{2})|)}{2\pi}(\partial_{x_2}\varphi^{XY}_{n_2 m_2}(\boldsymbol{\rho}_{2}))\label{HXX} , \\
&&H_{k,\left(n_{1} m_{1}\right)\left(n_{2} m_{2}\right)}^{X Y}=\int d \boldsymbol{\rho}_{1} d \boldsymbol{\rho}_{2}(\partial_{x_1}\varphi^{XY}_{n_1 m_1}(\boldsymbol{\rho}_{1}))\frac{K_0(|k(\boldsymbol{\rho}_{1}-\boldsymbol{\rho}_{2})|)}{2\pi}(\partial_{y_2}\varphi^{XY}_{n_2 m_2}(\boldsymbol{\rho}_{2})), \\
&&H_{k,\left(n_{1} m_{1}\right)\left(n_{2} m_{2}\right)}^{Y X}=\int d \boldsymbol{\rho}_{1} d \boldsymbol{\rho}_{2}(\partial_{y_1}\varphi^{XY}_{n_1 m_1}(\boldsymbol{\rho}_{1}))\frac{K_0(|k(\boldsymbol{\rho}_{1}-\boldsymbol{\rho}_{2})|)}{2\pi}(\partial_{x_2}\varphi^{XY}_{n_2 m_2}(\boldsymbol{\rho}_{2}), \\
&&H_{k,\left(n_{1} m_{1}\right)\left(n_{2} m_{2}\right)}^{Y Y}=\int d \boldsymbol{\rho}_{1} d \boldsymbol{\rho}_{2}(\partial_{y_1}\varphi^{XY}_{n_1 m_1}(\boldsymbol{\rho}_{1}))\frac{K_0(|k(\boldsymbol{\rho}_{1}-\boldsymbol{\rho}_{2})|)}{2\pi}(\partial_{y_2}\varphi^{XY}_{n_2 m_2}(\boldsymbol{\rho}_{2})),\label{HYY}
\end{eqnarray}
\end{widetext}
where $K_\alpha$ is the modified Bessel function of the second kind, $\boldsymbol{\rho}=x\hat{x}+y\hat{y}$, and $\varphi^{XY}_{n m}(\boldsymbol{\rho})=\mathcal{F}^X(x)\mathcal{F}^Y(y)\psi_n^X(x)\psi_m^Y(y)$. Note that we have relations $A_{k,\left(n_{1} m_{1}\right)\left(n_{2} m_{2}\right)}=\left(A_{k,\left(n_{2} m_{2}\right)\left(n_{1} m_{1}\right)}\right)^{*}$ and $B_{k,\left(n_{1} m_{1}\right)\left(n_{2} m_{2}\right)}=B_{k,\left(n_{2} m_{2}\right)\left(n_{1} m_{1}\right)}$.

As we consider the case where the thickness $d$ and the width $w$ are small such that the exchange energy difference $D_\mathrm{ex}K_{k,(n_1m_1)}^2-D_\mathrm{ex}K_{k,(n_2m_2)}^2$ [with $(n_1,m_1)\neq(n_2,m_2)$] is large as compared to the off-diagonal components [elements of $A_{k,(n_1m_1)(n_2m_2)}$ or $B_{k,(n_1,m_1)(n_2m_2)}$ with $(n_1m_1)\neq (n_2,m_2)$] of the Hamiltonian, we apply the block-diagonal approximation~\cite{Kalinikos_1986}. Note that we can go beyond the block-diagonal approximation with the procedure using a paraunitary matrix presented in  Ref.~[\onlinecite{colpa1978diagonalization}]. Under this block-diagonal approximation, we obtain
\begin{widetext}
\begin{eqnarray}
&&\mathcal{H}_{\mathrm{m}}=\frac{1}{2} \int \frac{d k}{2 \pi} \sum_{n m}\left[a_{k,(n,m)}^{*}\thinspace \thinspace a_{-k,(n,m)}\right]\left[\begin{array}{ll}A_{k,(n, m)} & B_{k,(n, m)} \\ B_{k,(n, m)}^{*} & A_{k,(n, m)}\end{array}\right]\left[\begin{array}{c}a_{k,(n,m)} \\ a_{-k,(n,m)}^{*}\end{array}\right],\\
&&A_{k,(n, m)} = A_{k,(n m)(n m)} ;\quad B_{k,(n, m)} = B_{k,(n m)(n m)}.
\end{eqnarray}
The Hamiltonian above can be diagonalized by the standard $2\times2$ Bogoliubov transformation:
\begin{eqnarray}
&&\beta_{k,(n, m)} = \lambda_{k,(n, m)} a_{k,(n, m)}+\mu_{k,(n, m)} a_{-k,(n, m)}^{*},\label{BogoStandard_B}\\
&&\beta_{-k,(n, m)}^{*} = \mu_{k,(n, m)}^{*} a_{k,(n, m)}+\lambda_{k,(n, m)} a_{-k,(n, m)}^{*},\label{BogoStandard_Bstar}\\
&&\lambda_{k,(n, m)}=\sqrt{\frac{A_{k,(n, m)}+\omega_{k,(n, m)}}{2 \omega_{k,(n,m)}}};\quad \mu_{k,(n, m)}=\frac{B_{k,(n, m)}}{\left|B_{k,(n, m)}\right|} \sqrt{\frac{A_{k,(n, m)}-\omega_{k,(n, m)}}{2 \omega_{k,(n, m)}}}\label{BogoWG},\\
&&\omega_{k,(n, m)}=\sqrt{A_{k,(n, m)}^{2}-\left|B_{k,(n, m)}\right|^{2}},
\end{eqnarray}
\end{widetext}
and we obtain
\begin{eqnarray}
\mathcal{H}_{\mathrm{m}}=\sum_{nm} \int \frac{d k}{2 \pi} \omega_{k,(n, m)} \beta_{k,(n, m)}^{*} \beta_{k,(n, m)}.
\end{eqnarray}
Now we limit our discussion to the subspace with $(n,m)=(0,0)$ that gives the lowest energy magnon band, for which magnetization dynamics is uniform across $x$-$y$ plane in the ferromagnet. After promoting the classical complex canonical variables to the quantum creation and annihilation operators via $\beta_{k,(0,0)}\rightarrow\sqrt{\hbar}\hat{\beta}_{k,(0,0)}$ and $\beta^*_{k,(0,0)}\rightarrow\sqrt{\hbar}\hat{\beta}^\dagger_{k,(0,0)}$, we obtain
\begin{eqnarray}
\mathcal{H}_{\mathrm{m}}=\int \frac{d k}{2 \pi} \hbar\omega_{k,(0,0)} \beta_{k,(0,0)}^\dagger \beta_{k,(0,0)},\label{HmagSimp}
\end{eqnarray}
which is presented in the main text. Here, $\hbar\omega_{k,(00)}$ is the magnon energy and $\beta_{k,(00)}$ is the normal mode magnon annihilation operator satisfying $[\beta_{k,(00)},\beta^\dag_{k',(00)}]=2\pi\delta(k-k')$. For calculating the dispersion relation in the main text, we numerically evaluate Eqs.~(\ref{HXX})-(\ref{HYY}). In the subspace with $(n,m)=(0,0)$, $\psi_n^X(x)$ and $\psi_m^Y(y)$ are constant functions, so the derivatives only act on $\mathcal{F}^X(x)$ and $\mathcal{F}^Y(y)$, resulting in the surface integrals and the evaluation is simpler. Beyond the diagonal approximation $(n_1,m_1)=(n_2,m_2)$ made for Eq.~(\ref{HmagFullMatrix}), we can diagonalize the full Hamiltonian via the Bogoliubov transformation with the paraunitary matrix~\cite{colpa1978diagonalization} after a truncation of large wavenumber modes, which is used in the magnetic bar calculations in Sec.~II and Appendix C.

\subsection{NV-magnon coupling}
The coupling strength between magnons and NV centers is obtained by applying the same Bogoliubov transformation in the interaction Hamiltonian Eq.~(\ref{Hint}). Up to the quadratic order in $\mathbf{m}(\mathbf{r})$, we obtain
\begin{widetext}
\begin{eqnarray}
\mathcal{H}_\mathrm{int}=\left.\sum_{i=1,2} \gamma \mu_{0} \mathbf{S}_{\mathrm{NV}_{i}} \cdot\left[\mathbf{H}_{\mathrm{d}}(\mathbf{r})+\nabla \int d \mathbf{r}^{\prime} G\left(\mathbf{r}-\mathbf{r}^{\prime}\right)\left(\nabla^{\prime} \cdot \vec{\mathcal{M}}\left(\mathbf{r}^{\prime}\right) \mathcal{F}\left(\mathbf{r}^{\prime}\right)-\frac{\partial_{z}^{\prime} \mathcal{F}\left(\mathbf{r}^{\prime}\right) \vec{\mathcal{M}}^{2}\left(\mathbf{r}^{\prime}\right)}{2 M_\mathrm{s}} \right)\right]\right|_{\mathbf{r}=\mathbf{r}_{i}}.\label{Hint3term}
\end{eqnarray}
In the infinitely long waveguide case, we have $\mathbf{H}_{\mathrm{d}}(\mathbf{r})=0$. Up to the lowest order (linear order) in $\mathbf{m}(\mathbf{r})$, we obtain
\begin{eqnarray}
&&\mathcal{H}_{\mathrm{int}}=\sum_{i=1,2} \left.\gamma \mu_{0} \mathbf{S}_{\mathrm{NV}_{i}} \cdot \mathbf{h}(\mathbf{r})\right|_{\mathbf{r}=\mathbf{r}_{i}},\\
&&\mathbf{h}(\mathbf{r}) = \nabla \int d \mathbf{r}^{\prime} G\left(\mathbf{r}-\mathbf{r}^{\prime}\right)\left(\nabla^{\prime} \cdot \vec{\mathcal{M}}\left(\mathbf{r}^{\prime}\right) \mathcal{F}\left(\mathbf{r}^{\prime}\right)\right).\label{h_def_without(2)}
\end{eqnarray}
As the NV axis is set $\hat{n}_\mathrm{NV}=\hat{z}$, the rotating-wave term comes from the perpendicular contribution $\mathbf{h}_{\perp}(\mathbf{r})={h}_{x}(\mathbf{r})\hat{x}+{h}_{y}(\mathbf{r})\hat{y}$. Using the Bogoliubov transformation~(\ref{BogoWG}), we obtain
\begin{eqnarray}
\mu_{0} \gamma \mathbf{h}_{\perp}(\mathbf{r})=\frac{\sqrt{2 \omega_{M} \omega_{d}}}{\sqrt{w / d^{2}}} \frac{1}{4} \sum_{n m} \int \frac{d k}{2 \pi} e^{i k z}[\widehat{e}_{+}\thinspace \thinspace\widehat{e}_{-}]\left[\begin{array}{cc}\Gamma_{k, n m}^{-,+} & \Gamma_{k, n m}^{-,-} \\ \Gamma_{k, n m}^{+,+} & \Gamma_{k, n m}^{+,-}\end{array}\right]\left[\begin{array}{cc}\lambda_{k,(n, m)} & -\mu_{k,(n, m)} \\ -\mu_{k,(n, m)}^{*} & \lambda_{k,(n, m)}\end{array}\right]\left[\begin{array}{l}\beta_{k,(n, m)} \\ \beta_{-k,(n, m)}^{\dagger}\end{array}\right],
\end{eqnarray}
where $\omega_d=\mu_0(\hbar\gamma)^2/(\hbar d^3)$ and
\begin{eqnarray}
&&\Gamma_{k, n m}^{-,+}=\left(\Gamma_{k, n m}^{X X}+\Gamma_{k, n m}^{Y Y}+i\left(\Gamma_{k, n m}^{X Y}-\Gamma_{k, n m}^{Y X}\right)\right),\\
&& \Gamma_{k, n m}^{-,-}=\left(\Gamma_{k, n m}^{X X}-\Gamma_{k, n m}^{Y Y}-i\left(\Gamma_{k, n m}^{X Y}+\Gamma_{k, n m}^{Y X}\right)\right),\\
&&\Gamma_{k, n m}^{+,+}=\left(\Gamma_{k, n m}^{X X}-\Gamma_{k, n m}^{Y Y}+i\left(\Gamma_{k, n m}^{X Y}+\Gamma_{k, n m}^{Y X}\right)\right),\\
&&\Gamma_{k, n m}^{+,-}=\left(\Gamma_{k, n m}^{X X}+\Gamma_{k, n m}^{Y Y}-i\left(\Gamma_{k, n m}^{X Y}-\Gamma_{k, n m}^{Y X}\right)\right).
\end{eqnarray}
Here, $\Gamma_{k, n m}^{X X}$, $\Gamma_{k, n m}^{X Y}$, $\Gamma_{k, n m}^{Y X}$, and $\Gamma_{k, n m}^{Y Y}$ are functions of $\boldsymbol{\rho}$, and they are given by
\begin{eqnarray}
&&\Gamma_{k, n m}^{X X}=-\int d \boldsymbol{\rho}^{\prime}|k|\left(\widehat{\boldsymbol{\rho}-\boldsymbol{\rho}^{\prime}}\right)_{x} \frac{K_{1}\left(\left|k\left(\boldsymbol{\rho}-\boldsymbol{\rho}^{\prime}\right)\right|\right)}{2 \pi} \partial_x^{\prime}\tilde{\varphi}_{nm}^{XY}(\boldsymbol{\rho}^{\prime})\label{GammaXX},\\
&&\Gamma_{k, n m}^{X Y}=-\int d \boldsymbol{\rho}^{\prime}|k|\left(\widehat{\boldsymbol{\rho}-\boldsymbol{\rho}^{\prime}}\right)_{x} \frac{K_{1}\left(\left|k\left(\boldsymbol{\rho}-\boldsymbol{\rho}^{\prime}\right)\right|\right)}{2 \pi} \partial_y^{\prime}\tilde{\varphi}_{nm}^{XY}(\boldsymbol{\rho}^{\prime})\label{GammaXY},\\
&&\Gamma_{k, n m}^{Y X}=-\int d \boldsymbol{\rho}^{\prime}|k|\left(\widehat{\boldsymbol{\rho}-\boldsymbol{\rho}^{\prime}}\right)_{y} \frac{K_{1}\left(\left|k\left(\boldsymbol{\rho}-\boldsymbol{\rho}^{\prime}\right)\right|\right)}{2 \pi} \partial_x^{\prime}\tilde{\varphi}_{nm}^{XY}(\boldsymbol{\rho}^{\prime})\label{GammaYX},\\
&&\Gamma_{k, n m}^{Y Y}=-\int d \boldsymbol{\rho}^{\prime}|k|\left(\widehat{\boldsymbol{\rho}-\boldsymbol{\rho}^{\prime}}\right)_{y} \frac{K_{1}\left(\left|k\left(\boldsymbol{\rho}-\boldsymbol{\rho}^{\prime}\right)\right|\right)}{2 \pi} \partial_y^{\prime}\tilde{\varphi}_{nm}^{XY}(\boldsymbol{\rho}^{\prime}),\label{GammaYY}
\end{eqnarray}
\end{widetext}
where $\tilde{\varphi}_{nm}^{XY}=\sqrt{dw}\varphi_{nm}^{XY}$ is a dimensionless function and $\widehat{\boldsymbol{\rho}-\boldsymbol{\rho}^{\prime}}=(\boldsymbol{\rho}-\boldsymbol{\rho}^{\prime})/|\boldsymbol{\rho}-\boldsymbol{\rho}^{\prime}|$. We consider the external field range $\gamma H_\mathrm{ext}<D_\mathrm{NV}$, where NV center's ground state is $|g\rangle=|S_z=0\rangle$ and the first excited state is $|e\rangle=|S_z=-\hbar\rangle$. In the NV center's subspace spanned by $\{|g\rangle,|e\rangle\}$, we can write
\begin{equation}
\mathcal{H}_\mathrm{NV}=\sum_{i=1,2}\frac{\hbar\omega_{\mathrm{NV}}}{2}\sigma^z_{\mathrm{NV}_i},\label{NVLower}
\end{equation} 
where $\omega_\mathrm{NV}=D_\mathrm{NV}-\gamma H_\mathrm{ext}$, $\sigma_\mathrm{NV}^z=|e\rangle\langle e|-|g\rangle\langle g|$, and we drop a constant shift in energy. We also have $S_\mathrm{NV}^+=\sqrt{2}\hbar\sigma_\mathrm{NV}^-$ and $S_\mathrm{NV}^-=\sqrt{2}\hbar\sigma_\mathrm{NV}^+$, where $\sigma^+_{\mathrm{NV}}=|e\rangle \langle g|$, and $\sigma^-_{\mathrm{NV}}=|g\rangle \langle e|$. Under the rotating wave approximation, we obtain
\begin{widetext}
\begin{eqnarray}
\mathcal{H}_\mathrm{int}\approx \hbar \sum_{i=1,2} \frac{\sqrt{\omega_{M} \omega_{d}}}{\sqrt{w / d^{2}}} \sum_{n m} \int \frac{d k}{2 \pi}\left.\frac{1}{2}\left(\Gamma_{k, n m}^{+,+} \lambda_{k,(n, m)}-\Gamma_{k, n m}^{+,-} \mu_{k,(n, m)}^{*}\right)\right|_{\boldsymbol{\rho}=\boldsymbol{\rho}_{i}} \sigma_{\mathrm{NV}_{i}}^{+} \beta_{k,(n, m)} e^{i k z_{i}}+\mathrm{H.c.}.
\end{eqnarray}
Limiting our discussion to the subspace with $(n,m)=(0,0)$, we obtain
\begin{eqnarray}
&&\mathcal{H}_\mathrm{int}=\hbar\frac{\sqrt{\omega_M\omega_d}}{\sqrt{w/d^2}}\sum_{i=1,2}\int\frac{dk}{2\pi}g({\bm{\rho}}_i,k)\sigma_{\mathrm{NV}_i}^+\beta_{k,(0,0)}e^{ikz_i}+\mathrm{H.c.},\label{HintWG}\\
&&g({\bm{\rho}}_i,k)=\left.\left(\left(\Gamma_{k, n m}^{+,+} / 2\right) \lambda_{k,(n, m)}-\left(\Gamma_{k, n m}^{+,-} / 2\right) \mu_{k,(n, m)}^{*}\right)\right|_{\boldsymbol{\rho}=\boldsymbol{\rho}_{i}},
\end{eqnarray}
\end{widetext}
which is presented in the main text. Here, $g({\bm{\rho}}_i,k)$ is the dimensionless coupling. To calculate the spatial distribution of the dimensionless coupling, we evaluate Eqs.~(\ref{GammaXX})-(\ref{GammaYY}) numerically.

\subsection{Effective NV-NV Hamiltonian}
The NV-NV interaction mediated by magnons can be calculated via the Schrieffer-Wolff transformation~\cite{bravyi2011schrieffer}, $\mathcal{H}\rightarrow D\mathcal{H} D^\dagger$ with $D=\exp( S-S^\dagger)$. Here, Eqs.~(\ref{HmagSimp}), (\ref{NVLower}), and (\ref{HintWG}) are used in $\mathcal{H}=\mathcal{H}_0+\mathcal{H}_\mathrm{int}$ with $\mathcal{H}_0=\mathcal{H}_\mathrm{NV}+\mathcal{H}_\mathrm{m}$. We pick
\begin{equation}
S=\frac{\sqrt{\omega_{M} \omega_{d}}}{\sqrt{w / d^{2}}} \sum_{i=1,2} \int \frac{d k}{2 \pi} \frac{g({\bm{\rho}}_i,k) \sigma_{\mathrm{NV}_{i}}^{+} \beta_{k,(0,0)} e^{i k z_{i}}}{\omega_{\mathrm{NV}}-\omega_{k,(0,0)}},\label{SWGen}
\end{equation} 
such that $[S-S^\dagger,\mathcal{H}_0]=-\mathcal{H}_\mathrm{int}$. Noting that we can write $S-S^\dagger=(i/\hbar)\int_{-\infty}^0d\tau \mathcal{H}_\mathrm{int}(\tau)$, where $\mathcal{H}_\mathrm{int}(\tau)$ is the interaction Hamiltonian in the interaction picture, we obtain the following effective Hamiltonian
\begin{eqnarray}
\mathcal{H}_\mathrm{eff}=\frac{1}{2}[S-S^\dagger,\mathcal{H}_\mathrm{int}]=\frac{i}{2\hbar}\int_{-\infty}^0 d\tau [\mathcal{H}_\mathrm{int}(\tau),\mathcal{H}_\mathrm{int}],\nonumber\\ \label{HeffFull}
\end{eqnarray}
which is related to the linear response theory. This effective Hamiltonian includes the Lamb shift, the Stark shift, and the NV-NV interaction. The NV-NV interaction contribution is, assuming $\boldsymbol{\rho}_1=\boldsymbol{\rho}_2$ and writing $g(k)=g({\bm{\rho}}_i,k)$,
\begin{eqnarray}
&&\mathcal{H}_\mathrm{eff}^\mathrm{NV-NV}=-\hbar\left(g_{\mathrm{eff}} \sigma_{\mathrm{NV}_{1}}^{+} \sigma_{\mathrm{NV}_{1}}^{-}+\mathrm{H.c.}\right),\label{geffWFHamil}\\
&&g_{\mathrm{eff}} = \frac{\omega_{M} \omega_{d}}{w / d^{2}} \int \frac{d k}{2 \pi}|g(k)|^{2} \frac{\exp \left[i k\left(z_{1}-z_{2}\right)\right]}{\omega_{k,(0,0)}-\omega_{\mathrm{NV}}},\label{geffWG}
\end{eqnarray}
which is presented in the main text. Here, $g_\mathrm{eff}$ is the effective NV-NV coupling strength. The entangling gate rate presented in Fig.~\ref{fig2}(e) is based on the inverse of the time required for the $\sqrt{i\mathrm{SWAP}}$ gate, $\tau_{\sqrt{i\mathrm{SWAP}}}=\pi/(4|g_\mathrm{eff}|)$, under the interaction Hamiltonian~(\ref{geffWFHamil}).

Analytic expression of $g_\mathrm{eff}$ presented in the main text is obtained by the following approximations. We first expand the dispersion $\omega_{k,(0,0)}$ around the two energy minimum at $k=\pm k_\mathrm{min}$ and approximate $g(k)\approx g(k_\mathrm{min})$. Secondly, we also approximate the curvature to be exchange dominated, i.e., $\omega_{l,(0,0)}\approx \omega_{k_\mathrm{min},(0,0)}+D_\mathrm{ex}(k\mp k_\mathrm{min})^2$. Then we obtain, after writing $\Delta\omega=\omega_{k_\mathrm{min},(0,0)}-\omega_\mathrm{NV}$,
\begin{widetext}
\begin{eqnarray}
g_{\mathrm{eff}} &\approx& \frac{\omega_{M} \omega_{d}}{w / d^{2}}\left|g\left(k_{\min }\right)\right|^{2}\left(\int_{-\infty}^{\infty} \frac{d k}{2 \pi} \frac{\exp \left[i k\left(z_{1}-z_{2}\right)\right]}{D_{\mathrm{ex}}\left(k-k_{\min }\right)^{2}+\Delta \omega}+\int_{-\infty}^{\infty} \frac{d k}{2 \pi} \frac{\exp \left[i k\left(z_{1}-z_{2}\right)\right]}{D_{\operatorname{ex}}\left(k+k_{\min }\right)^{2}+\Delta \omega}\right)\nonumber,\\
&=&\frac{\omega_{M} \omega_{\bar{d}}}{\Delta \omega}\left|g\left(k_{\min }\right)\right|^{2} \cos \left(k_{\min } \delta z\right) \exp \left[\delta z / \xi_{0}\right],\label{geffAnalitic}
\end{eqnarray}
\end{widetext}
where $\xi_0=\sqrt{D_\mathrm{ex}/\Delta\omega}$, $\delta z=|z_1-z_2|$ and $\omega_{\bar{d}}=\mu_0(\gamma\hbar)^2/(\hbar dw\xi_0)$. Note that the circle dots in Fig.~\ref{fig2}(e) are obtained by the numerical evaluation of Eq.~(\ref{geffWG}), while the solid curves are obtained from the analytical expression~(\ref{geffAnalitic}), thus showing the great agreement between them.

To evaluate how good the perturbation is, we consider one NV case and recall the wave function modification in the first order perturbation
\begin{eqnarray}
|n^{(1)}\rangle=\frac{1}{E_n^{(0)}-\mathcal{H}_0}\mathcal{H}_\mathrm{int}|n^{0}\rangle=\sum_{k(\neq n)}\frac{\langle k^{(0)}|\mathcal{H}_\mathrm{int}|n^{(0)}\rangle}{E_n^{(0)}-E_k^{(0)}}|k^{(0)}\rangle,\nonumber\\
\end{eqnarray}
where $|n^{(0)}\rangle$ and $E_n^{(0)}$ are the unperturbed eigenstate and eigenenergy. The fraction of the finite magnon-number state contribution in the original ground state $|n^{(0)}\rangle=|g\rangle|0\rangle_\mathrm{m}$ is, where $|0\rangle_\mathrm{m}$ is the magnon vacuum,
\begin{eqnarray}
\left\|\left|n^{(1)}\right\rangle \right\|^{2}&=&\sum_{k(\neq n)}\left|\frac{\left\langle k^{(0)}|\mathcal{H}_\mathrm{int}| n^{(0)}\right\rangle}{E_{n}^{(0)}-E_{k}^{(0)}}\right|^{2},\nonumber\\
&=&\frac{\omega_{M} \omega_{d}}{w / d^{2}} \int \frac{d k}{2 \pi} \frac{|g(k)|^{2}}{\left(\omega_{k,(0,0)}-\omega_{\mathrm{NV}}\right)^{2}}.
\end{eqnarray}
Under the geometry presented in the red curve in Fig.~\ref{fig2}(e), we obtain $\|\left|n^{(1)}\right\rangle \|^{2}\approx10^{-3}\ll 1$, which indicates the perturbation theory is valid.

To estimate the corresponding cooperativity of the red solid curve in Fig.~\ref{fig2}(e), we assume the waveguide has a length $l$ as in~\cite{flebus2019entangling}. By discretizing the integral $\int dk$ using the periodic boundary condition and rescaling the creation/annihilation operators via $\bar{\beta}_{k,(0,0)}=\beta_{k,(0,0)}/\sqrt{l}$ to have a correct commutation relation for the discretized modes, $[\bar{\beta}_{k,(0,0)},\tilde{\beta}^\dagger_{k',(0,0)}]=\delta_{k,k'}$, the interaction Hamiltonian becomes
\begin{eqnarray}
&&\mathcal{H}_{\mathrm{int}}= \sum_{i=1,2} \sum_{k} \hbar\bar{g}(k) \sigma_{\mathrm{NV}_{i}}^{+} \bar{\beta}_{k,(0,0)} e^{i k z_{i}}+\mathrm{H.c.},\\
&&\bar{g}(k) =\frac{\sqrt{\omega_{M} \omega_{d}}}{\sqrt{l w / d^{2}}} g(k).
\end{eqnarray}
As we are mostly using magnons with $|k|\approx k_\mathrm{min}$ in the virtual-magnon mediated NV-NV coupling, it is reasonable to calculate the equivalent cooperativity with $\bar{g}=\bar{g}(k_\mathrm{min})$:
\begin{eqnarray}
{\mathcal{C}}_{\mathrm{eq}}=\frac{\bar{g}^{2}}{\alpha \omega_{\min }/ T_{2}^{*}}.
\end{eqnarray}
Under the geometry presented in the red curve in Fig.~\ref{fig2}(e), and using the NV center's coherence time~\cite{herbschleb2019ultra} $T_2^*=1\ \mathrm{ms}$ and the Gilbert damping parameter of YIG~\cite{tabuchi2014hybridizing} $\alpha=10^{-5}$, we obtain $\bar{g}\approx130\ \mathrm{kHz}$ and ${\mathcal{C}}_{\mathrm{eq}}\approx3700$.

\subsection{Temperature independence of the effective NV-NV coupling mediated by virtual magnons}

{
Here we show that up to second order in perturbation theory, the NV-NV coupling mediated by the virtual magnons is insensitive to the temperature. For simplicity, here we only consider the case where two NV centers are coupled to a common single $k$-magnon mode with coupling strength $g_k$ for both NV centers, i.e., {$\mathcal{H}_0=\hbar\omega_\mathrm{NV}(\sigma^z_{\mathrm{NV}_1}+\sigma^z_{\mathrm{NV}_2})/2+\hbar(\omega_{\rm{NV}}+\Delta_k)a_k^\dagger a_k$, $\mathcal{H}_\mathrm{int}=\hbar [g_k(\sigma^+_{\mathrm{NV}_1}+\sigma^+_{\mathrm{NV}_2})a_k+\mathrm{H.c.}]$, and $[a_k,a_{k}^\dagger]=1$}, although the discussion can be generalized to a multi-mode or a waveguide case. To demonstrate that, we calculate through the transition matrix formalism the rate $T_{\left|e_{1}g_{2}n_k\right\rangle \rightarrow\left|g_{1}e_{2}n_k\right\rangle }$ from an initial pure state $\left|e_{1}g_{2}n_k\right\rangle $ ({$|n_k\rangle=(a_k^\dagger)^{n_k}|0\rangle/\sqrt{n_k! }$ with $n_k=0,1,2,\cdots$}) to the final state $\left|g_{1}e_{2}n_k\right\rangle $,
\begin{align}
T_{\left|e_{1}g_{2}n_{k}\right\rangle \rightarrow\left|g_{1}e_{2}n_{k}\right\rangle } =\frac{1}{\hbar}\sum_{i}\frac{\left\langle g_{1}e_{2}n_{k}\right|{\cal H}_\mathrm{int}\left|i\right\rangle \left\langle i\right|{\cal H}_\mathrm{int}\left|e_{1}g_{2}n_{k}\right\rangle }{E_{\left|g_{1}e_{2}n_{k}\right\rangle }-E_{\left|i\right\rangle }},
\end{align}
where $\left|i\right\rangle $ represent the whole set of intermediates many-body
states and $E_{|i\rangle}$ is the energy of the state $|i\rangle$ without interaction. The transition is only non-null for $\left|i\right\rangle =|{\mathrm{NV\ states}}\rangle\otimes\left|n_{k}\pm1\right\rangle$, yielding
\begin{eqnarray}
&&T_{\left|e_{1}g_{2}n_{k}\right\rangle \rightarrow\left|g_{1}e_{2}n_{k}\right\rangle }\nonumber\\
&&=\frac{1}{\hbar}\frac{\left\langle g_{1}e_{2}n_{k}\right|{\cal H}_\mathrm{int}\left|g_{1}g_{2}n_{k}+1\right\rangle \left\langle g_{1}g_{2}n_{k}+1\right|{\cal H}_\mathrm{int}\left|e_{1}g_{2}n_{k}\right\rangle }{E_{\left|g_{1}e_{2}n_{k}\right\rangle }-E_{\left|g_{1}g_{2}n_{k}+1\right\rangle }}\nonumber\\
&&\ \ +\frac{1}{\hbar}\frac{\left\langle g_{1}e_{2}n_{k}\right|{\cal H}_\mathrm{int}\left|e_{1}e_{2}n_{k}-1\right\rangle \left\langle e_{1}e_{2}n_{k}-1\right|{\cal H}_\mathrm{int}\left|e_{1}g_{2}n_{k}\right\rangle }{E_{\left|g_{1}e_{2}n_{k}\right\rangle }-E_{\left|e_{1}e_{2}n_{k}-1\right\rangle }},\nonumber\\
&&=\frac{1}{\hbar}\frac{\hbar g_k\sqrt{n_{k}+1}\hbar g^{*}_k\sqrt{n_{k}+1}}{E_{\left|g_{1}e_{2}n_{k}\right\rangle }-E_{\left|g_{1}g_{2}n_{k}+1\right\rangle }}+\frac{1}{\hbar}\frac{\hbar g^{*}_k\sqrt{n_{k}}\hbar g_k\sqrt{n_{k}}}{E_{\left|g_{1}e_{2}n_{k}\right\rangle }-E_{\left|e_{1}e_{2}n_{k}-1\right\rangle }}.\nonumber\\
\end{eqnarray}
By identifying $E_{\left|g_{1}e_{2}n_{k}\right\rangle }-E_{\left|g_{1}g_{2}n_{k}+1\right\rangle }=-\hbar\Delta_k$
and $E_{\left|g_{1}e_{2}n_{k}\right\rangle }-E_{\left|e_{1}e_{2}n_{k}-1\right\rangle }=\hbar\Delta_k$,
we obtain
\begin{equation}
T_{\left|e_{1}g_{2}n_{k}\right\rangle \rightarrow\left|g_{1}e_{2}n_{k}\right\rangle }=\frac{\left(n_{k}+1\right)|g_{k}|^{2}}{-\Delta_k}+\frac{n_{k}|g_{k}|^{2}}{\Delta_k}=-\frac{|g_{k}|^{2}}{\Delta_k},    
\end{equation}
thus first proving the insensitivity to the initial magnon state $\left|n_{k}\right\rangle$. Moreover, we recall that for finite temperature we do not have the pure initial state $\left|e_{1}g_{2}n_{k}\right\rangle$  for a specific $n_{k}$ but rather a statistic mix of them, given by the quantum thermal state $\rho_{0}=Z^{-1}\sum_{n_k}e^{-\beta\hbar n_{k}\omega_{k}} \left|e_{1}g_{2}n_{k}\right\rangle \left\langle e_{1}g_{2} n_{k}\right|$, $Z=\sum_{n_{k}}e^{-\beta\hbar n_{k}\omega_{k}}$ with the inverse temperature $\beta=1/k_{\rm{B}} T$ and $\omega_k=\omega_{\rm{NV}}+\Delta_k$. Finally, using the linearity of the quantum evolution it is straightforward to prove the temperature independence of the off-resonance transition $\left|e_{1}g_{2}\right\rangle \rightarrow\left|g_{1}e_{2}\right\rangle$.
}

\section{Finite length ferromagnetic bar}
\subsection{Diagonalization of the magnon Hamiltonian}
The NV-magnon coupling strength is even stronger under the magnon confinement effect where the ferromagnet length $l$ is finite. To diagonalize the magnon Hamiltonian Eq.~(\ref{SimpHmagAll}), in the same way as in Sec.~I, we expand the canonical variables as
\begin{eqnarray}
&&a(\mathbf{r})=\sum_{n m p} f_{n}^{X}(x) f_{m}^{Y}(y) f_{p}^{Z}(z) a_{(n m p)},\label{YIGbar_a}\\
&&a^{*}(\mathbf{r})=\sum_{n m p} f_{n}^{X}(x) f_{m}^{Y}(y) f_{p}^{Z}(z) a_{(n m p)}^{*},\label{YIGbar_adag}\\
&&f_{p}^{Z}(z)=\sqrt{ \mathcal{F}^{Z}(z)}\psi_p^Z(z),\\
&&\psi_p^Z(z)=\sqrt{\frac{2}{\left(1+\delta_{p, 0}\right) l}} \cos \left(\kappa_{p}^{z} z\right),
\end{eqnarray}
where $\kappa_p^Z=p\pi/l$, $p=0,1,\cdots$, and $\mathcal{F}^Z(z)=\Theta(z)\Theta(l-z)$. Note that we have $\mathcal{F}(\mathbf{r})=\mathcal{F}^X(x)\mathcal{F}^Y(y)\mathcal{F}^Z(z)$. After simplification and writing $\mu=(nmp)$, the magnon Hamiltonian Eq.~(\ref{SimpHmagAll}) with corresponding parameters become
\begin{eqnarray}
&&\mathcal{H}_\mathrm{m}=\frac{\omega_{M}}{2} \sum_{\mu_{1} \mu_{2}}\left[a_{\mu_{1}}^{*}\thinspace \thinspace a_{\mu_{1}}\right]\left[\begin{array}{ll}A_{\mu_{1} \mu_{2}} & B_{\mu_{1} \mu_{2}} \\ B_{\mu_{1} \mu_{2}}^{*} & A_{\mu_{1} \mu_{2}}^{*}\end{array}\right]\left[\begin{array}{l}a_{\mu_{2}}  \\ a_{\mu_{2}}^{*}\end{array}\right],\nonumber\\ \label{HmatrixBar} \\
&&A_{\mu_{1} \mu_{2}} = \widetilde{\Delta}_{\mu_{1}} \delta_{\mu_{1} \mu_{2}}-\mathcal{N}_{\mu_{1} \mu_{2}}+H_{\mu_{1} \mu_{2}}^{00} \label{s90},\\
&&B_{\mu_{1} \mu_{2}} = H_{\mu_{1} \mu_{2}}^{01} \label{s91},\\
&&\widetilde{\Delta}_{(n m p)} = (\omega_{H}+D_{\mathrm{ex}} K_{(n m p)}^{2})/\omega_{M}, \label{s92}\\
&&\mathcal{N}_{\mu_{1} \mu_{2}} =- \int d \mathbf{r} \tilde{H}_{\mathrm{d}}^{z}(\mathbf{r}) f_{\mu_{1}}^{X Y Z}(\mathbf{r}) f_{\mu_{2}}^{X Y Z}(\mathbf{r}),\label{DemagInt1}
\end{eqnarray}
where $K_{(n m p)}^{2}=\left(\kappa_{n}^{X}\right)^{2}+\left(\kappa_{m}^{Y}\right)^{2}+\left(\kappa_{p}^{Z}\right)^{2}$, $f_\mu^{XYZ}(\mathbf{r})=f_n^X(x)f_m^Y(y)f_p^Z(z)$, $\tilde{H}_{\mathrm{d}}^{z}(\mathbf{r})$ is a dimensionless demagnetization field
\begin{equation}
\widetilde{H}_{\mathrm{d}}^{z}(\mathbf{r}) = \frac{H_{\mathrm{d}}^{z}(\mathbf{r})}{M_{s}}=\frac{1}{M_{s}} \partial_{z} \int d \mathbf{r}^{\prime} G\left(\mathbf{r}-\mathbf{r}^{\prime}\right)\left(\nabla^{\prime} \cdot \mathbf{M}_{0}\left(\mathbf{r}^{\prime}\right)\right),\label{DemagInt2}
\end{equation}
and $H_{\mu_{1} \mu_{2}}^{00}$ and $H_{\mu_{1} \mu_{2}}^{01}$ are given by
\begin{eqnarray}
&&H_{\mu_{1} \mu_{2}}^{00}=\frac{1}{2}\left(H_{\mu_{1} \mu_{2}}^{X X}+H_{\mu_{1} \mu_{2}}^{Y Y}+i\left(H_{\mu_{1} \mu_{2}}^{X Y}-H_{\mu_{1} \mu_{2}}^{Y X}\right)\right),\nonumber\\ \\
&&H_{\mu_{1} \mu_{2}}^{01}=\frac{1}{2}\left(H_{\mu_{1} \mu_{2}}^{X X}-H_{\mu_{1} \mu_{2}}^{Y Y}-i\left(H_{\mu_{1} \mu_{2}}^{X Y}+H_{\mu_{1} \mu_{2}}^{Y X}\right)\right).\nonumber\\
\end{eqnarray}
Here, $H_{\mu_{1} \mu_{2}}^{X X}$, $H_{\mu_{1} \mu_{2}}^{X Y}$, $H_{\mu_{1} \mu_{2}}^{Y X}$, and $H_{\mu_{1} \mu_{2}}^{Y Y}$ are given by
\begin{widetext}
\begin{eqnarray}
&&H_{\mu_{1}\mu_{2}}^{X X}=\int d \mathbf{r}_{1} d \mathbf{r}_{2}\left(\partial_{x_{1}} \varphi_{\mu_1}^{X Y Z}\left(\mathbf{r}_{1}\right)\right) G\left(\mathbf{r}_{1}-\mathbf{r}_{2}\right)\left(\partial_{x_{2}} \varphi_{\mu_2}^{X Y Z}\left(\mathbf{r}_{2}\right)\right)\label{HXXbar},\\
&&H_{\mu_{1}\mu_{2}}^{X Y}=\int d \mathbf{r}_{1} d \mathbf{r}_{2}\left(\partial_{x_{1}} \varphi_{\mu_1}^{X Y Z}\left(\mathbf{r}_{1}\right)\right) G\left(\mathbf{r}_{1}-\mathbf{r}_{2}\right)\left(\partial_{y_{2}} \varphi_{\mu_2}^{X Y Z}\left(\mathbf{r}_{2}\right)\right)\label{HXYbar},\\
&&H_{\mu_{1}\mu_{2}}^{Y X}=\int d \mathbf{r}_{1} d \mathbf{r}_{2}\left(\partial_{y_{1}} \varphi_{\mu_1}^{X Y Z}\left(\mathbf{r}_{1}\right)\right) G\left(\mathbf{r}_{1}-\mathbf{r}_{2}\right)\left(\partial_{x_{2}} \varphi_{\mu_2}^{X Y Z}\left(\mathbf{r}_{2}\right)\right)\label{HYXbar},\\
&&H_{\mu_{1}\mu_{2}}^{Y Y}=\int d \mathbf{r}_{1} d \mathbf{r}_{2}\left(\partial_{y_{1}} \varphi_{\mu_1}^{X Y Z}\left(\mathbf{r}_{1}\right)\right) G\left(\mathbf{r}_{1}-\mathbf{r}_{2}\right)\left(\partial_{y_{2}} \varphi_{\mu_2}^{X Y Z}\left(\mathbf{r}_{2}\right)\right)\label{HYYbar},
\end{eqnarray}
\end{widetext}
where $\varphi^{XYZ}_{n m p}(\mathbf{r})=\mathcal{F}(\mathbf{r})\psi_n^X(x)\psi_m^Y(y)\psi_p^Z(z)$. Note that we have relations $A_{\mu_1\mu_2}=A^*_{\mu_2\mu_1}$ and $B_{\mu_1\mu_2}=B_{\mu_2\mu_1}$.

Finally, the Hamiltonian Eq.~(\ref{HmatrixBar}) can be written in the matrix form
\begin{eqnarray}
\mathcal{H}_{\mathrm{m}}=\frac{\omega_{M}}{2}\left[\boldsymbol{\alpha}^{*}\thinspace \thinspace \boldsymbol{\alpha}\right] \hat{\mathbf{H}}\left[\begin{array}{c}\boldsymbol{\alpha} \\ \boldsymbol{\alpha}^{*}\end{array}\right],
\end{eqnarray}
where $\boldsymbol{\alpha}=[a_{\mu_0}\thinspace \thinspace a_{\mu_1}\thinspace \thinspace \cdots]$, $\boldsymbol{\alpha}^*=[a^*_{\mu_0}\thinspace \thinspace a^*_{\mu_1}\thinspace \thinspace \cdots]$, and we transpose $\boldsymbol{\alpha}$ or $\boldsymbol{\alpha}^*$ if necessary. No confusion is expected for the column or row vectors for $\boldsymbol{\alpha}$ and $\boldsymbol{\alpha}^\dagger$ as in Refs.~[\onlinecite{colpa1978diagonalization}] and [\onlinecite{shindou2013topological}]. This Hamiltonian matrix can be diagonalized by the paraunitary matrix~\cite{colpa1978diagonalization} $\mathbf{T}$ via
\begin{eqnarray}
&&\left[\begin{array}{c}\boldsymbol{\alpha} \\ \boldsymbol{\alpha}^{*}\end{array}\right]=\mathbf{T}\left[\begin{array}{c}\boldsymbol{\beta} \\ \boldsymbol{\beta}^{*}\end{array}\right],\\
&&\mathcal{H}_{\mathrm{m}}=\frac{\omega_{M}}{2}\left[\boldsymbol{\beta}^{*}\thinspace \thinspace \boldsymbol{\beta}\right] \left[\begin{array}{cc}\mathbf{E}&  \mathbf{O}\\ \mathbf{O} & \mathbf{E}\end{array}\right]\left[\begin{array}{c}\boldsymbol{\beta} \\ \boldsymbol{\beta}^{*}\end{array}\right],
\end{eqnarray}
where, $\boldsymbol{\beta}=[\beta_{\mu_0}\thinspace \thinspace \beta_{\mu_1}\thinspace \thinspace \cdots]$ and $\boldsymbol{\beta}^*=[\beta^*_{\mu_0}\thinspace \thinspace \beta^*_{\mu_1}\thinspace \thinspace \cdots]$ are normal mode magnon complex canonical variables, and $\omega_M\mathbf{E}=\mathrm{diag}[\omega_{\mu_0},\omega_{\mu_1},\cdots]$ is a diagonal matrix whose entries are magnon eigenfrequencies with $0\leq\omega_{\mu_0}\leq\omega_{\mu_1}\leq\cdots$. The paraunitary matrix $\mathbf{T}$ satisfies
\begin{eqnarray}
&&\mathbf{T}^\dagger\boldsymbol{\sigma}_3\mathbf{T}=\boldsymbol{\sigma}_3,\\ &&\boldsymbol{\sigma}_3=\mathrm{diag}[+1,+1,\cdots,+1,-1,-1,\cdots,-1].
\end{eqnarray}
Based on Ref.~[\onlinecite{colpa1978diagonalization}], one can find the  paraunitary matrix $\mathbf{T}$ using a method based on the Cholesky decomposition. The outline of the method is shown in the following.
\begin{enumerate}
\item Firstly, we decompose $\hat{\mathbf{H}}$ into a product of an upper triangle matrix $\mathbf{K}$ and its Hermitian conjugate using the Cholesky decomposition
\begin{equation}
\hat{\mathbf{H}}=\mathbf{K}^{\dagger} \mathbf{K}.
\end{equation}
\item Next, we define a new Hermitian matrix $\mathbf{W}=\mathbf{K}\boldsymbol{\sigma}_3\mathbf{K}^\dagger$ and diagonalize this matrix with a unitary matrix $\mathbf{U}$:
\begin{equation}
\mathbf{U}^{\dagger} \mathbf{W} \mathbf{U}=\left[\begin{array}{cc}\mathbf{E} & \mathbf{O} \\ \mathbf{O} & -\mathbf{E}\end{array}\right].
\end{equation}
Note that one can find $\mathbf{U}$ such that the right-hand side becomes the desired form, which is proven in Ref.~[\onlinecite{colpa1978diagonalization}].
\item Lastly, we define the following matrix $\widetilde{\mathbf{T}}$:
\begin{equation}
\widetilde{\mathbf{T}}=\mathbf{K}^{-1} \mathbf{U}\left[\begin{array}{cc}\mathbf{E}^{1 / 2} & \mathbf{O} \\ \mathbf{O} & -\mathbf{E}^{1 / 2}\end{array}\right]=\left[\begin{array}{cc}\widetilde{\mathbf{T}}^{p p} & \widetilde{\mathbf{T}}^{p n} \\ \widetilde{\mathbf{T}}^{n p} & \widetilde{\mathbf{T}}^{n n}\end{array}\right].
\end{equation}
Then the desired paraunitary matrix is 
\begin{equation}
\mathbf{T}=\left[\begin{array}{cc}\mathbf{T}^{p p} & \mathbf{T}^{p n} \\ \mathbf{T}^{n p} & \mathbf{T}^{n n}\end{array}\right]=\left[\begin{array}{cc}\widetilde{\mathbf{T}}^{p p} & \left(\widetilde{\mathbf{T}}^{n p}\right)^{*} \\ \widetilde{\mathbf{T}}^{n p} & \left(\widetilde{\mathbf{T}}^{p p}\right)^{*}\end{array}\right].
\end{equation}
\end{enumerate}

To obtain the eigenfrequencies of the magnons for the finite magnetic bar case, we restrict our discussion for $(n,m)=(0,0)$ and consider $p=0,1,\cdots,N$, where $p=N$ is the highest $z$-directional wavenumber to be taken into account and we truncated the sum. We set $\mu_0=(000)$, $\mu_{1}=(001)$, $\cdots$, $\mu_N=(00N)$. After the above Bogoliubov transformation with the paraunitary matrix, we obtain
\begin{equation}
\mathcal{H}_\mathrm{m}=\sum_{p=0,1,\cdots}\omega_{(00p)}\beta^*_{(00p)}\beta_{(00p)},
\end{equation}
with corresponding transformation given by
\begin{eqnarray}
&&\left[\begin{array}{c}a_{(000)} \\ a_{(001)} \\ \vdots \\ a_{(00 N)}\end{array}\right]=\mathbf{T}^{p p}\left[\begin{array}{c}\beta_{(000)} \\ \beta_{(001)} \\ \vdots \\ \beta_{(00 N)}\end{array}\right]+\mathbf{T}^{p n}\left[\begin{array}{c}\beta_{(000)}^* \\ \beta_{(001)}^* \\ \vdots \\ \beta_{(00 N)}^*\end{array}\right],\\
&&\left[\begin{array}{c}a^*_{(000)} \\ a^*_{(001)} \\ \vdots \\ a^*_{(00 N)}\end{array}\right]=(\mathbf{T}^{p n})^*\left[\begin{array}{c}\beta_{(000)} \\ \beta_{(001)} \\ \vdots \\ \beta_{(00 N)}\end{array}\right]+(\mathbf{T}^{p p})^*\left[\begin{array}{c}\beta_{(000)}^* \\ \beta_{(001)}^* \\ \vdots \\ \beta_{(00 N)}^*\end{array}\right].\nonumber\\
\end{eqnarray}
To calculate the magnon eigenfrequencies, we evaluate numerically Eqs.~(\ref{DemagInt1}), (\ref{DemagInt2}) and (\ref{HXXbar})-(\ref{HYYbar}). After promoting $\beta_\mu\rightarrow\sqrt{\hbar}\hat{\beta}_\mu$ and  $\beta^*_\mu\rightarrow\sqrt{\hbar}\hat{\beta}^\dagger_\mu$, we obtain
\begin{equation}
\mathcal{H}_\mathrm{m}=\sum_{p=0,1,\cdots}\hbar\omega_{(00p)}\beta^\dagger_{(00p)}\beta_{(00p)},\label{Hmagbar}
\end{equation}
which is presented in the main text. 

\subsection{NV-magnon coupling}
The coupling strength between magnons and NV centers is obtained by applying the same Bogoliubov transformation with the paraunitary matrix $\mathbf{T}$ [Eq.~(\ref{Hint3term})]. Although the demagnetization field $\mathbf{H}_\mathrm{d}$ contribution in (\ref{Hint3term}) is not negligible when NV centers are placed near the two edges of the ferromagnetic bar, we verify it is small in the calculations for Figs.~\ref{fig4}(d) and \ref{fig5}. In the same way as in Sec.~I, the perpendicular component of the fringing field $\mathbf{h}_\perp$ is given by
\begin{widetext}
\begin{eqnarray}
\gamma \mu_{0} \mathbf{h}_{\perp}(\mathbf{r})=\sqrt{2 \omega_{M} \omega_{d w l}} \frac{1}{4}[\widehat{e}_{+}\thinspace \thinspace\widehat{e}_{-}]\left[\begin{array}{ll}{\left[\Gamma_{\mu_{0}}^{-,+} \cdots \Gamma_{\mu_{N}}^{-,+}\right]} & {\left[\Gamma_{\mu_{0}}^{-,-} \cdots \Gamma_{\mu_{N}}^{-,-}\right]} \\ {\left[\Gamma_{\mu_{0}}^{+,+}\cdots \Gamma_{\mu_{N}}^{+,+}\right]} & {\left[\Gamma_{\mu_{0}}^{+,-}\cdots \Gamma_{\mu_{N}}^{+,-}\right]}\end{array}\right]\left[\begin{array}{cc}\mathbf{T}^{p p} & \mathbf{T}^{p n} \\ \mathbf{T}^{n p} & \mathbf{T}^{n n}\end{array}\right]\left[\begin{array}{c}\boldsymbol{\beta} \\ \boldsymbol{\beta}^\dagger\end{array}\right],\nonumber\\
\end{eqnarray}
\end{widetext}
where $\omega_{dwl}=\mu_0(\gamma\hbar)^2/(\hbar wld)$ and
\begin{eqnarray}
&&\Gamma_{\mu}^{-,+}=\left(\Gamma_{\mu}^{X X}+\Gamma_{\mu}^{Y Y}+i\left(\Gamma_{\mu}^{X Y}-\Gamma_{\mu}^{Y X}\right)\right),\\
&&\Gamma_{\mu}^{-,-}=\left(\Gamma_{\mu}^{X X}-\Gamma_{\mu}^{Y Y}-i\left(\Gamma_{\mu}^{X Y}+\Gamma_{\mu}^{Y X}\right)\right),\\
&&\Gamma_{\mu}^{+,+}=\left(\Gamma_{\mu}^{X X}-\Gamma_{\mu}^{Y Y}+i\left(\Gamma_{\mu}^{X Y}+\Gamma_{\mu}^{Y X}\right)\right),\\
&&\Gamma_{\mu}^{+,-}=\left(\Gamma_{\mu}^{X X}+\Gamma_{\mu}^{Y Y}-i\left(\Gamma_{\mu}^{X Y}-\Gamma_{\mu}^{Y X}\right)\right).
\end{eqnarray}
Here $\Gamma_{\mu}^{X X}$, $\Gamma_{\mu}^{X Y}$, $\Gamma_{\mu}^{Y X}$, and $\Gamma_{\mu}^{Y Y}$ are functions of $\mathbf{r}$, and they are given by
\begin{eqnarray}
&&\Gamma_{\mu}^{X X}=\int d \mathbf{r}^{\prime} \frac{-\left(\mathbf{r}-\mathbf{r}^{\prime}\right)_{x}}{4 \pi\left|\mathbf{r}-\mathbf{r}^{\prime}\right|^{3}} \partial_{x}^{\prime} \tilde{\varphi}_{\mu}^{X Y Z}\left(\mathbf{r}^{\prime}\right)\label{GammaXXbar},\\
&&\Gamma_{\mu}^{X Y}=\int d \mathbf{r}^{\prime} \frac{-\left(\mathbf{r}-\mathbf{r}^{\prime}\right)_{x}}{4 \pi\left|\mathbf{r}-\mathbf{r}^{\prime}\right|^{3}} \partial_{y}^{\prime} \tilde{\varphi}_{\mu}^{X Y Z}\left(\mathbf{r}^{\prime}\right)\label{GammaXYbar},\\
&&\Gamma_{\mu}^{Y X}=\int d \mathbf{r}^{\prime} \frac{-\left(\mathbf{r}-\mathbf{r}^{\prime}\right)_{y}}{4 \pi\left|\mathbf{r}-\mathbf{r}^{\prime}\right|^{3}} \partial_{x}^{\prime} \tilde{\varphi}_{\mu}^{X Y Z}\left(\mathbf{r}^{\prime}\right)\label{GammaYXbar},\\
&&\Gamma_{\mu}^{Y Y}=\int d \mathbf{r}^{\prime} \frac{-\left(\mathbf{r}-\mathbf{r}^{\prime}\right)_{y}}{4 \pi\left|\mathbf{r}-\mathbf{r}^{\prime}\right|^{3}} \partial_{y}^{\prime} \tilde{\varphi}_{\mu}^{X Y Z}\left(\mathbf{r}^{\prime}\right)\label{GammaYYbar},
\end{eqnarray}
where $\tilde{\varphi}_\mu^{XYZ}=\sqrt{wld}\varphi_\mu^{XYZ}$ is a dimensionless function.

In the same way as in Sec.~I, and under the rotating-wave approximation, we obtain the NV-magnon interaction Hamiltonian in the form of the Jaynes-Cummings model
\begin{eqnarray}
&&\mathcal{H}_\mathrm{int}=\sum_{i=1,2}\sum_{\mu=(00p)}\hbar g_\mu(\mathbf{r}_i)\sigma_{\mathrm{NV}_i}^+\beta_\mu+\mathrm{H.c.}\label{Hintbar}\\
&&g_{(00p)}(\mathbf{r}_i) =\sqrt{\omega_{M} \omega_{d w l}}\times\nonumber\\ &&\quad \quad \sum_{q=0,1, \cdots, N}\left.\left[\left(\Gamma_{(00 q)}^{+,+}/2\right)\left[\mathrm{\bf{T}}^{p p}\right]_{q p}+\left(\Gamma_{(00 q)}^{+,-}/2\right)\left[\mathrm{\bf{T}}^{n p}\right]_{q p}\right]\right|_{\mathbf{r}=\mathbf{r}_{i}},\nonumber\\\label{YIGbarCoupling}
\end{eqnarray}
which is presented in the main text. To calculate the spatial distribution of the dimensionless coupling, we evaluate numerically Eqs.~(\ref{GammaXXbar})-(\ref{GammaYYbar}).

\subsection{Effective NV-NV Hamiltonian}
When we introduce a detuning between the target mode frequency $\omega_{(00p)}$ and the NV frequency $\omega_\mathrm{NV}$,  we obtain an effective Hamiltonian in the same way as in Sec.~I. Now the total Hamiltonian  $\mathcal{H}=\mathcal{H}_0+\mathcal{H}_\mathrm{int}$ with $\mathcal{H}_0=\mathcal{H}_\mathrm{NV}+\mathcal{H}_\mathrm{m}$ is given by Eqs.~(\ref{NVLower}), (\ref{Hmagbar}), and (\ref{Hintbar}). For the the Schrieffer-Wolff transformation, we choose
\begin{eqnarray}
S=\sum_{i=1,2 \atop \mu=(00 p)} \frac{g_{\mu}\left(\mathbf{r}_{i}\right) \sigma_{\mathrm{NV}_{i}}^{+} \beta_{\mu}}{\omega_{\mathrm{NV}_{i}}-\omega_{\mu}},
\end{eqnarray}
in the same way as in Eq.~(\ref{SWGen}). Following Eq.~(\ref{HeffFull}), we obtain
\begin{eqnarray}
\mathcal{H}_\mathrm{eff}&=&\hbar \sum_{i, \mu} \frac{\left|g_{\mu}\left(\mathbf{r}_{i}\right)\right|^{2}}{\omega_{\mathrm{NV}}-\omega_{\mu}}\left(|e\rangle_{i}\langle e|+\sigma_{\mathrm{NV}_{i}}^{z} \beta_{\mu}^{\dagger} \beta_{\mu}\right)\nonumber\\
&&+\frac{\hbar}{2} \sum_{i, \mu \neq v}\left( \frac{g_{\mu}\left(\mathbf{r}_{i}\right) g_{v}^{*}\left(\mathbf{r}_{i}\right)}{\omega_{\mathrm{NV}}-\omega_{\mu}} \sigma_{\mathrm{NV}_{i}}^{z} \beta_{v}^{\dagger} \beta_{\mu}+\mathrm{H.c.}\right)\nonumber\\
&&+\hbar\sum_{\mu}\left(\frac{g_{\mu}\left(\mathbf{r}_{1}\right) g_{\mu}^{*}\left(\mathbf{r}_{2}\right)}{\omega_{\mathrm{NV}}-\omega_{\mu}} \sigma_{\mathrm{NV}_{1}}^{+} \sigma_{\mathrm{NV}_{2}}^{-}+\mathrm{H.c.}\right),\nonumber\\
\end{eqnarray}
where the first right hand side term is the Lamb shift and the Stark shift, respectively. The interaction Hamiltonian between the two NV centers is given by the last right hand side term. If we detune the NV frequency from the mode frequency for $\mu=(00p)$ by $\omega_{\mathrm{NV}}=\omega_{(00 p)}-\Delta \omega$ and if we only consider the effect from the mode $\mu=(00p)$, we obtain
\begin{eqnarray}
&&\mathcal{H}_\mathrm{eff}^\mathrm{NV-NV}=-\hbar\left(g_{\mathrm{eff}} \sigma_{\mathrm{NV}_{1}}^{+} \sigma_{\mathrm{NV}_{2}}^{-}+\mathrm{H.c.}\right),\\
&&g_{\mathrm{eff}}=g_{(00 p)}\left(\mathbf{r}_{1}\right) g_{(00 p)}^{*}\left(\mathbf{r}_{2}\right) / \Delta \omega.\label{geffbar}
\end{eqnarray}
In Fig.~\ref{fig4}(c), we focus on the magnon mode with $p=5$ and plot the bare coupling $g_{(00p)}(\mathbf{r})$. In Fig.~\ref{fig4}(d), we use Eq.~(\ref{geffbar}) focusing on the magnon mode with $p=5$ and plot the effective coupling strength $g_\mathrm{eff}$.

\section{Transduction and virtual-magnon exchange protocols}
\subsection{Governing equations for numerical simulations}
The comparison between the two entanglement protocols discussed in the main manuscript is performed with the Lindblad master equation simulation~\cite{lindblad1976,breuer2002theory} focusing only on the magnon mode with $\mu=(00p)$, $p=5$, as presented in Eq.~(\ref{mainLindblad}). The total Hamiltonian $\mathcal{H}=\mathcal{H}_\mathrm{NV}+\mathcal{H}_\mathrm{m}+\mathcal{H}_\mathrm{int}$ to be used is given by Eqs.~(\ref{NVLower}), (\ref{Hmagbar}), and (\ref{Hintbar}),
\begin{eqnarray}
&&\mathcal{H}_\mathrm{NV}=\sum_{i=1,2}\frac{\hbar\omega_{\mathrm{NV}}}{2}\sigma^z_{\mathrm{NV}_i}, \\
&&\mathcal{H}_\mathrm{m}=\hbar\omega_{(00p)}\beta^\dagger_{(00p)}\beta_{(00p)},\\
&&\mathcal{H}_\mathrm{int}=\sum_{i=1,2}\hbar g_{(00p)}(\mathbf{r}_i)\sigma_{\mathrm{NV}_i}^+\beta_\mu+\mathrm{H.c.},
\end{eqnarray}
where we only considered a single magnon mode $\mu=(00p)=(005)$. The identification $\kappa=\alpha\omega_\mu$ presented in Sec.~V is appropriate as the dissipation term in the LLG equation $\left.\partial_t\mathbf{M}\right|_{\mathrm{diss}}=+(\alpha/M_{\mathrm{s}})\mathbf{M}\times \partial_t\mathbf{M}$ results in $\partial_t\beta_\mu\approx-i\omega_\mu\beta_\mu-\alpha\omega_\mu\beta_\mu$, which is consistent with the master equation result $\partial_t\langle a\rangle=-i\omega_\mu\langle a\rangle-\kappa\langle a\rangle$ when considering only the Boson Hamiltonian. More specifically, as we are considering the case where the equilibrium magnetization is along the $z$-axis, the linearized equation of motion yields  $\partial_t\mathbf{m}(\mathbf{r})=\left.\partial_t\mathbf{m}(\mathbf{r})\right|_{\mathrm{coh}}+\alpha\hat{z}\times\partial_t\mathbf{m}(\mathbf{r})$, where $\left.\partial_t\mathbf{m}(\mathbf{r})\right|_{\mathrm{coh}}$ is the coherent evolution part described by Eq.~(\ref{LLGcoherent}). This leads to 
\begin{eqnarray}
&&\partial_t a(\mathbf{r})=-i\frac{\mathcal{H}}{\delta a^*(\mathbf{r})}-i\alpha\partial_t a(\mathbf{r}),\\
&&\partial_t a^*(\mathbf{r})=+i\frac{\mathcal{H}}{\delta a(\mathbf{r})}+i\alpha\partial_t a^*(\mathbf{r}).
\end{eqnarray}
The positive frequency solutions [solutions with $\sim\mathrm{exp}(-i\omega t)$] are obtained by finding nontrivial solutions of
\begin{eqnarray}
-i\omega a(\mathbf{r})&=&-i\frac{\mathcal{H}}{\delta a^*(\mathbf{r})}-\alpha\omega a(\mathbf{r})\nonumber\\
&=&-i\frac{\delta}{\delta a^*(\mathbf{r})}[\mathcal{H}-i\alpha\omega\int d\mathbf{r} a^*(\mathbf{r})a(\mathbf{r})],\\
-i\omega a^*(\mathbf{r})&=&+i\frac{\mathcal{H}}{\delta a(\mathbf{r})}+\alpha\omega a^*(\mathbf{r})\nonumber\\
&=&+i\frac{\delta}{\delta a(\mathbf{r})}[\mathcal{H}-i\alpha\omega\int d\mathbf{r} a^*(\mathbf{r})a(\mathbf{r})].
\end{eqnarray}
Here we notice that one can write $\mathcal{H}-i\alpha\omega\int d\mathbf{r} a^*(\mathbf{r})a(\mathbf{r})=\left.\mathcal{H}\right|_{\omega_H\rightarrow \omega_H-i\alpha\omega}$ [See Eqs.~(\ref{HmagFullMatrix})--(\ref{s40}) and Eqs.~(\ref{HmatrixBar})--(\ref{s92})], i.e., the Gilbert damping term can be included in the external magnetic field contribution via $\omega_H\rightarrow \omega_H-i\alpha\omega$~\cite{stancil2009spin,gurevich1996magnetization}. To find $\omega$ that gives nontrivial solution, we firstly set $\alpha=0$ and obtain $\omega=\omega_{\mu}$. Then we obtain the solution in the case $\alpha\neq 0$ ($\alpha\ll1$) as~\cite{stancil2009spin,gurevich1996magnetization}
\begin{equation}
\omega=\omega_\mu-i\alpha \omega \frac{\partial\omega_\mu}{\partial\omega_H}.
\end{equation}
As we can see from Fig.~\ref{fig4}(b), we have $(\partial \omega_{\mu}/\partial \omega_H)\approx 1$, so we obtain $\omega\approx\omega_\mu/(1+i\alpha)\approx\omega_\mu-i\alpha\omega_\mu+\mathcal{O}(\alpha^2)$, yielding $\partial_t\beta_\mu\approx -i(\omega_\mu-i\alpha\omega_\mu)\beta_\mu$. In the simulation presented in Fig.~\ref{fig5}, the two NV centers are placed at $(x_1,y_1,z_1)=(d+h,w, 400\text{ nm})$ and $(x_2,y_2,z_2)=(d+h,w, 400\text{ nm}+\delta z)$ with $\delta z=2.2\ \mu\mathrm{m}$, which results in $g_{(005)}(\mathbf{r}_1)=g$ and $g_{(005)}(\mathbf{r}_2)=-g$ with $g=2\pi\times517\ \mathrm{ kHz}$. The simulation is performed under the field $H_\mathrm{c}$, which gives the magnon frequency $\omega_{(005)}\approx2\pi\times 2.78\ \mathrm{GHz}$. Moreover, we solve the Lindblad equation in the rotating frame with frequency $\omega_{(005)}$ for the transduction protocol and with frequency $\omega_\mathrm{NV}$ for the virtual-magnon exchange protocol. As the NV center's longitudinal relaxation time $T_1$ is longer than both $T_2^{*}$ and $1/(\alpha \omega_m)$, we do not include its corresponding terms $\mathcal{D}[\sigma^-_{\mathrm{NV}_i}]$ and $\mathcal{D}[\sigma^+_{\mathrm{NV}_i}]$ in the current simulation.

As shown in the left schematic of Fig.~\ref{fig5}(a), idler frequencies of $\mathrm{NV}_1$ and $\mathrm{NV}_2$ in the transduction protocol are $\omega_{\mathrm{NV}_1}=\omega_\mathrm{m}+\delta\omega_\mathrm{idle}$ and $\omega_{\mathrm{NV}_2}=\omega_\mathrm{m}-\delta\omega_\mathrm{idle}$, respectively. The detuning $\delta\omega_\mathrm{idle}=2\pi\times 5\ \mathrm{MHz}$ is chosen as the neighboring frequencies around $\omega_{(005)}$ are separated by more than $2\pi\times10\ \mathrm{MHz}$ from $\omega_{(005)}$, as shown in the Fig.~\ref{fig4}(b). The $i$SWAP gate time is $\tau_{i\mathrm{SWAP}}=\pi/(2g)$. Starting from the initial state $|g\rangle_1|e\rangle_2$, the fidelity is calculated as the state overlap between the $\mathrm{NV}$ state and the expected entangled state $|\psi\rangle\propto\frac{1}{\sqrt{2}}(|g\rangle_1|e\rangle_2+e^{-i\delta\omega_\mathrm{idle}\tau_{i\mathrm{SWAP}}}|e\rangle_1|g\rangle_2)$. On the other hand, the detuning in the virtual-magnon exchange protocol is $\omega_\mathrm{NV}=\omega_\mathrm{m}-\Delta\omega$ with $\Delta\omega=2\pi\times 3\ \mathrm{MHz}$, and the fidelity is calculated as the state overlap with $|\psi\rangle=\frac{1}{\sqrt{2}}(|g\rangle_1|e\rangle_2-i|e\rangle_1|g\rangle_2)$.

The indicator of the violation of the Bell inequality presented in Fig.~\ref{fig5} is calculated following Refs.~[\onlinecite{horodecki1995violating}] and [\onlinecite{bartkiewicz2013entanglement}] as
\begin{eqnarray}
&&\text { CHSH violation } = \max [0, \mathcal{M}(\rho)-1],\\
&&\mathcal{M}(\rho)=\max _{j<k}\left\{h_{j}+h_{k}\right\},
\end{eqnarray}
where $h_j\ (j=1,2,3)$ are eigenvalues of the matrix $\mathbf{U}=\mathbf{T}^{T}\mathbf{T}$ with $T_{ij}=\mathrm{Tr}[\rho(\sigma_i\otimes\sigma_j)]$. When $\mathrm{(CHSH\ violation)}>0$, the Clauser-Horne-Shimony-Holt (CHSH) form of Bell inequality is violated. As shown in Fig.~\ref{fig5}, this is stricter condition than the inseparability of the two-qubit state captured by the entanglement negativity~\cite{vidal2002computable}, $\mathcal{N}>0$.

\begin{figure*}[t]
\includegraphics[scale=1]{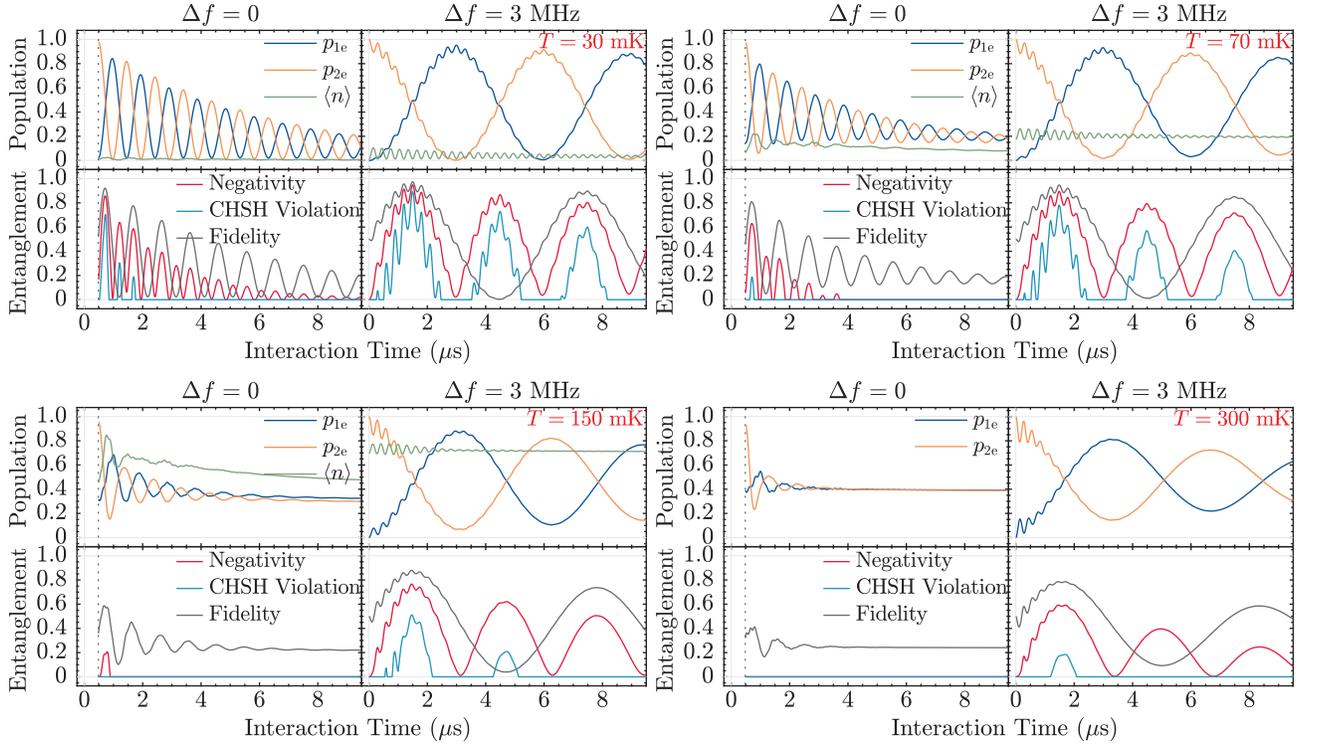}
\caption{Temperature dependence of the two entangling protocols presented in Fig.~\ref{fig5}, where the $T=70\ \mathrm{mK}$ simulation presented on the right-top corner is the same as Fig.~\ref{fig5}.}
\label{figS1} 
\end{figure*}

\subsection{Supplementary simulations}
In Fig.~\ref{figS1}, we show the temperature dependence of the two entanglement protocols as mentioned in the main text. While we only present the case with $T=70\ \mathrm{mK}$ case in Fig.~\ref{fig5}, here we present simulations under $T=30\ \mathrm{mK}$, $70\ \mathrm{mK}$, $150\ \mathrm{mK}$, and $300\ \mathrm{mK}$. As the virtual-magnon exchange protocol does not populate the magnon level in the limit $\Delta\omega/g\rightarrow \infty$, i.e., magnons are only created virtually, it is observed that this protocol is robust against the thermal fluctuations. At the same time, as shown in the simulation under  $T=30\ \mathrm{mK}$, transduction protocols improves drastically from $T=70\ \mathrm{mK}$ compared to the virtual-magnon exchange protocol.

\begin{figure*}[t]
\includegraphics[scale=1]{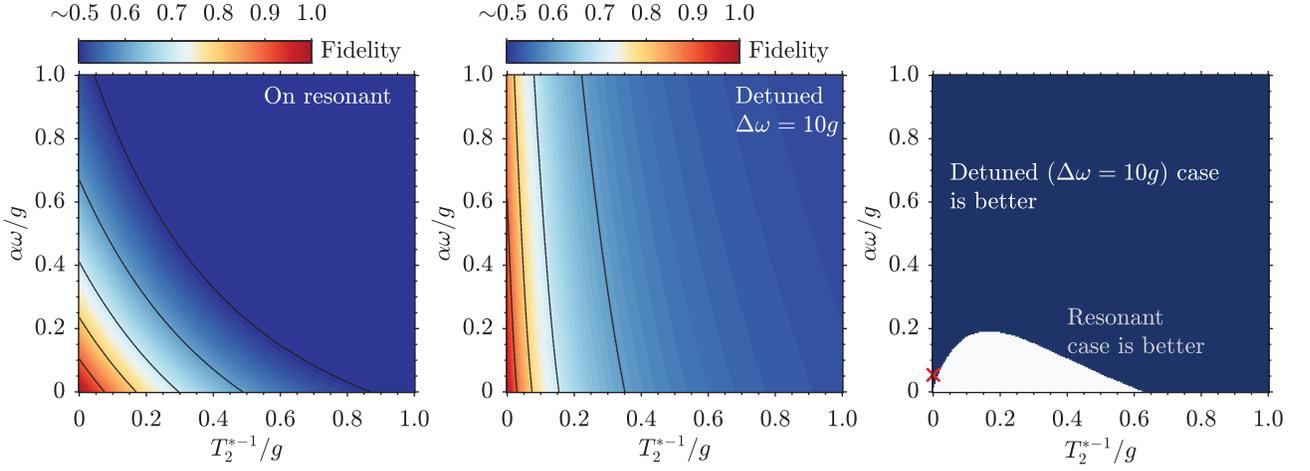}
\caption{Comparison between the transduction (on-resonant) and the virtual-magnon exchange (detuned) protocol of entanglement generation. Maximum fidelity is calculated for each protocol under different Gilbert damping parameter $\alpha$ and NV center's coherence time $T_2^*$, where $\omega=\omega_{(005)}$. Contours indicate $\mathrm{Fidelity}=0.5$, $0.6$, $0.7$, $0.8$, and $0.9$. A phase diagram for which protocol gives better fidelity is presented on the rightmost figure, where the red cross marker represents the parameters used in Fig.~\ref{fig5}. We choose $\Delta\omega=10g$ for this simulation. For the simplicity of the numerical simulation, we turn on and off  the coupling strength instead of inserting the idling frequency $\delta\omega_\mathrm{idle}$.}
\label{figS2} 
\end{figure*}

\begin{figure*}[t]
\includegraphics[scale=1]{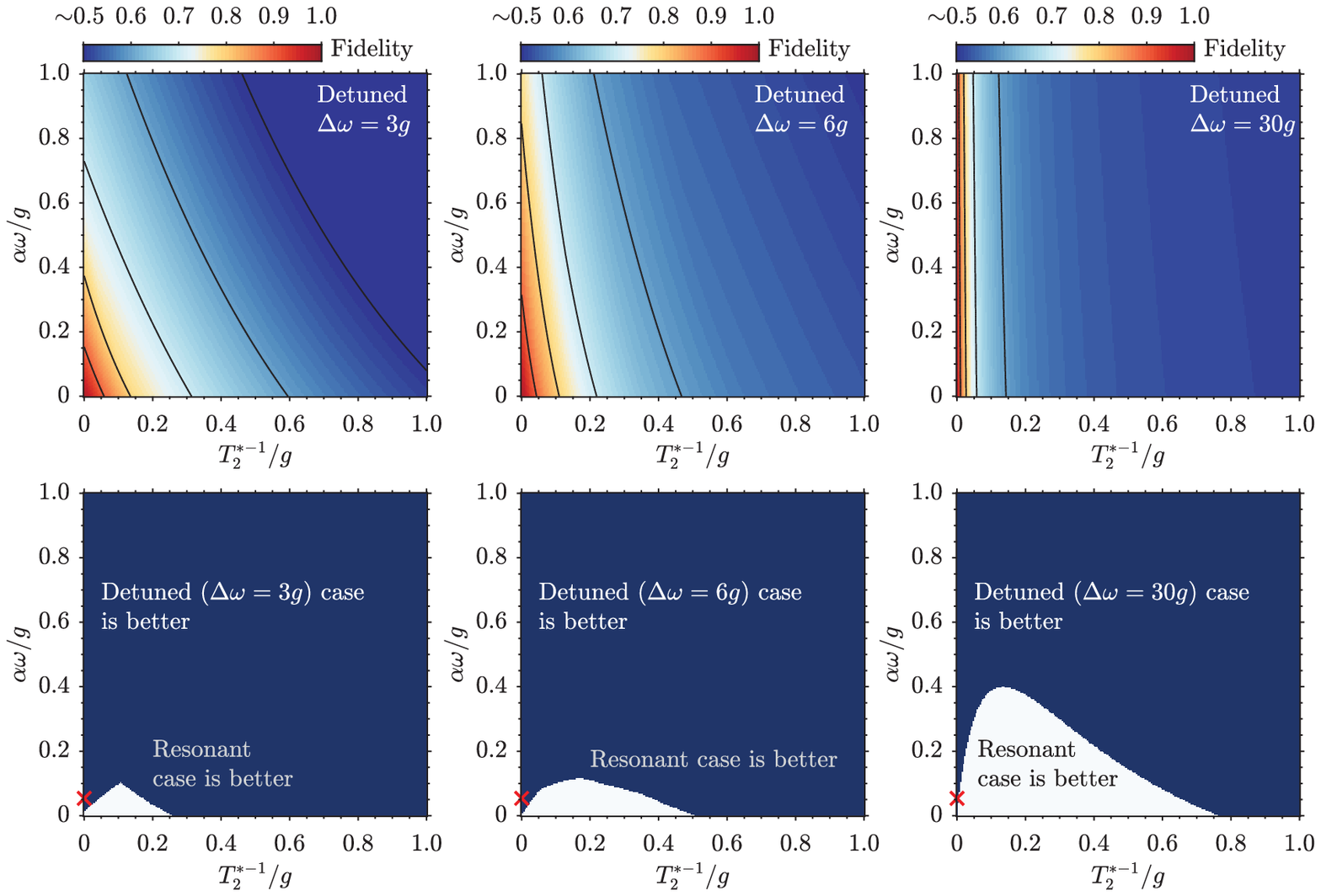}
\caption{Detuning $\Delta\omega$ dependence of the final entangled state's fidelity, as shown in the middle figure in Fig.~\ref{figS2}. The lower figures show the corresponding phase diagrams as in the rightmost figure in Fig.~\ref{figS2} for the corresponding detuning frequency values. }
\label{figS3} 
\end{figure*}

To explore the parameters $\alpha$ and $T_2^*$ dependence of the fidelity on the final entangled state for each protocols, we show in Fig.~\ref{figS2} the parameter dependence of the fidelity at $T=0$. The rightmost figure in Fig.~\ref{figS2} shows the phase diagram for which protocol gives better fidelity, where maximum fidelity from each protocols are compared. In the virtual-magnon exchange protocol denoted as detuned, we choose $\Delta\omega=10g$. To simplify the numerical calculation, fidelity at times $t=(\mathrm{integer})\times\frac{\pi}{\sqrt{2+(\Delta\omega/g)^2}}/g$ are evaluated for the virtual-magnon exchange protocol, which gives approximately optimal fidelity (see small oscillations observed in the $\Delta f=3\ \mathrm{MHz}$ cases in Fig.~\ref{figS1}). For the transduction protocol, fidelity is evaluated at the time after $\tau_{i\mathrm{SWAP}}/2$ interaction time of entangling $\mathrm{NV}_2$ and magnons followed by $\tau_{i\mathrm{SWAP}}$ $i$SWAP-gate time between $\mathrm{NV}_1$ and magnons. Here, the coupling strength $g_\mu(\mathbf{r}_i)$ is controlled to be $g_\mu(\mathbf{r}_i)=0$ for non-interacting duration instead of inserting idling frequency $\delta\omega_\mathrm{idle}$, for simplicity. As the resulting fidelity in the virtual-magnon exchange protocol depends on the amount of the detuning $\Delta\omega/g$, we show in Fig.~\ref{figS3} the same simulation as in Fig.~\ref{figS2} under multiple detuning values. As shown in the right-top figure in Fig.~\ref{figS3}, when the detuning is large $\Delta\omega/g=30$, higher fidelity entangled state can be created even when the magnon damping $\alpha\omega$ is not very small. This is because magnons are only excited virtually in the virtual-magnon exchange protocol.

\begin{figure*}[t]
\includegraphics[scale=1]{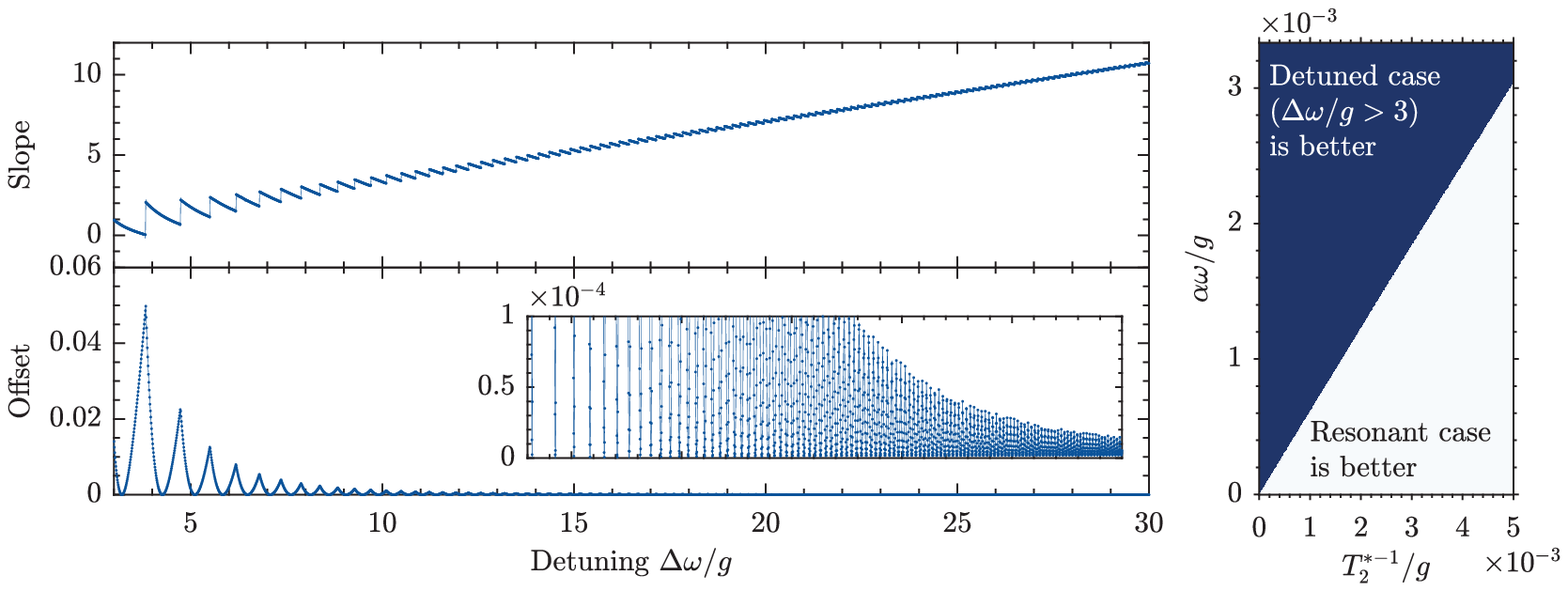}
\caption{Small $\alpha$ and $T_2^{*-1}$ behavior of the boundary curves between the two regions as shown in the lower figures in Fig.~\ref{figS3} under multiple detuning values. The boundary is approximately $\alpha\omega/g=(\mathrm{slope})\times(T_2^{*-1}/g)+(\mathrm{offset})$, and the slope and its offset are shown on the left-top and left-bottom figures. The right figure shows the parameter region where there exist a detuning value in $\Delta\omega>3g$ where the virtual-magnon exchange protocol gives better fidelity than the transduction protocol.}
\label{figS4} 
\end{figure*}

As indicated from the phase diagrams presented in Figs.~\ref{figS2} and \ref{figS3}, in the regions where $\alpha$ and $T_2^{*-1}$ are both sufficiently small, the transduction protocol is better when $\alpha\omega$ is much smaller than $T_2^{*-1}$. On the other hand, virtual-magnon exchange protocol is better when $T_2^{*-1}$ is much smaller than $\alpha\omega$. This tradeoff comes from the fact that the transduction protocol is the faster in gate operation but populate real magnons that are sensitive to the magnon damping, while virtual-magnon exchange protocol is slower in gate operation but it does not populate magnon states and hence the protocol is insensitive to the magnon damping. In Fig.~\ref{figS4}, we present the behavior of the boundary line between the two regions for small $\alpha$ and $T_2^{*-1}$, where the boundary can be approximated to $\alpha\omega/g=(\mathrm{slope})\times(T_2^{*-1}/g)+(\mathrm{offset})$. We note that the offset has nodes for detuning values
\begin{eqnarray}
\Delta\omega/g=\frac{2\sqrt{2}(2n-1)}{\sqrt{4n-1}},\quad n=1,2,\cdots.\label{DeltaCond}
\end{eqnarray}
This comes from the small and fast oscillation on top of the slow envelope oscillation observed in the virtual magnon exchange protocol of Fig.~\ref{figS1}. The virtual-magnon exchange protocol without the magnon damping and the NV decoherence gives a perfect entangled state only when the condition represented by Eq.~(\ref{DeltaCond}) is satisfied. Under this condition, fidelity in the region $\alpha\omega/g\ll1$ and $T_2^{*-1}/g\ll1$ is calculated as
\begin{widetext}
\begin{eqnarray}
(\mathrm{Fidelity})=1&&-\frac{(4n-1)^{3/2}\pi}{16\sqrt{2}n^2}(\alpha\omega/g)-\frac{\sqrt{4n-1}(-3+24n-80n^2+128n^3+256n^4)\pi}{1024\sqrt{2}n^4}(T_2^{*-1}/g).\label{FidOffRes}
\end{eqnarray}
\end{widetext}
On the other hand, fidelity in the transduction protocol in the region $\alpha\omega/g\ll1$ and $T_2^{*-1}/g\ll1$ is calculated as
\begin{eqnarray}
(\mathrm{Fidelity})=1-\frac{\pi-1}{2}(\alpha\omega/g)-\frac{15\pi}{32}(T_2^{*-1}/g).\nonumber\\ \label{FidOnRes}
\end{eqnarray}
Combining Eqs.~(\ref{FidOffRes}) and (\ref{FidOnRes}), we obtain the slope value of the boundary line shown in Fig.~\ref{figS4} for detunings $\Delta\omega/g$ that give zero offset. When $\Delta\omega/g$ is large, the asymptotic behavior of the slope is
\begin{eqnarray}
(\mathrm{slope})\sim\frac{\pi}{4(\pi-1)}(\Delta\omega/g)\approx0.367(\Delta\omega/g),
\end{eqnarray} 
which matches with the numerical simulation presented in Fig.~\ref{figS4}. However, note that in real magnonic system the detuning is limited by the neighboring mode's frequency separation.

\begin{figure*}[t]
\includegraphics[scale=1]{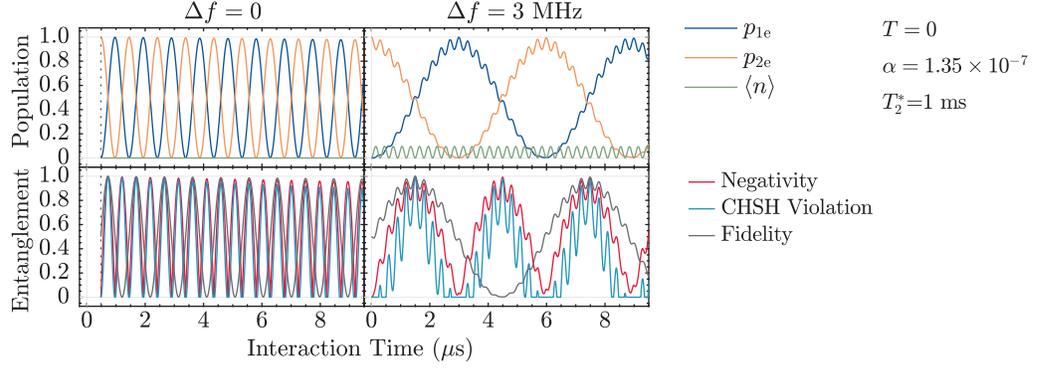}
\caption{The same simulation as in Fig.~\ref{fig5} under Gilbert damping parameter $\alpha=1.35\times 10^{-7}$, which makes the two protocols comparable. Temperature is $T=0$ and we controlled the strength of the coupling instead of inserting the idling frequency $\delta\omega_\mathrm{idle}$ for consistency with Figs.~\ref{figS2}, \ref{figS3}, and \ref{figS4}.}
\label{figS5} 
\end{figure*}

Based on the simulation in Fig.~\ref{figS4}, the boundary line under the detuning $\Delta\omega=2\pi\times3\ \mathrm{MHz}$ is numerically obtained as $(\alpha\omega/g)=1.24\times10^{-4}+1.95(T_2^{*-1}/g)$. The Gilbert damping parameter $\alpha$ that makes the two protocol comparable is $\alpha=1.35\times 10^{-7}$. In Fig.~\ref{figS5}, we show the same simulation as in Fig.~\ref{fig5} with parameters $T=0$ and $\alpha=1.35\times 10^{-7}$ where we see comparable entanglement values for both protocols, although the transduction protocol is faster in gate operation. For consistency with the analysis presented in Figs.~\ref{figS2}-\ref{figS4}, the coupling strength $g$ was turned on and off as a function of time instead of inserting the idling frequency $\delta\omega_\mathrm{idle}$.

\section{Magnon-originated NV center decoherence}
\subsection{Higher order magnon contribution}
In this section, we will estimate the decay and decoherence of NV centers due to the interaction with magnon modes with $\mu\neq(005)$ at field $H_c$, which were not taken into account in the Lindblad simulation in the main text. Based on the interaction Hamiltonian Eq.~(\ref{Hintbar}), as the modes with $\mu\neq(005)$ are well separated in frequency, they do not affect the decay and decoherence of NV centers as long as the linewidth $\alpha\omega_\mu$ is small. Here we go beyond the linear order interaction, and consider the following NV-magnon interaction (see Eq.~(\ref{Hint3term})),
\begin{widetext}
\begin{eqnarray}
\mathcal{H}_{\text {int }}&=&\left.\gamma \mu_{0} \mathbf{S}_{\mathrm{NV}} \cdot\left[\mathbf{H}_{\mathrm{d}}(\mathbf{r})+\nabla \int d \mathbf{r}^{\prime} G\left(\mathbf{r}-\mathbf{r}^{\prime}\right)\left(\nabla^{\prime} \cdot \vec{\mathcal{M}}\left(\mathbf{r}^{\prime}\right) \mathcal{F}\left(\mathbf{r}^{\prime}\right)-\frac{\partial_{Z}^{\prime} \mathcal{F}\left(\mathbf{r}^{\prime}\right) \vec{\mathcal{M}}^{2}\left(\mathbf{r}^{\prime}\right)}{2 M_\mathrm{s}}\right)\right]\right|_{\mathbf{r}=\mathbf{r}_{\mathrm{NV}}}\nonumber ,\\
&=&\left.\gamma \mu_{0} \mathbf{S}_{\mathrm{NV}} \cdot\left[\mathbf{H}_{\mathrm{d}}(\mathbf{r})+\mathbf{h}(\mathbf{r})+\mathbf{h}_{2}(\mathbf{r})\right]\right|_{\mathbf{r}=\mathbf{r}_{\mathrm{NV}}}\nonumber , \\
&=&\left.\gamma \mu_{0} \mathbf{S}_{\mathrm{NV}} \cdot\left[\mathbf{h}(\mathbf{r})+\delta \mathbf{h}_{2}(\mathbf{r})\right]\right|_{\mathbf{r}=\mathbf{r}_{\mathrm{NV}}}+\mathrm{const.},
\end{eqnarray}
\end{widetext}
where $\mathbf{h}(\mathbf{r})$ is provided in Eq.~(\ref{h_def_without(2)}) and we define
\begin{eqnarray}
&&\mathbf{h}_{2}(\mathbf{r}) \equiv-\nabla \int d \mathbf{r}^{\prime} G\left(\mathbf{r}-\mathbf{r}^{\prime}\right) \frac{\partial_{z}^{\prime} \mathcal{F}\left(\mathbf{r}^{\prime}\right) \vec{\mathcal{M}}^{2}\left(\mathbf{r}^{\prime}\right)}{2 M_\mathrm{s}},\\
&&\delta \mathbf{h}_{2}(\mathbf{r})=\mathbf{h}_{2}(\mathbf{r})-\left\langle\mathbf{h}_{2}(\mathbf{r})\right\rangle.
\end{eqnarray}
Here, the average is taken with the magnon thermal state $\rho_\mathrm{m}=\mathrm{exp}[-\sum_\mu \hbar\omega_\mu \beta_\mu^\dagger \beta_\mu/k_\mathrm{B} T]$, i.e., $\langle\cdots\rangle=\mathrm{Tr}[\cdots\rho_\mathrm{m}]$. In the NV center's subspace spanned by $\{|g\rangle,|e\rangle\}$, we can write
\begin{eqnarray}
&&\mathcal{H}_{\text {int }}=\hbar\left(\sigma_{\mathrm{NV}}^{-} b^{+}+\sigma_{\mathrm{NV}}^{+} b^{-}+\frac{1}{2} \sigma_{\mathrm{NV}}^{z} b_{z}\right),\\
&&\left.b^{\pm} = \frac{\gamma \mu_{0}}{\sqrt{2}}\left(h^{\mp}(\mathbf{r})+\delta h_{2}^{\mp}(\mathbf{r})\right)\right|_{\mathbf{r}=\mathbf{r}_{\mathrm{NV}}},\\
&&b^{z} =-\left.\gamma \mu_{0}\left(h^{z}(\mathbf{r})+\delta h_{2}^{z}(\mathbf{r})\right)\right|_{\mathbf{r}=\mathbf{r}_\mathrm{NV}}.
\end{eqnarray}
Assuming a Markovian magnon bath, the NV center's longitudinal decay rates ($1/T_1$) are
\begin{eqnarray}
&&\Gamma_{|e\rangle \rightarrow|g\rangle}^{1}=\int d t e^{+i \omega_\mathrm{NV} t}\left\langle b^{-}(t) b^{+}(0)\right\rangle,\\
&&\Gamma_{|g\rangle \rightarrow|e\rangle}^{1}=\int d t e^{-i \omega_{\mathrm{NV}} t}\left\langle b^{+}(t) b^{-}(0)\right\rangle.
\end{eqnarray}
Under the same assumption, the NV center's decoherence rate ($1/T_2^*$) is related to the $\omega\approx0$ region of $S(\omega)$ with
\begin{eqnarray}
S(\omega)=\int d t e^{-i \omega t}\left\langle b^{z}(t) b^{z}(0)\right\rangle,
\end{eqnarray}
where the Ramsey decoherence follows
\begin{eqnarray}
\rho_{eg}\sim\mathrm{exp} \left[\frac{1}{2}\int \frac{d\omega}{2\pi}S(\omega)\left(\frac{\mathrm{sin}(\omega t/2)}{\omega/2}\right)^2\right].
\end{eqnarray}

The longitudinal relaxation rate $\Gamma^1_{|e\rangle\rightarrow|g\rangle}$ will include terms like $\int d t e^{i \omega_\mathrm{NV}}\left\langle\beta_{\mu}(t) \beta^{\dagger}_{\mu}(0)\right\rangle$ and $\int d t e^{i \omega_\mathrm{NV}}\left\langle\beta_{\mu}(t) \beta_{v}^{\dagger}(t) \beta_{\mu}^\dagger(0) \beta_{v}(0)\right\rangle$. The former is the one-magnon decay contribution ($\omega_\mathrm{NV}=\omega_\mu$) and the latter is the two-magnon decay contribution ($\omega_\mathrm{NV}=\omega_\mu-\omega_\nu$). However, in our discretized magnon modes, the chances of having $\omega_\mathrm{NV}=\omega_\mu$ or $\omega_\mathrm{NV}=\omega_\mu-\omega_\nu$ are small, at least when the linewidth $\alpha\omega_\mu$ of magnons is narrow.

In contrast, for the decoherence that is obtained from $\omega\approx0$ part of $S(\omega)$, there is a big contribution from terms of the form $\int d t e^{-i \omega t}\left\langle\delta n_{\mu}(t) \delta n_{\mu}(0)\right\rangle$, where $\delta n_\mu=\beta_\mu^\dagger\beta_\mu-\langle\beta_\mu^\dagger\beta_\mu\rangle$. This  arises from the second-order noise correlation of $\delta h_2^{z}(\mathbf{r})$. Furthermore, we notice that this noise contribution is coming not only from the magnon mode with $\omega_\mu\approx\omega_\mathrm{NV}$, but also from high energy magnons up to $\omega_\mu<k_\mathrm{B}T/\hbar$. As the decoherence contribution is expected to be dominant, we estimate the order of its timescale. To simplify the calculation and to avoid the paraunitary matrix diagonalization of a large matrix, we approximate that $a_\mu$ is the normal mode, i.e. $a_\mu\sim e^{-i\omega_\mu t}$. We take $\omega_\mu=\omega_\mathrm{min}+DK_\mu^2$, where $\omega_\mathrm{min}$ is the minimum frequency obtained from the paraunitary matrix diagonalization in Sec.~II. Hence we write
\begin{eqnarray}
&&\gamma \mu_{0} \mathbf{h}_{2}(\mathbf{r})=\omega_{d w l} \sum_{\mu \mu^{\prime}} \Theta_{\mu \mu^{\prime}} a_{\mu}^{\dagger} a_{\mu^{\prime}},\\
&&\Theta_{\mu \mu^{\prime}}=d w l \int d \mathbf{r}^{\prime}\left[(-\nabla) G\left(\mathbf{r}-\mathbf{r}^{\prime}\right)\right]_{z}\nonumber\\
&&\quad \quad \quad \quad \quad \quad \quad  \times\partial_{z}^{\prime}\left[\mathcal{F}\left(\mathbf{r}^{\prime}\right) \psi_{\mu}(\mathbf{r}^{\prime}) \psi_{\mu^{\prime}}(\mathbf{r}^{\prime})\right].
\end{eqnarray}

\begin{figure}[t]
\includegraphics[scale=1]{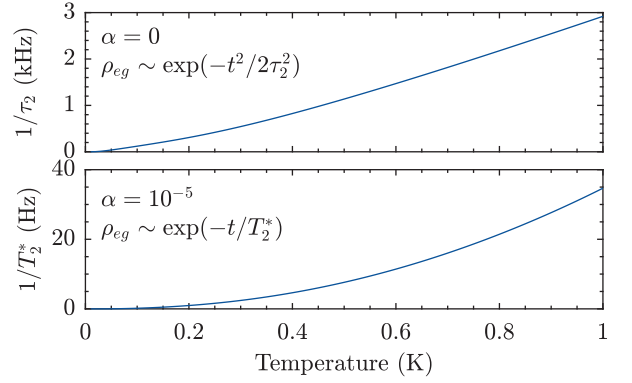}
\caption{Decoherence times calculated from  Eqs.~(\ref{T2model1}) and (\ref{T2model2}) as a function of temperature. The lower figure is calculated using $\alpha=10^{-5}$.}
\label{figS6} 
\end{figure}

The terms that affect the NV center's decoherence are the contributions from $\mu=\mu'$. Thus, to estimate the decoherence rate, we take
\begin{eqnarray}
&&\mathcal{H}_{\mathrm{int}}^{\mathrm{approx}}=-\frac{1}{2} \sigma_{\mathrm{NV}}^{z}\left(\hbar \omega_{d w l} \sum_{\mu} \Theta_{\mu \mu} \delta n_{\mu}\right),\\
&&\delta n_{\mu}=a_{\mu}^{\dagger} a_{\mu}-\left\langle a_{\mu}^{\dagger} a_{\mu}\right\rangle.
\end{eqnarray}
In the limit $\alpha\rightarrow0$ (although this is not compatible with the Markov approximation) we have $\langle\delta n_\mu (t)\delta n_\mu\rangle=\langle n_\mu^2\rangle-\langle n_\mu\rangle^2$ that yields
\begin{eqnarray}
&&S(\omega)=\omega_{d w l}^{2} \sum_{\mu}\left(\Theta_{\mu \mu}\right)^{2}\left(\left\langle n_{\mu}^{2}\right\rangle-\left\langle n_{\mu}\right\rangle^{2}\right) \cdot 2 \pi \delta(\omega),\nonumber\\ \\
&&\rho_{g e}(t) \sim \exp \left[-\frac{t^{2}}{2\left(\tau_{2}\right)^{2}}\right],\\
&&\frac{1}{\tau_{2}}=\omega_{d w l} \sqrt{\sum_{\mu}\left(\Theta_{\mu \mu}\right)^{2}\left(\left\langle n_{\mu}^{2}\right\rangle-\left\langle n_{\mu}\right\rangle^{2}\right)},\label{T2model1}
\end{eqnarray}
where $\tau_2$ is the decoherence timescale. This expression is acceptable as long as the magnon damping $2\alpha\omega_\mu$ is much smaller than $1/\tau_2$. When $2\alpha\omega_\mu$ is not small, we take $\langle\delta n_\mu (t)\delta n_\mu\rangle=(\langle n_\mu^2\rangle-\langle n_\mu\rangle^2)e^{-2\alpha\omega_\mu t}$ and obtain
\begin{eqnarray}
&&S(\omega)=\omega_{d w l}^{2} \sum_{\mu}\left(\Theta_{\mu \mu}\right)^{2}\left(\left\langle\hat{n}_{\mu}^{2}\right\rangle-\left\langle\hat{n}_{\mu}\right\rangle^{2}\right)  \frac{4\alpha\omega_\mu}{\omega^{2}+(2\alpha\omega_\mu)^{2}},\nonumber\\ \\
&&\rho_{g e}(t) \sim \exp \left[-\frac{1}{2} S(\omega=0) t\right]=\exp \left[-\frac{t}{T_{2}^{*}}\right],\\
&&\frac{1}{T_2^*}=\omega_{d w l}^{2} \sum_{\mu}\left(\Theta_{\mu \mu}\right)^{2}\left(\left\langle\hat{n}_{\mu}^{2}\right\rangle-\left\langle\hat{n}_{\mu}\right\rangle^{2}\right) \frac{1}{2 \alpha \omega_{\mu}},\label{T2model2}
\end{eqnarray}
where $T_2^*$ is the decoherence rate. In Fig.~\ref{figS6}, we show the two decoherence times from Eqs.~(\ref{T2model1}) and (\ref{T2model2}).

\subsection{Dispersive coupling contribution}

\begin{figure}[t]
\includegraphics[scale=1]{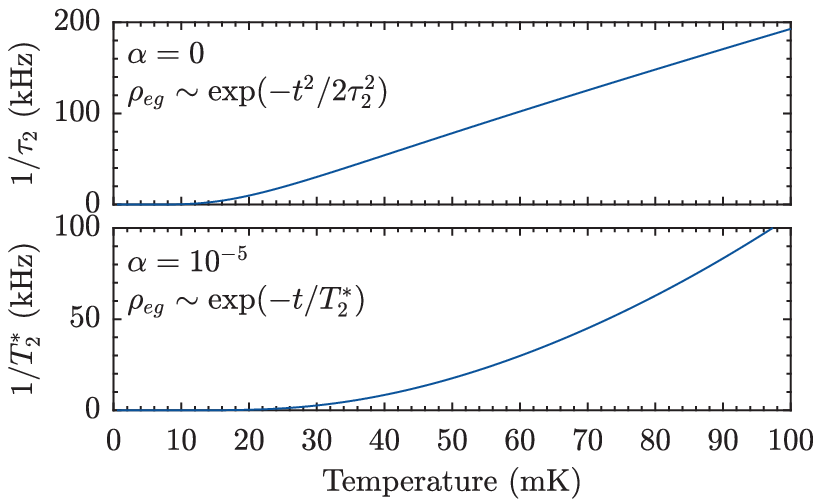}
\caption{Decoherence times calculated from Eqs.(\ref{T2model1Stark}) and (\ref{T2model2Stark}) as a function of temperature. The lower figure is calculated using $\alpha=10^{-5}$.}
\label{figS7} 
\end{figure}

\begin{figure*}[t]
\includegraphics[scale=1]{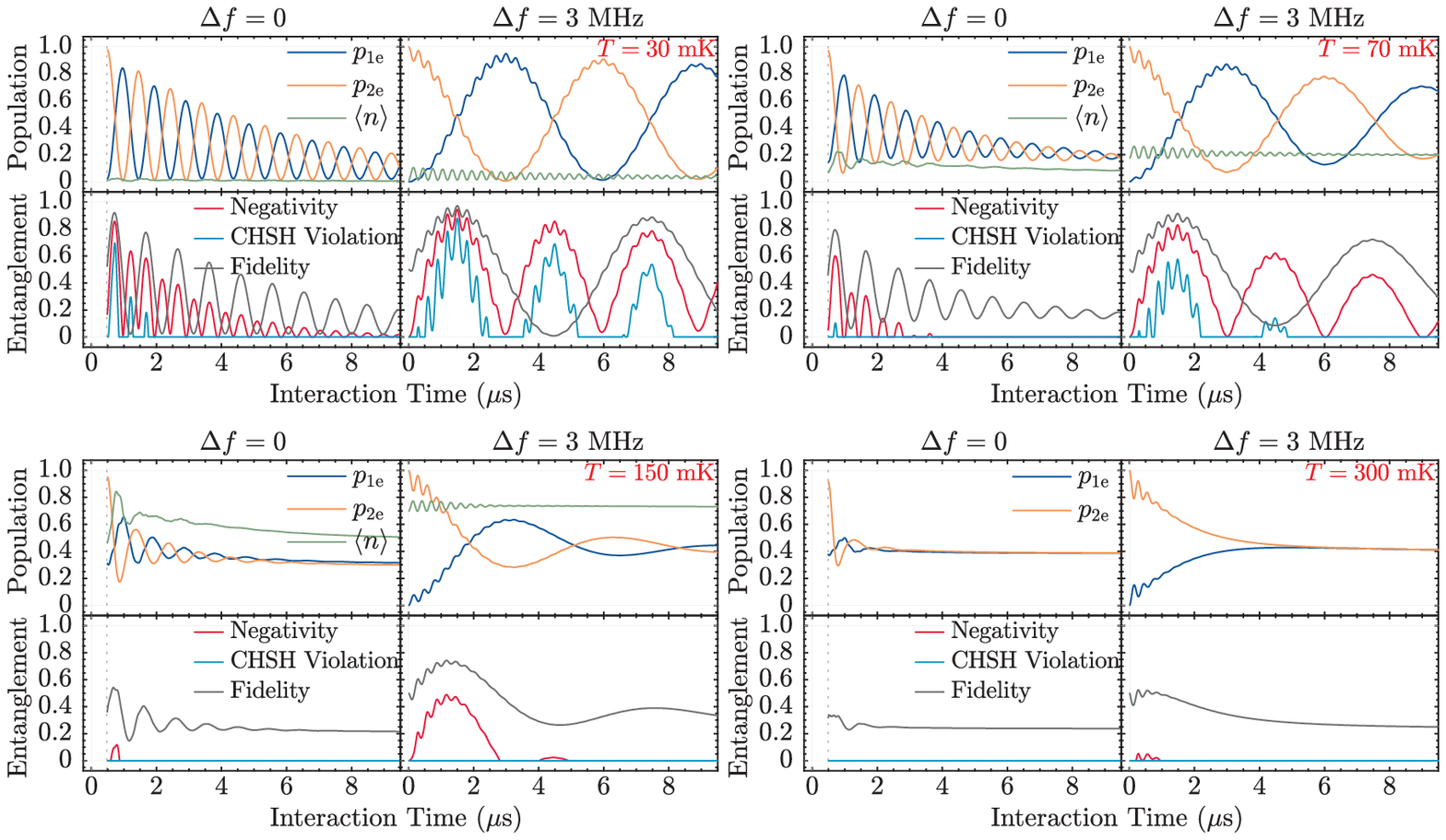}
\caption{Temperature dependence of the two entangling protocols presented in Fig.~\ref{figS1} with the NV centers' depahsing due to the magnon number fluctuations of the neighboring magnon modes calculated in Fig.~\ref{figS7}.}
\label{figS8} 
\end{figure*}

While the Hamiltonian Eq.~(\ref{Hintbar}) does not appear to cause a decoherence, after performing the Schrieffer-Wolff transformation in the dispersive regime ($|\omega_\mu-\omega_\mathrm{NV}|>\left|g_{\mu}\left(\mathbf{r}_\mathrm{NV}\right)\right|$), we obtain Eq.~(\ref{geffbar}), where we can securely affirm that the second term (Stark shift term) will cause the decoherence, as considered in Ref.~[\onlinecite{trifunovic2013long}]. In this section we calculate the decoherence due to this contribution. We consider the effect of 
\begin{eqnarray}
&&\mathcal{H}_\mathrm{eff}^\mathrm{dispersive}=\frac{1}{2}\sigma_{\mathrm{NV}}^{z}\sum_{\mu\neq(005)}\left(2\hbar \frac{\left|g_{\mu}\left(\mathbf{r}_\mathrm{NV}\right)\right|^{2}}{\omega_{\mathrm{NV}}-\omega_{\mu}} \delta n_\mu\right),\nonumber\\ \\
&&\delta n_\mu=\beta_{\mu}^{\dagger} \beta_{\mu}-\langle \beta_{\mu}^{\dagger} \beta_{\mu} \rangle.
\end{eqnarray}

We exclude $\mu=(005)$ in the sum as we are considering the field $H_c$ where $\omega_\mathrm{NV}$ and $\omega_{(005)}$ are on resonant. In the same way as in Eqs.~(\ref{T2model1}) and (\ref{T2model2}), we obtain
\begin{eqnarray}
&&\frac{1}{\tau_{2}}=\sqrt{\sum_{\mu\neq (005)}\left(\frac{2\left|g_{\mu}\left(\mathbf{r}_\mathrm{NV}\right)\right|^{2}}{\omega_{\mathrm{NV}}-\omega_{\mu}} \right)^{2}\left(\left\langle n_{\mu}^{2}\right\rangle-\left\langle n_{\mu}\right\rangle^{2}\right)}\label{T2model1Stark},\\
&&\frac{1}{T_2^*}=\sum_{\mu\neq(005)}\left(\frac{2\left|g_{\mu}\left(\mathbf{r}_\mathrm{NV}\right)\right|^{2}}{\omega_{\mathrm{NV}}-\omega_{\mu}} \right)^{2}\left(\left\langle\hat{n}_{\mu}^{2}\right\rangle-\left\langle\hat{n}_{\mu}\right\rangle^{2}\right) \frac{1}{2 \alpha \omega_{\mu}}.\nonumber\\ \label{T2model2Stark}
\end{eqnarray}

In Fig.~\ref{figS7}, we show the two decoherence times from Eqs.(\ref{T2model1Stark}) and (\ref{T2model2Stark}). From $T\leq70\ \mathrm{mK}$ and $\alpha=10^{-5}$ part of Figs.~\ref{figS6} and \ref{figS7}, the magnon induced decoherence time is $T_2^*>20\ \mu\mathrm{s}$, and it is expected that this dephasing contribution does not change the general trend of the result of the simulation presented in Fig.~\ref{fig5}.

In Fig.~\ref{figS8}, we show figures corresponding to Fig.~\ref{figS1} with the NV centers' dephasing rate calculated in Fig.~\ref{figS7}. The simulation confirms that the general tendency presented in Fig.~\ref{fig5} does not change due to the dephasing contribution calculated in Fig.~\ref{figS7}.

\section{Average Gate Fidelity for off-resonance protocol}
\begin{figure}[t]
\includegraphics[scale=1]{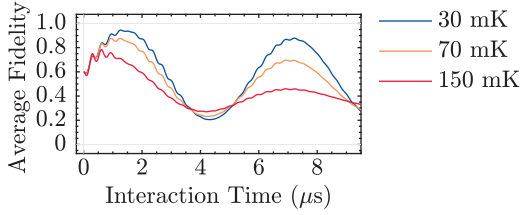}
\caption{Average gate fidelity as a function of the interaction time under multiple temperatures for the off-resonant protocol.}
\label{figS9} 
\end{figure}

To show that the magnon-mediated entanglement protocols can directly be extended to two-qubit gates, in this section we have calculated the average gate fidelity as a square-root-of-$i$SWAP gate for the off-resonant protocol under the same condition as in Fig.~\ref{fig5}. To calculate the average gate fidelity, we employ a method based on the entanglement fidelity $F_e$~\cite{nielsen2002simple}. For that we introduce two auxiliary qubits $\text{aux}_1$ and $\text{aux}_2$ and prepare the following maximally entangled state~\cite{Schumacher1996}
\begin{eqnarray}
|\phi\rangle=\frac{1}{\sqrt{4}}\left(\begin{array}{l}|e\rangle_{\text{NV}_1}|e\rangle_{\text{NV}_2}|e\rangle_{\text{aux}_1}|e\rangle_{\text{aux}_2}\\+|e\rangle_{\text{NV}_1}|g\rangle_{\text{NV}_2}|e\rangle_{\text{aux}_1}|g\rangle_{\text{aux}_2}\\ +|g\rangle_{\text{NV}_1}|e\rangle_{\text{NV}_2}|g\rangle_{\text{aux}_1}|e\rangle_{\text{aux}_2}\\+|g\rangle_{\text{NV}_1}|g\rangle_{\text{NV}_2}|g\rangle_{\text{aux}_1}|g\rangle_{\text{aux}_2}\end{array}\right),
\end{eqnarray}
as an initial qubit state. Then we evolve in time the NV and magnon states according to the Lindblad master equation of the previous sections, and calculate the fidelity $F_e$ as the state overlap between the calculated state and the desired state after the following gate
\begin{widetext}
\begin{eqnarray}
U_{\text{gate}}&=&\left.\mathrm{exp}[-i(\left|g_{\mathrm{eff}}\right|\left(\sigma_{\mathrm{NV}_{1}}^{+} \sigma_{\mathrm{NV}_{2}}^{-}+\text {H.c.}\right)-\left|g_{\mathrm{eff}}\right|\left(\sigma_{\mathrm{NV}_{1}}^{+} \sigma_{\mathrm{NV}_{1}}^{-}+\sigma_{\mathrm{NV}_{2}}^{+} \sigma_{\mathrm{NV}_{2}}^{-}\right))t]\right|_{t=\tau_{\sqrt{i\mathrm{SWAP}}}},\nonumber\\
&=&[|e e\rangle~|e g\rangle~|g e\rangle~|g g\rangle]_{\mathrm{NV}_{1} \mathrm{NV}_{2}}\left[\begin{array}{cccc}i & 0 & 0 & 0 \\ 0 & \frac{1+i}{2} & \frac{1-i}{2} & 0 \\ 0 & \frac{1-i}{2} & \frac{1+i}{2} & 0 \\ 0 & 0 & 0 & 1\end{array}\right]\left[\begin{array}{c}\langle e e| \\ \langle e g| \\ \langle g e| \\ \langle g g|\end{array}\right]_{\mathrm{NV}_{1} \mathrm{NV}_{2}},
\end{eqnarray}
\end{widetext}
where $\tau_{\sqrt{i\mathrm{SWAP}}}=\pi/(4|g_{\mathrm{eff}}|)$. As the square of $U_{\text{gate}}$ is equivalent to the $i$SWAP gate up to single-qubit operations, $U_{\text{gate}}$ can be thought of as a square-root-of-$i$SWAP gate. The average gate fidelity $\bar{F}$ is calculated via~\cite{nielsen2002simple}
\begin{eqnarray}
\bar{F}=\frac{d F_{\mathrm{e}}+1}{d+1},
\end{eqnarray}
where $d=4$. We have calculated the average gate fidelity under temperatures $T=30, 70,$ and $150$~mK, as shown in Fig.~\ref{figS9}, and have obtained $\bar{F}=0.94, 0.88,$ and $0.78$, respectively.

\section{Simulation under a larger Gilbert damping parameter}
\begin{figure}[t]
\includegraphics[scale=1]{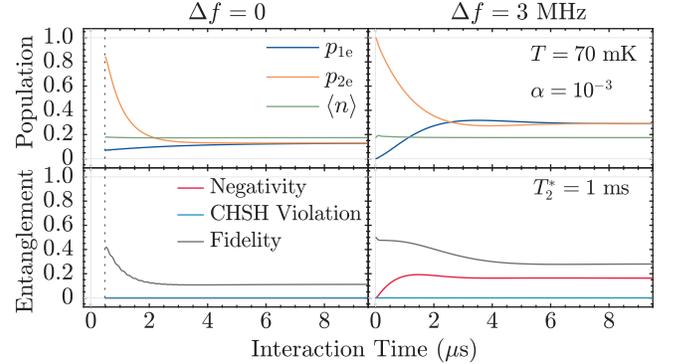}
\caption{Simulation analogous to Fig.~\ref{fig5} under a larger Gilbert damping parameter $\alpha=10^{-3}$.}
\label{figS10} 
\end{figure}

The Gilbert damping parameter $\alpha=10^{-5}$ that is observed in bulk YIG crystals~\cite{tabuchi2014hybridizing} would be optimistic for small YIG structures that we consider in this work. However, as one can calculate from Fig.~\ref{fig4}(c), we obtain a high cooperativity $\mathcal{C}\approx500$ even with a larger Gilbert damping parameter $\alpha=10^{-3}$. In Fig.~\ref{figS10}, we show a simulation analogous to the one presented in Fig.~\ref{fig5} with $\alpha=10^{-3}$. From this simulation, we find that the off-resonance protocol produces entangled states, as the entanglement negativity is larger than zero. However, this turns out to be not a useful entanglement as $(\mathrm{CHSH\ Violation})=0$ indicates that the state does not violate the Bell inequality. This happens because of the increased $T_1$ decay rate of NV centers due to the overlap of the broad magnon mode resonance with the NV-center's transition. Although the off-resonance protocol is less sensitive to the magnon decay, the detuning $\Delta \omega$ needs to be sufficiently larger than the linewidth of the magnon-mode resonance $\alpha\omega_{\mu}$ in order to suppress this decay contribution.

The resulting entangled mixed state presented in Fig.~S10 can be understood in the following way. As the interaction Hamiltonian is $\mathcal{H}_{\mathrm{int}}=ga(\sigma_{\mathrm{NV}_1}^{+}-\sigma_{\mathrm{NV}_2}^{+})+\mathrm{H.c.}$, we notice that $|D\rangle=(|g\rangle_{\mathrm{NV}_1}|e\rangle_{\mathrm{NV}_2}+|e\rangle_{\mathrm{NV}_1}|g\rangle_{\mathrm{NV}_2})/\sqrt{2}$ is a dark state with respect to the magnon mode, or alternatively, $|D\rangle$ is a state within a subspace that is free from the magnon-induced $T_1$ decay (decoherence free subspace), because $\mathcal{H}_{\mathrm{int}}|D\rangle|n_{\mathrm{m}}\rangle=0$ with a magnon number state $|n_{\mathrm{m}}\rangle$. Accordingly, the initial state of NV centers can be written as $|\psi_{\mathrm{init}}\rangle=|g\rangle_{\mathrm{NV}_1}|e\rangle_{\mathrm{NV}_2}=(|D\rangle+|B\rangle)/\sqrt{2}$, with $|B\rangle=(|g\rangle_{\mathrm{NV}_1}|e\rangle_{\mathrm{NV}_2}-|e\rangle_{\mathrm{NV}_1}|g\rangle_{\mathrm{NV}_2})/\sqrt{2}$, and  initial density operator $\rho_{\mathrm{init}}=|\psi_\mathrm{init}\rangle\langle \psi_\mathrm{init}|=(|D\rangle\langle D|+|D\rangle\langle B|+|B\rangle\langle D|+|B\rangle\langle B|)/2$. After the time evolution, the part related to $|D\rangle\langle D|$ remains constant as $|D\rangle$ is in the decoherence free subspace. Assuming that the system is at absolute zero temperature for simplicity, and  that the other terms eventually evolve to the ground state $|D\rangle\langle B|+|B\rangle\langle D|+|B\rangle\langle B|\rightarrow |00\rangle\langle00|$ due to the energy relaxation, where $|00\rangle=|g\rangle_{\mathrm{NV}_1}|g\rangle_{\mathrm{NV}_2}$, we obtain the final density operator
\begin{equation}
\rho_{\mathrm{fin}}=(|D\rangle\langle D |+|00\rangle\langle 00 |)/2.
\end{equation}
As the partial transpose of this density matrix has a negative eigenvalue $-(\sqrt{2}-1)/4$, we obtain the entanglement negativity of the final state $\mathcal{N}_{\mathrm{fin}}=(\sqrt{2}-1)/4$ and $\mathcal{N}_{\mathrm{fin}}/\mathcal{N}_{\mathrm{B}}=(\sqrt{2}-1)/2\approx 0.21$. This explains the lower-right panel of Fig.~\ref{figS10} with an additional note that at $T=70$~mK the final density operator that evolved from $|D\rangle\langle B|+|B\rangle\langle D|+|B\rangle\langle B|$ is no longer $|00\rangle\langle00|$, but rather a mixture of $|00\rangle\langle00|$, $|B\rangle\langle B|$, and $|11\rangle\langle11|$, where $|11\rangle=|e\rangle_{\mathrm{NV}_1}|e\rangle_{\mathrm{NV}_2}$.

\begin{figure}[t]
\includegraphics[scale=1]{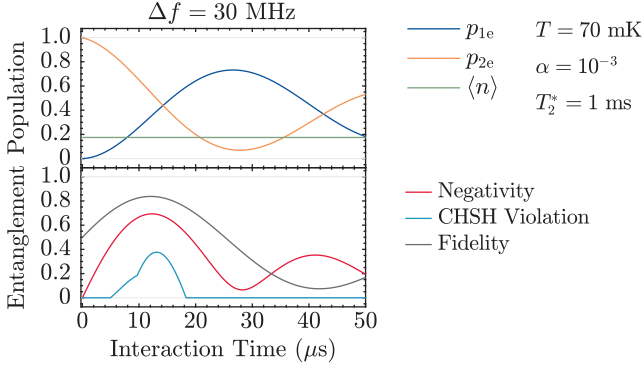}
\caption{Simulation analogous to Fig.~\ref{fig5} for the off resonance case under a larger Gilbert damping parameter $\alpha=10^{-3}$ and larger frequency detuning $\Delta f=30$~MHz.}
\label{figS11} 
\end{figure}

To mitigate the magnon-induced $T_1$ decay in the case of the larger Gilbert damping parameter, one can make the detuning $\Delta f$ larger. Although in our case this is limited by the frequency spacing of the neighboring magnon modes [see Fig.~\ref{fig4}(b)], we show in Fig.~\ref{figS11} the simulation with a larger detuning value $\Delta f=30$~MHz. We note, however, that this is not possible for the magnonic system we have considered in the main text as the neighboring magnon-mode frequency separations are smaller than $30$~MHz [See Fig.~4(b)] in the main text. Conversely, this simulation clarifies that the system will make useful entanglement that can violate the Bell inequality. This implies that to improve the quality of the resulting entanglement further optimization on the length $l$ of the magnetic bar structure is needed, as it defines the frequency spacing of magnon modes.

\section{Longitudinal decay of NV center due to the coupling to magnon modes}
In this section, we evaluate the longitudinal decay contribution of the magnon modes on the NV center placed on top of the YIG bar under the conditions shown in Fig.~\ref{fig5}. Although in the case where two NV centers exist, there are collective decay contribution (Purcell relaxation or Purcell decay~\cite{Benito2019}) described by extra Lindblad terms e.g., $\mathcal{L}[ \sigma^{\pm}_{\mathrm{NV}_1}+\sigma^{\pm}_{\mathrm{NV}_2}]$, we do not take this effect into account for simplicity.

\begin{figure}[t]
\includegraphics[scale=1]{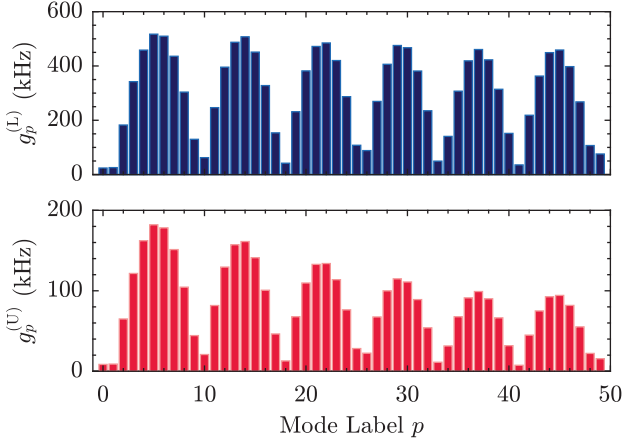}
\caption{NV-magnon coupling strength $g_{p}^{(X)}$ {as a function of the mode label $p$} for the {lower} ($|0\rangle\leftrightarrow |-1\rangle$, $X=\rm{L}$) and the {upper} ($|0\rangle\leftrightarrow |+1\rangle$, $X=\rm{U}$) NV center's transitions.}
\label{figS12} 
\end{figure}

In order to also take into account the effect of NV center's upper frequency transition {($|0\rangle\leftrightarrow |+1\rangle$)} {on the longitudinal NV center decay}, we redefine the coupling in Eq.~(\ref{YIGbarCoupling}) as (with $X=\mathrm{L,U}$ {representing} the {lower} and the {upper} {frequency} transitions of the NV centers, respectively):
\begin{eqnarray}
&&\mathcal{H}^{(X)}_{\mathrm{int}}=\sum_{p}\hbar g_p^{(X)}\sigma^{+}_{\mathrm{NV}(X)}\beta_{(00p)}+\mathrm{H.c.},\\
&&g_p^{(\mathrm{L})}=\sqrt{\omega_{M} \omega_{d w l}}\times\nonumber\\
&&\quad \sum_{q=0,1, \cdots, N}\left.\left[\left(\Gamma_{(00 q)}^{+,+}/2\right)\left[\mathrm{T}^{p p}\right]_{q p}+\left(\Gamma_{(00 q)}^{+,-}/2\right)\left[\mathrm{T}^{n p}\right]_{q p}\right]\right|_{\mathbf{r}=\mathbf{r}_{\mathrm{NV}}},\nonumber\\ \\
&&g_p^{(\mathrm{U})}=\sqrt{\omega_{M} \omega_{d w l}}\times\nonumber\\
&&\quad\sum_{q=0,1, \cdots, N}\left.\left[\left(\Gamma_{(00 q)}^{-,+}/2\right)\left[\mathrm{T}^{p p}\right]_{q p}+\left(\Gamma_{(00 q)}^{-,-}/2\right)\left[\mathrm{T}^{n p}\right]_{q p}\right]\right|_{\mathbf{r}=\mathbf{r}_{\mathrm{NV}}},\nonumber\\
\end{eqnarray}
where $\sigma^{+}_{\mathrm{NV}(\mathrm{L})}=|-1\rangle \langle 0|$ and $\sigma^{+}_{\mathrm{NV}(\mathrm{U})}=|+1\rangle \langle 0|$. Under the condition where the NV center is placed at the cross marker position in Fig.~\ref{fig4}(c), the coupling strength as a function of the {magnon} mode label $p$ is shown in Fig.~\ref{figS12}. The difference {in strength} between $g_{p}^{(\mathrm{L})}$ and $g_{p}^{(\mathrm{U})}$ is due to the {smaller character of} circular polarization of the magnetic field generated by {our length} magnon modes~\cite{rustagi2020}.

\begin{figure}[t]
\includegraphics[scale=1]{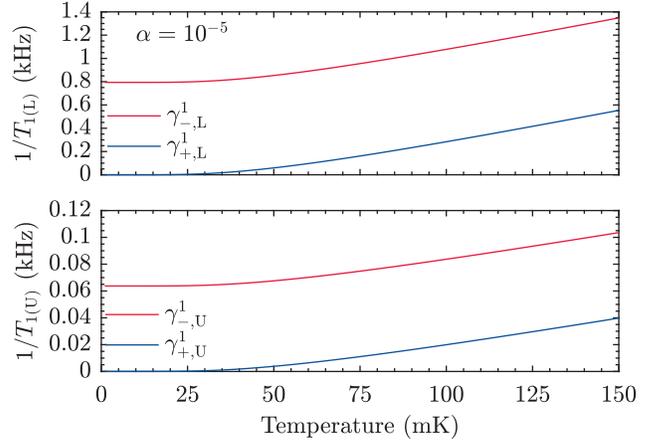}
\caption{Longitudinal ($T_1$) decay rates of NV centers for the NV center's lower and upper transitions due to the coupling to magnon modes other than $p=5$, under the condition where NV center's lower transition frequency is on-resonant to the magnon mode with $p=5$ as calculated in Fig.~\ref{fig5}. The Gilbert damping parameter $\alpha=10^{-5}$ is used.}
\label{figS13} 
\end{figure}

Now we consider the contribution {of the coupling with magnon modes other than $p=5$} to the NV center's longitudinal ($T_1$) decay rates. When we write $B^{-}=\sum_p \hbar g_p^{(X)}\beta_{(00p)}$ and $B^{+}=(B^{-})^{\dagger}$ under the Markov approximation, we obtain the dissipation (non-Hermitian evolution) terms
\begin{eqnarray}
&&\partial_t\rho|_{\mathrm{diss}}=\gamma_{-,{X}}^{1}\mathcal{L}[\sigma_{\mathrm{NV}(X)}^{-}]\rho+\gamma_{+,{X}}^{1}\mathcal{L}[\sigma_{\mathrm{NV}(X)}^{+}]\rho,\\
&&\gamma_{-,{X}}^{1}=\int_{-\infty}^{\infty} d t e^{i \Omega_{X} t}\left\langle B^{-}(t) B^{+}(0)\right\rangle/\hbar^2,\\
&&\gamma_{+,{X}}^{1}=\int_{-\infty}^{\infty} d t e^{i \Omega_{X} t}\left\langle B^{+}(0) B^{-}(t)\right\rangle/\hbar^2,
\end{eqnarray}
where $B^{\pm}(t)$ is {written} in the interaction picture, $\langle \cdots \rangle=Tr[\rho_{\mathrm{m}}\cdots]$, $\rho_{\mathrm{m}}$ is the thermal magnon density operator, and $\Omega_{\mathrm{L}}=D_{\mathrm{NV}}-\gamma H_\mathrm{ext}$ for the {lower} frequency transition ($X=\mathrm{L}$) and $\Omega_{\mathrm{U}}=D_\mathrm{NV}+\gamma H_\mathrm{ext}$ for the upper frequency transition ($X=\mathrm{U}$), respectively. Assuming $\left\langle\beta_{\mu}(t) \beta_{v}^{\dagger}(0)\right\rangle=\left\langle\beta_{\mu}(0) \beta_{v}^{\dagger}(0)\right\rangle e^{-i \omega_{\mu} t-|\kappa| t}$ and $\left\langle\beta_{\mu}^{\dagger}(0) \beta_{v}(t)\right\rangle=\left\langle\beta_{\mu}^{\dagger}(0) \beta_{v}(0)\right\rangle e^{-i \omega_{\mu} t-|\kappa| t}$ with $\kappa=\alpha\omega_{\mu}$, we obtain~\cite{Blais2004,Benito2019}
\begin{eqnarray}
\gamma_{-,{X}}^{1}&=&\sum_{\substack{\mu=(00p)\\p=0,1,\cdots}}\left|g_{p}^{(X)}\right|^{2} \frac{\left(n_{\mathrm{B}}\left(\omega_{\mu}\right)+1\right)\cdot2 \kappa}{\left(\Omega_{{X}}-\omega_{\mu}\right)^{2}+\kappa^{2}},\nonumber\\
&\approx& \sum_{\substack{\mu=(00p)\\p=0,1,\cdots}}\left|g_{p}^{(X)}\right|^{2} \frac{\left(n_{\mathrm{B}}\left(\omega_{\mu}\right)+1\right)\cdot 2 \kappa}{\left(\Omega_{{X}}-\omega_{\mu}\right)^{2}},\\
\gamma_{+,{X}}^{1}&=&\sum_{\substack{\mu=(00p)\\p=0,1,\cdots}}\left|g_{p}^{(X)}\right|^{2}  \frac{n_{\mathrm{B}}\left(\omega_{\mu}\right)\cdot2 \kappa}{\left(\Omega_{{X}}-\omega_{\mu}\right)^{2}+\kappa^{2}},\nonumber\\
&\approx& \sum_{\substack{\mu=(00p)\\p=0,1,\cdots}}\left|g_{p}^{(X)}\right|^{2}   \frac{n_{\mathrm{B}}\left(\omega_{\mu}\right)\cdot 2\kappa}{\left(\Omega_{{X}}-\omega_{\mu}\right)^{2}},
\end{eqnarray}
where $n_{\mathrm{B}}(\omega)=[\mathrm{exp}(\hbar \omega/k_{\mathrm{B}} T)-1]^{-1}$ is the Bose-Einstein distribution function and we have approximated $(\Omega_{{\mathrm{L}/\mathrm{U}}}-\omega_\mu)\gg \kappa$ to obtain the last expressions. Note that for the lower frequency transition, we do not include $p=5$ in the summation as this is the on-resonant magnon mode and its effect is directly included in the simulation in Fig.~\ref{fig5}. With the Gilbert damping parameter $\alpha=10^{-5}$, we evaluated the above  expression and obtained Fig.~\ref{figS13}. As the calculated relaxation time is much longer than the time scale that is simulated in Fig.~5, this $T_1$ decay contribution from magnon modes other than $p=5$ is negligible for the condition we considered.

\section{NV-magnon coupling strength under parametric driving of magnon modes}
We comment that the discretized magnon-mode levels studied in Figs.~\ref{fig4} and \ref{fig5} may enable a control of the NV-magnon coupling strength under parametric driving of magnons via the squeezing effect that is studied in cavity quantum electrodynamics~\cite{Leroux2018}. When a modulated external magnetic field is applied along the $z$-axis, $h^{\mathrm{mod}}_z(t)\hat{z}$, we obtain the additional magnon Hamiltonian $\mathcal{H}_{\mathrm{m}}^{\mathrm{mod}}=-\mu_{0}h^{\mathrm{mod}}_z(t)\int d{\bf r}\delta M_z({\bf r})$, which includes terms proportional to $h_{z}^{\mathrm{mod}}(t)\beta_{\mu}^2+\mathrm{H.c.}$ due to the Bogoliubov transformation. In this respect, the control of the NV-magnon coupling strength can be performed by modulating $h_z(t)$ with a frequency near $2\omega_{\mu}$ in analogy to the parametric excitations of magnons under parallel pumping~\cite{stancil2009spin}.

\section{Periodic modulation of the external magnetic field}
In Fig.~\ref{fig5}, we have considered a modulation of the NV-center transition frequencies with respect to the magnon-mode frequency to generate entanglement between NV centers. Alternatively, the NV-center or the magnon-mode frequency can also be controlled by a periodic modulation of the external magnetic field $h_{z}^{\mathrm{mod}}(t)$ with frequency near the detuning frequency $\delta \omega$~\cite{Oliver1653,XufengPRL2020}. In Ref.~[\onlinecite{XufengPRL2020}], interaction between photons in a microwave cavity and magnos in a bulk YIG sphere under a periodic modulation of the $z$-directional external field is experimentally studied with a use of Floquet theory. In Ref.~[\onlinecite{Oliver1653}] and others, it has been studied that the periodic modulation of qubit transition frequencies results in sideband transitions known as Landau-Zener-Stückelberg interference. Although these may enable different protocols of entangling NV centers under the ac modulation of the external magnetic field, this is beyond the scope of this work.

\section{Effect of nonuniform local magnetic field at YIG}
When we consider the case where multiple NV centers are placed on top of the YIG waveguide, we mentioned in the main text that one can use local magnetic field to change the NV centers' frequencies with respect to the magnon mode's lowest frequency. We note, however, that there would be an unavoidable and undesirable local magnetic field ${\bf{h}}^{(2)}_{\mathrm{ext}}({\bf{r}})$ at the underlying YIG location, the effect of which can be captured by an additional magnon Hamiltonian
\begin{eqnarray}
\mathcal{H}_{\text{m}}^{(2)}=-\mu_0\int d{\bf{r}} {\bf{h}}^{(2)}_{\mathrm{ext}}({\bf{r}})\cdot {\bf{M}}({\bf{r}}).
\end{eqnarray}
Although we do not fully study the nontrivial effect of $\mathcal{H}_{\text{m}}^{(2)}$ on the magnon transport properties in the YIG waveguides and bars, as the effect can be mitigated by using local electric field~\cite{electricfieldNV2011} or strain~\cite{PhysRevLett.113.020503} instead, we note that it can directly be calculated for the finite-length YIG bar case through the diagonalization of the magnon Hamiltonian (Sec.~IV). Alternatively, in the following subsections, we briefly discuss a perturbative approach to consider the effect of the nonuniform local magnetic field on our YIG bar and waveguide cases. To this end, we consider the $z$-directional magnetic field contribution only, as we only need a $z$-directional magnetic field to shift NV centers' frequencies. Therefore, what we consider in this section is the effect of the following Hamiltonian
\begin{eqnarray}
\mathcal{H}_{\text{m}}^{(2)}&=&-\mu_0\int d{\bf{r}} {h}^{(2)}_{z,\mathrm{ext}}({\bf{r}}) \delta M_z ({\bf{r}}),\nonumber\\
&=&\gamma\mu_0 h_{\mathrm{ext}}\int d{\bf{r}} {\bar{h}}({\bf{r}}) a^*({\bf{r}})a({\bf{r}}),\nonumber\\
&=&\omega_{h}\int d{\bf{r}} {\bar{h}}({\bf{r}}) a^*({\bf{r}})a({\bf{r}}),
\end{eqnarray}
where we write ${h}^{(2)}_{z,\mathrm{ext}}({\bf{r}})=h_{\mathrm{ext}}{\bar{h}}({\bf{r}})$ with a dimensionless function ${\bar{h}}({\bf{r}})$ describing the position dependence of the nonuniform magnetic field and $\omega_{h}=\gamma\mu_0 h_{\mathrm{ext}}$ is the frequency scale corresponding to the strength of the local nonuniform magnetic field.

\subsection{Perturbative approach to the YIG waveguide case}
In this section, we consider the case of the infinitely long YIG waveguide. Following the expansion Eqs.~(\ref{WGexp_a}) and (\ref{WGexp_astar}), using the Bogoliubov transformation Eqs.~(\ref{BogoStandard_B})-(\ref{BogoWG}), and considering the magnon modes with $(n,m)=(0,0)$ only, we obtain
\begin{widetext}
\begin{eqnarray}
\mathcal{H}_{\text{m}}^{(2)}=\hbar\omega_{h} \int \frac{d k}{2 \pi} \int \frac{d k^{\prime}}{2 \pi} \mathcal{D}_{k-k^{\prime}}\left(-\mu_{k,(0,0)}^{*} \beta_{-k,(0,0)}+\lambda_{k,(0,0)} \beta_{k,(0,0)}^{\dagger}\right)\left(\lambda_{k^{\prime},(0,0)} \beta_{k^{\prime},(0,0)}-\mu_{k^{\prime},(0,0)} \beta_{-k^{\prime},(0,0)}^{\dagger}\right),\nonumber\\
\end{eqnarray}
\end{widetext}
where
\begin{eqnarray}
\mathcal{D}_{k-k^{\prime}} = \int d \mathbf{r} {\bar{h}}({\bf{r}}) e^{-i\left(k-k^{\prime}\right) z}\left(f_{0}^{X}(x) f_{0}^{Y}(y)\right)^{2},
\end{eqnarray}
has the dimension of length. To know the effect of this additional perturbative Hamiltonian $\mathcal{H}_{\text{m}}^{(2)}$ on the NV-NV effective coupling strength, we can consider how $\mathcal{H}_{\text{m}}^{(2)}$ modifies the advanced Greens functions [see Eq.(\ref{HeffFull})], defined by
\begin{eqnarray}
i G_A(t)=-\theta(-t)\langle[\beta_{k,(0,0)}(t),\beta^\dagger_{k',(0,0)}(0)]\rangle_{\mathrm{Heis}},\label{AdvGreenFct}
\end{eqnarray}
where the subindex {``}Heis{''} indicates that the operators inside the bracket are in the Heisenberg picture, i.e., {the} dynamics {of our system} is governed by $\mathcal{H}_{\mathrm{m}}+\mathcal{H}_{\mathrm{m}}^{(2)}$. To evaluate the effect of $\mathcal{H}_{\text{m}}^{(2)}$ perturbatively, one can use a standard diagrammatic perturbation theory. For example{,} at $T=0$, one can calculate the left-hand side of Eq.~(\ref{AdvGreenFct}) by first calculating the time-ordered Green's function~\cite{coleman2015introduction}, 
\begin{eqnarray}
i G(t)&=&\langle \mathcal{T} \beta_{k,(0,0)}(t)\beta^\dagger_{k',(0,0)}(0)\rangle_{\mathrm{Heis}},\nonumber\\
&=&\frac{\langle 0|\mathcal{T}S(\infty)\beta_{k,(0,0)}(t)\beta^\dagger_{k',(0,0)}(0)|0\rangle}{\langle 0| S(\infty)|0\rangle} ,\label{ManyBodyTheory}\\
S(\infty)&=&\mathcal{T}\mathrm{exp}\left[-\frac{i}{\hbar}\int_{-\infty}^{+\infty}dt' \mathcal{H}_{\text{m}}^{(2)}(t')\right],
\end{eqnarray}
where $\mathcal{T}$ represents the time-ordered product and operators without subindex ``Heis'' are in the interaction picture. We have considered a standard treatment of gradually turning on and off the interaction $\mathcal{H}_{\mathrm{m}}^{(2)}$ at infinitely early and late times. Then we obtain the retarded Green's function by shifting the position of the pole in the frequency domain. The lowest order contribution in Eq.~(\ref{ManyBodyTheory}) is
\begin{eqnarray}
&&\langle 0 | \mathcal{T}\beta_{k,(0,0)}(t)\beta^{\dagger}_{k',(0,0)}(0)|0\rangle=\theta(t)e^{-i\omega_{k,(0,0)}}\cdot 2\pi\delta(k-k')\nonumber\\
&&=\int\frac{d\omega}{2\pi}e^{-i\omega t} iG_0(\omega,k)\cdot 2\pi\delta(k-k'),
\end{eqnarray}
where we defined
\begin{eqnarray}
G_0(\omega,k)=\frac{1}{\omega-\omega_{k,(0,0)}+i0}.
\end{eqnarray}
The next order contribution is, using Wick's theorem,
\begin{widetext}
\begin{eqnarray}
&&\left\langle\left. \mathcal{T}\left(-i \int_{-\infty}^{\infty} d t^{\prime} \mathcal{H}_{\mathrm{m}}^{(2)}\left(t^{\prime}\right)\right) \beta_{k,(0,0)}(t) \beta_{k^{\prime},(0,0)}^{\dagger}(0)\right\rangle_{\text {conn}}\right.\nonumber\\
&&=(-i) \int_{-\infty}^{\infty} d t^{\prime} \omega_{h} \int \frac{d k_{1}}{2 \pi} \int \frac{d k_{2}}{2 \pi} \widetilde{\mathcal{D}}_{k_{1}, k_{2}}\left\langle 0\left|\mathcal{T} \beta_{k,(0,0)}(t) \beta_{k_{1},(0,0)}^{\dagger}\left(t^{\prime}\right)\right|0\right\rangle\left\langle 0\left|\mathcal{T} \beta_{k_{2},(0,0)}\left(t^{\prime}\right) \beta_{k^{\prime},(0,0)}^{\dagger}(0)\right|0\right\rangle\nonumber\\
&&=(-i) \int_{-\infty}^{\infty} d t^{\prime} \omega_{h} \widetilde{\mathcal{D}}_{k, k^{\prime}}\left(\int \frac{d \omega}{2 \pi} e^{-i \omega\left(t-t^{\prime}\right)} i G_{0}(\omega, k)\right)\left(\int \frac{d \omega^{\prime}}{2 \pi} e^{-i \omega^{\prime} t^{\prime}} i G_{0}\left(\omega^{\prime}, k^{\prime}\right)\right)\nonumber\\
&&=\int \frac{d \omega}{2 \pi} e^{-i \omega t} i G_{0}(\omega, k) \omega_{h} \widetilde{\mathcal{D}}_{k, k^{\prime}} G_{0}\left(\omega, k^{\prime}\right),
\end{eqnarray}
\end{widetext}
where {the subindex ``}conn{''} indicates the connected diagrams and 
\begin{eqnarray}
\widetilde{\mathcal{D}}_{k, k^{\prime}} = \mathcal{D}_{k-k^{\prime}}( \lambda_{k,(0, 0)} \lambda_{k^{\prime},(0,0)}+ \mu_{k,(0,0)} \mu_{k^{\prime},(0,0)}^{*}),\nonumber\\
\end{eqnarray}
which has the dimension of length. Therefore, we obtain
\begin{eqnarray}
&&iG(\omega)=\int dt e^{i\omega t} iG(t)\nonumber\\
&&\approx iG_0(\omega,k)2\pi\delta(k-k')+iG_0(\omega,k)\omega_{h}\widetilde{\mathcal{D}}_{k, k^{\prime}} G_{0}\left(\omega, k^{\prime}\right).\nonumber\\ \label{lowestWick}
\end{eqnarray}
According to Eqs.~(\ref{HeffFull}) and (\ref{geffWG}), the effective NV-NV interaction is related to the $\omega=\omega_{\mathrm{NV}}$ contribution of the Green's function $G_A(\omega_{\mathrm{NV}})$. Assuming $\widetilde{\mathcal{D}}_{k, k^{\prime}}$ will contribute to the effective NV-NV coupling on the same order as $2\pi \delta(k-k')$ in Eq.~(\ref{lowestWick}) for simplicity to evaluate the scale of the contribution of the perturbation and as they have the same dimension of length, and using $\omega_{h} G_0(\omega_{\mathrm{NV}},k')\sim \omega_{h}/(\omega_{\mathrm{NV}}-\omega_{k',(0,0)})\sim \omega_{h}/(\omega_{\mathrm{min}}-\omega_{\mathrm{NV}})$, the effect of the local magnetic field $h_{\mathrm{ext}}$ on the NV-NV effective coupling, based on Eqs.~(\ref{HeffFull}),(\ref{geffWG}), and (\ref{lowestWick}), is given by
\begin{eqnarray}
\frac{g_{\mathrm{eff}}-\left.g_{\mathrm{eff}}\right|_{h_{\mathrm{ext}}=0}}{\left.g_{\mathrm{eff}}\right|_{h_{\mathrm{ext}}=0}}\sim\omega_{h}G_0(\omega_{\mathrm{NV}},k')\sim \frac{\omega_{h}}{\omega_{\mathrm{min}}-\omega_{\mathrm{NV}}},\nonumber\\
\end{eqnarray}
although further investigation is needed for the full comparison of the two terms in Eq.~(\ref{lowestWick}) as well as for higher order terms.

\subsection{Perturbative approach to the YIG bar case}
In the case of the YIG bar, with the use of Eqs.~(\ref{YIGbar_a}),(\ref{YIGbar_adag}), the Hamiltonian $\mathcal{H}_{\text{m}}^{(2)}$ can be written as
\begin{eqnarray}
&&\mathcal{H}_{\text{m}}^{(2)}=\frac{\omega_{h}}{2} \sum_{\mu_{1}, \mu_{2}}\left[a_{\mu_{1}}^{*}\  a_{\mu_{1}}\right]\left[\begin{array}{cc}{\left[\overline{\mathbf{h}}_{\text {ext}}\right]_{\mu_{1} \mu_{2}}} & \mathbf{O} \\ \mathbf{O} & {\left[\overline{\mathbf{h}}_{\text {ext }}\right]_{\mu_{1} \mu_{2}}}\end{array}\right]\left[\begin{array}{c}a_{\mu_{2}} \\ a_{\mu_{2}}^{*}\end{array}\right],\nonumber\\ \\
&&\left[\overline{\mathbf{h}}_{\text {ext }}\right]_{\mu_{1} \mu_{2}} =\int d \mathbf{r} \bar{h}(\mathbf{r}) f_{\mu_{1}}^{X Y Z}(\mathbf{r}) f_{\mu_{2}}^{X Y Z}(\mathbf{r}).
\end{eqnarray}
Now we define the perturbation Hamiltonian matrix $\lambda\mathbf{V}$ as
\begin{eqnarray}
\lambda[\mathbf{V}]_{\mu_{1} \mu_{2}} \equiv \omega_{h}\left[\begin{array}{cc}\left[\overline{\mathbf{h}}_{\mathrm{ext}}\right]_{\mu_{1} \mu_{2}} & \mathbf{O} \\ \mathbf{O} & {\left[\overline{\mathbf{h}}_{\mathrm{ext}}\right]_{\mu_{1} \mu_{2}}}\end{array}\right].
\end{eqnarray}

In the following, we will consider the effect of $\lambda\mathbf{V}$ in the expansion with the order $\lambda$ for the case of the diagonalization with a paraunitary matrix. We want to diagonalize the total Hamiltonian matrix $\hat{\mathbf{H}}=\hat{\mathbf{H}}_0+\lambda\mathbf{V}$ in the form
\begin{eqnarray}
&&\mathbf{T}^{\dagger} \mathbf{H} \mathbf{T}=\mathbf{\Lambda}=\left[\begin{array}{ll}\mathbf{E} & \mathbf{O} \\ \mathbf{O} & \mathbf{E}\end{array}\right],\label{PertProb1}\\
&&\mathbf{T}^{\dagger} \boldsymbol{\sigma}_{3} \mathbf{T}=\boldsymbol{\sigma}_{3}\label{PertProb2},
\end{eqnarray}
and we assume we know this expansion in the case with $\lambda=0$ as
\begin{eqnarray}
&&\mathbf{T}_0^{\dagger} \mathbf{H}_0 \mathbf{T}_0=\mathbf{\Lambda}_0,\\
&&\mathbf{T}_0^{\dagger} \boldsymbol{\sigma}_{3} \mathbf{T}_0=\boldsymbol{\sigma}_{3}.
\end{eqnarray}
Based on these, we expand the perturbed paraunitary $\mathbf{T}$ and eigenvalues $\mathbf{\Lambda}$ matrices as
\begin{eqnarray}
&&\mathbf{T}=\mathbf{T}_0+\lambda \mathbf{T}_1+\cdots.\\
&&\mathbf{\Lambda}=\mathbf{\Lambda}_0+\lambda \mathbf{\Lambda}_1+\cdots.
\end{eqnarray}
Substituting these into Eqs.~(\ref{PertProb1}) and (\ref{PertProb2}), and taking leading order terms in $\lambda$, we obtain
\begin{eqnarray}
&&\mathbf{\Lambda}_{1}=\sum_{i}|i\rangle\left[\mathbf{T}_{0}^{\dagger} \mathbf{V} \mathbf{T}_{0}\right]_{i i}\langle i|,\label{pertEvalues}\\
&&\mathbf{T}_{1}=-\mathbf{T}_{0} \boldsymbol{\sigma}_{3} \sum_{i \neq j}|i\rangle \frac{\left[\mathbf{T}_{0}^{\dagger} \mathbf{V T}_{0}\right]_{i j}}{\left[\boldsymbol{\sigma}_{3} \mathbf{\Lambda}_{0}\right]_{i i}-\left[\boldsymbol{\sigma}_{3} \mathbf{\Lambda}_{0}\right]_{j j}}\langle j|+\mathbf{T}_{0} \boldsymbol{\sigma}_{3} \mathbf{D},\nonumber\\
\end{eqnarray}
where $\mathbf{D}$ is an arbitrary diagonal matrix with purely-imaginary entries. This is due to the degrees of freedom of the paraunitary matrix $\mathbf{T}\rightarrow \mathbf{T} \mathrm{exp}[i\lambda\times(\mathrm{real\ diagonal\ matrix})]$, which we encounter in the unitary diagonalization case as well. Therefore, we simply set $\mathbf{D}=0$ and obtain
\begin{eqnarray}
&&\mathbf{T}_{1}=\mathbf{T}_0\mathbf{L},\\
&&\mathbf{L}=-\boldsymbol{\sigma}_{3} \sum_{i \neq j}|i\rangle \frac{\left[\mathbf{T}_{0}^{\dagger} \mathbf{V T}_{0}\right]_{i j}}{\left[\boldsymbol{\sigma}_{3} \mathbf{\Lambda}_{0}\right]_{i i}-\left[\boldsymbol{\sigma}_{3} \mathbf{\Lambda}_{0}\right]_{j j}}\langle j|.\label{pertLmatrix}
\end{eqnarray}

As $|\left[\boldsymbol{\sigma}_{3} \mathbf{\Lambda}_{0}\right]_{i i}-\left[\boldsymbol{\sigma}_{3} \mathbf{\Lambda}_{0}\right]_{j j}|=\omega_\mu+\omega_\nu\gg\omega_{h}$ when $[\boldsymbol{\sigma}]_{ii}[\boldsymbol{\sigma}]_{jj}=-1$, we approximately {neglect} the off-block-diagonal sector of $\mathbf{L}$, and write
\begin{eqnarray}
\mathbf{L}\approx\left[\begin{array}{cc}\mathbf{L}^{p p} & \mathbf{O} \\ \mathbf{O} & \mathbf{L}^{n n}\end{array}\right].
\end{eqnarray}
Then the perturbed paraunitary matrix becomes
\begin{eqnarray}
\left[\begin{array}{cc}\mathbf{T}^{p p} & \mathbf{T}^{p n} \\ \mathbf{T}^{n p} & \mathbf{T}^{n n}\end{array}\right]\approx\left[\begin{array}{cc}\mathbf{T}_{0}^{p p} & \mathbf{T}_{0}^{p n} \\ \mathbf{T}_{0}^{n p} & \mathbf{T}_{0}^{n n}\end{array}\right]+\left[\begin{array}{ll}\mathbf{T}_{0}^{p p} & \mathbf{T}_{0}^{p n} \\ \mathbf{T}_{0}^{n p} & \mathbf{T}_{0}^{n n}\end{array}\right]\left[\begin{array}{cc}\mathbf{L}^{p p} & \mathbf{O} \\ \mathbf{O} & \mathbf{L}^{n n}\end{array}\right].\nonumber\\
\end{eqnarray}
Based on Eq.~(\ref{YIGbarCoupling}), we obtain the coupling strength as
\begin{eqnarray}
g_{(00 p)}&=&g_{(00 p)}^{0}+\sum_{q=0,1, \cdots, N} g_{(00 q)}^{0}\left[\mathbf{L}^{p p}\right]_{q p},\\
g^{0}_{(00p)}&=&\sqrt{\omega_{M} \omega_{d w l}}\times\nonumber\\
&&\sum_{q=0,1, \cdots, N}\left[\left(\Gamma_{(00 q)}^{+,+} / 2\right)\left[\mathbf{T}_{0}^{p p}\right]_{q p}+\left(\Gamma_{(00 q)}^{+,-} / 2\right)\left[\mathbf{T}_{0}^{n p}\right]_{q p}\right],\nonumber\\
\end{eqnarray}
where $g^0_{(00p)}$ is the coupling strength we obtained without the perturbation Hamiltonian $\mathcal{H}_{\text{m}}^{(2)}$. From Eqs.~(\ref{pertEvalues}) and (\ref{pertLmatrix}) {with $\mathbf{\Lambda}_1\sim\omega_{h}$ and $\mathbf{L}\sim \omega_{h}/(\omega_{\nu}-\omega_{\mu})$}, we find the following scaling behavior for the change in the magnon mode frequency and the NV-magnon coupling strength due to the local nonuniform magnetic field $h_{\mathrm{ext}}$, 
\begin{eqnarray}
&&\omega_{\mu}-\left.\omega_{\mu}\right|_{h_{\mathrm{ext}}=0}\sim \omega_{h},\\
&&\frac{g_\mu-\left.g_{\mu}\right|_{h_{\mathrm{ext}}=0}}{\left.g_{\mu}\right|_{h_{\mathrm{ext}}=0}}\sim \frac{ \omega_{h}}{\omega_{\nu(\neq\mu)}-\omega_{\mu}}\label{coupscale},
\end{eqnarray}
although Eq.~(\ref{coupscale}) {strongly} depends on how much the additional magnetic field mixes different normal magnon modes, {described by} the off-diagonal components of $\mathbf{T}_{0}^{\dagger} \mathbf{V T}_{0}$.

\normalem
\bibliography{main.bib}
\end{document}